\documentclass[sigconf, 9pt]{acmart}
\acmYear{2025}\copyrightyear{2025}
\setcopyright{acmlicensed}
\acmConference[SIGCOMM '25]{ACM SIGCOMM 2025 Conference}{September 8--11, 2025}{Coimbra, Portugal}
\acmBooktitle{ACM SIGCOMM 2025 Conference (SIGCOMM '25), September 8--11, 2025, Coimbra, Portugal}
\acmDOI{10.1145/3718958.3750465}
\acmISBN{979-8-4007-1524-2/25/09}

\usepackage[english]{babel}
\usepackage{blindtext}

\usepackage{hyperref}
\hypersetup{
  colorlinks=true,
  linkcolor=blue,
  filecolor=magenta,
  urlcolor=cyan,
}

\usepackage{threeparttable}
\usepackage{tikz}
\usepackage{color}
\usepackage{xcolor}
\definecolor{strawberry}{rgb}{1.0, 0.26, 0.64}
\definecolor{ruby}{rgb}{0.88, 0.07, 0.37}
\definecolor{princetonorange}{rgb}{1.0, 0.56, 0.0}

\usepackage{xurl}
\usepackage{enumitem}
\usepackage{epsfig,endnotes}
\usepackage[export]{adjustbox}
\usepackage{tabularx}
\usepackage{appendix}

\usepackage{lineno}
\usepackage{pifont}
\usepackage{algorithm}
\usepackage[noend]{algpseudocode}
\algnewcommand{\LineComment}[1]{\Statex \hspace{\algorithmicindent}\textcolor{gray}{\(\triangleright\) #1}}
\algnewcommand{\FuncInput}[1]{\Statex \textcolor{black}{\(\triangleright\) \textbf{input}\xspace#1}}
\algnewcommand{\FuncOutput}[1]{\Statex \textcolor{black}{\(\triangleright\) \textbf{output}\xspace#1}}
\usepackage{graphicx}
\usepackage{float}
\usepackage{xspace}
\usepackage{multirow}
\usepackage{comment}
\usepackage{fancybox, fancyvrb, calc}
\usepackage{caption}
\usepackage{subcaption}
\usepackage{amsmath}
\usepackage{amsthm}
\theoremstyle{definition}

\usepackage{amssymb}
\usepackage{mathtools}
\usepackage{natbib}
\usepackage{pbox}
\usepackage{wrapfig}
\usepackage{textcomp}
\usepackage[utf8]{inputenc}
\setlist{nolistsep}
\usepackage{hhline}
\usepackage{booktabs}

\usepackage{rotating}
\usepackage{hhline}

\usepackage{epstopdf}
\usepackage[normalem]{ulem}
\usepackage{footnote}
\usepackage{tablefootnote}

\newcommand{\cut}[1]{}

\newcommand{\paraspace}{\vspace{0.03in}}
\newcommand{\parab}[1]{\paraspace\noindent{\bf #1} }

\newcommand{\eg}{e.g., }

\newcommand{\ie}{i.e., }

\newcommand{\ct}[1]{{\texttt{#1}}}

\usepackage{hyperref}
\newcommand{\secref}[1]{\hyperref[#1]{\S\ref*{#1}}}
\newcommand{\figref}[1]{\hyperref[#1]{Figure~\ref*{#1}}}
\newcommand{\tabref}[1]{\hyperref[#1]{Table~\ref*{#1}}}
\newcommand{\algoref}[1]{\hyperref[#1]{Algorithm~\ref*{#1}}}

\usepackage{pifont}

\makeatletter
\def\@mkauthors@iii{
\author@bx@wd=\textwidth\relax
\advance\author@bx@wd by -\author@bx@sep\relax
\ifnum\@ACM@authorsperrow>0\relax
\divide\author@bx@wd by \@ACM@authorsperrow\relax
\else
\ifcase\num@authorgroups
\relax 
\or  
\or  
\divide\author@bx@wd by \num@authorgroups\relax
\or  
\divide\author@bx@wd by \num@authorgroups\relax
\or  
\divide\author@bx@wd by 2\relax
\else 
\divide\author@bx@wd by 3\relax
\fi
\fi
\advance\author@bx@wd by -\author@bx@sep\relax
\gdef\@currentauthors{}
\gdef\@currentaffiliation{}
\def\@author##1{\ifx\@currentauthors\@empty
\gdef\@currentauthors{\par##1}
\else
\g@addto@macro\@currentauthors{\par##1}
\fi
\gdef\and{}}
\def\email##1##2{\ifx\@currentaffiliation\@empty
\gdef\@currentaffiliation{\bgroup
\mathchardef\UrlBreakPenalty=10000\nolinkurl{##2}\egroup}
\else
\g@addto@macro\@currentaffiliation{\par\bgroup
\mathchardef\UrlBreakPenalty=10000\nolinkurl{##2}\egroup}
\fi}
\def\affiliation##1##2{\ifx\@currentaffiliation\@empty
\gdef\@currentaffiliation{
  \setkeys{@ACM@affiliation@}{obeypunctuation=false}
  \setkeys{@ACM@affiliation@}{##1}\@ACM@resetaffil
##2\@ACM@checkaffil}
\else
\g@addto@macro\@currentaffiliation{\par
  \setkeys{@ACM@affiliation@}{obeypunctuation=false}
  \setkeys{@ACM@affiliation@}{##1}\@ACM@resetaffil
##2\@ACM@checkaffil}
\fi
\global\let\and\@typeset@author@bx
}
\hsize=\textwidth
\global\setbox\mktitle@bx=\vbox{\noindent
\unvbox\mktitle@bx\par\medskip\leavevmode
\lineskip=0.85pc\relax\centering\hspace*{-1em}
\addresses\let\and\@typeset@author@bx\and\par\bigskip}}
\makeatother

\definecolor{mygreen}{rgb}{0,0.8,0.4}
\newcommand{\sigcomm}[1]{#1}
\newcommand{\arxiv}[1]{#1}

\newcommand{\sys}{{\sc{MixNet}}\xspace}
\newcommand{\sysbf}{\bfseries{\scshape{MixNet}}\xspace}
\newcommand{\copilot}{{\sc{MixNet}}\xspace-\textsc{Copilot}}
\newcommand{\mixtral}{Mixtral 8$\times$7B\xspace}
\newcommand{\mixtrallarge}{Mixtral 8$\times$22B\xspace}
\newcommand{\llamamoe}{LLaMA-MoE\xspace}
\newcommand{\qwenmoe}{Qwen-MoE\xspace}
\newcommand{\deepseek}{DeepSeek-R1\xspace}



\begin{document}
\date{}

\title{\textsc{MixNet}: A Runtime Reconfigurable Optical-Electrical Fabric for Distributed Mixture-of-Experts Training}

\newcommand{\hkust}{Hong Kong University of Science and Technology}
\author{Xudong Liao}
\affiliation{
  \institution{\hkust{}}
  \city{}
  \country{}
}
\email{}

\author{Yijun Sun}
\affiliation{
  \institution{\hkust{}}
  \city{}
  \country{}
}
\email{}

\author{Han Tian}
\affiliation{
  \institution{\hkust{}}
  \city{}
  \country{}
}
\email{}

\author{Xinchen Wan}
\affiliation{
  \institution{\hkust{}}
  \city{}
  \country{}
}
\email{}

\author{Yilun Jin}
\affiliation{
  \institution{\hkust{}}
  \city{}
  \country{}
}
\email{}

\author{Zilong Wang}
\affiliation{
  \institution{\hkust{}}
  \city{}
  \country{}
}
\email{}

\author{Zhenghang Ren}
\affiliation{
  \institution{\hkust{}}
  \city{}
  \country{}
}
\email{}

\author{Xinyang Huang}
\affiliation{
  \institution{\hkust{}}
  \city{}
  \country{}
}
\email{}

\author{Wenxue Li}
\affiliation{
  \institution{\hkust{}}
  \city{}
  \country{}
}
\email{}

\author{Kin Fai Tse}
\affiliation{
  \institution{\hkust{}}
  \city{}
  \country{}
}
\email{}

\author{Zhizhen Zhong}
\affiliation{
  \institution{Massachusetts Institute of Technology}
  \city{}
  \country{}
}
\email{}

\author{Guyue Liu}
\affiliation{
  \institution{Peking University}
  \city{}
  \country{}
}
\email{}

\author{Ying Zhang}
\affiliation{
  \institution{Meta}
  \city{}
  \country{}
}
\email{}

\author{Xiaofeng Ye}
\affiliation{
  \institution{EmbedWay}
  \city{}
  \country{}
}
\email{}

\author{Yiming Zhang}
\affiliation{
  \institution{Xiamen University}
  \city{}
  \country{}
}
\email{}

\author{Kai Chen}
\affiliation{
  \institution{\hkust{}}
  \city{}
  \country{}
}
\email{}

\renewcommand{\shortauthors}{Xudong Liao et al.}
\renewcommand{\shorttitle}{\sys{}: A Runtime Reconfigurable Optical-Electrical Fabric}
\begin{abstract}
Mixture-of-Expert (MoE) models outperform conventional models by selectively activating different subnets, named \emph{experts}, on a per-token basis. This gated computation generates \emph{dynamic} communications that cannot be determined beforehand, challenging the existing GPU interconnects that remain \emph{static} during distributed training. In this paper, we advocate for a first-of-its-kind system, called \sys{}, that unlocks topology reconfiguration \emph{during} distributed MoE training. 
Towards this vision, we first perform a production measurement study and show that the MoE dynamic communication pattern has \emph{strong locality}, alleviating the need for global reconfiguration. Based on this, we design and implement a \emph{regionally reconfigurable high-bandwidth domain} that augments existing electrical interconnects using optical circuit switching (OCS), achieving scalability while maintaining rapid adaptability. We build a fully functional \sys prototype with commodity hardware and a customized collective communication runtime. Our prototype trains state-of-the-art MoE models with \emph{in-training} topology reconfiguration across 32 A100 GPUs. Large-scale packet-level simulations show that \sys{} achieves performance comparable to a non-blocking fat-tree fabric while boosting the networking cost efficiency (e.g., performance per dollar) of four representative MoE models by 1.2$\times$--1.5$\times$ and 1.9$\times$--2.3$\times$ at 100 Gbps and 400 Gbps link bandwidths, respectively.

\end{abstract}

\maketitle

\section{Introduction}\label{sec:intro}

Mixture-of-Experts (MoE) models~\cite{shazeer2017outrageously, switch-transformers, mixtral, llama-moe, qwen-moe, grok, xverse-moe, deepseek2, deepseekv3,deepseek-r1,qwen25} have gained significant traction in the machine learning community to improve the performance of large language models (LLMs)~\cite{grok, mixtral}. Unlike traditional methods that scale LLMs by stacking dense layers, which leads to a linear increase in computational costs as model sizes expand, MoE models utilize multiple parallel expert layers and activate only a subset of them based on the input token for each training iteration (e.g., xAI discloses that 25\% weights are active in Grok-1~\cite{grok}, \sigcomm{while DeepSeek-V3~\cite{deepseekv3} only activates 37 billion parameters over total 671 billion parameters}). This dynamic approach enables models to grow to large sizes without a proportional cost increase in computation.

However, such dynamic expert activations require \emph{all-to-all} communications in and out of expert layers in each training iteration. 
Among parallelization strategies, \emph{expert parallelism (EP)}, which assigns expert layers to different GPUs, requires a high volume of traffic that is comparable to that of \emph{tensor parallelism (TP)}, and much larger than other parallelisms. Moreover, the token-specific activation of experts in EP results in \emph{temporally non-deterministic} and \emph{spatially non-uniform} communication patterns that vary across training iterations, challenging existing GPU interconnects. 

\arxiv{Today's GPU interconnects contain intra-server \emph{scale-up networks} (e.g.,  NVSwitch~\cite{nvswitch} or NVLink~\cite{nvlink}) and inter-server \emph{scale-out networks} (e.g., Ethernet or Infiniband).
Both of them are currently dimensioned with uniform and static network topologies (e.g., fully-connected crossbar topology for scale-up networks~\cite{nvswitch}, and Clos-style fat-tree for scale-out networks~\cite{fattree,jupiter-rising}). When accommodating the temporal and spatial variations of MoE communication patterns, these fabrics contain over-provisioned full bisection bandwidth that is mostly under-utilized.}
\sigcomm{Some recent proposals on the use of optical circuit switching (OCS) perform topology reconfiguration for spatially non-uniform traffic distributions. However, they assume stable temporal patterns such that no reconfigurations occur during the entire training process~\cite{topoopt, tpuv4}.} 
As a result, these interconnect architectures face bottlenecks, leading to inefficient resource usage and slowdowns in distributed MoE training.

Therefore, to fully unlock the computational advantages of MoE models, we need to design a novel GPU interconnect fabric that is adaptable to dynamic all-to-all communication patterns at runtime. 
\sigcomm{Achieving such adaptability requires the topology to be reconfigurable \emph{during} the distributed MoE training process.}
This is very challenging because today's commodity OCS technologies face fundamental trade-offs between \emph{low reconfiguration latency} (to enable reconfiguration during training) and \emph{high scalability} (to interconnect tens of thousands of GPUs) (more details in \tabref{tab:ocs}).

To understand the problem space, we first perform a comprehensive measurement study in a production GPU cluster to investigate the real-world communication patterns of distributed MoE training. 
Our measurements reveal that although EP generates substantial variability during training, its dynamic range is strictly within an MoE block, creating \emph{strong locality} for all-to-all traffic on a global scale (\S\ref{sec:ideal}).

\begin{figure*}[!t]
\vspace{-0.5em}
    \begin{subfigure}[t]{0.37\linewidth}
        \centering
        \includegraphics[height=12.3em]{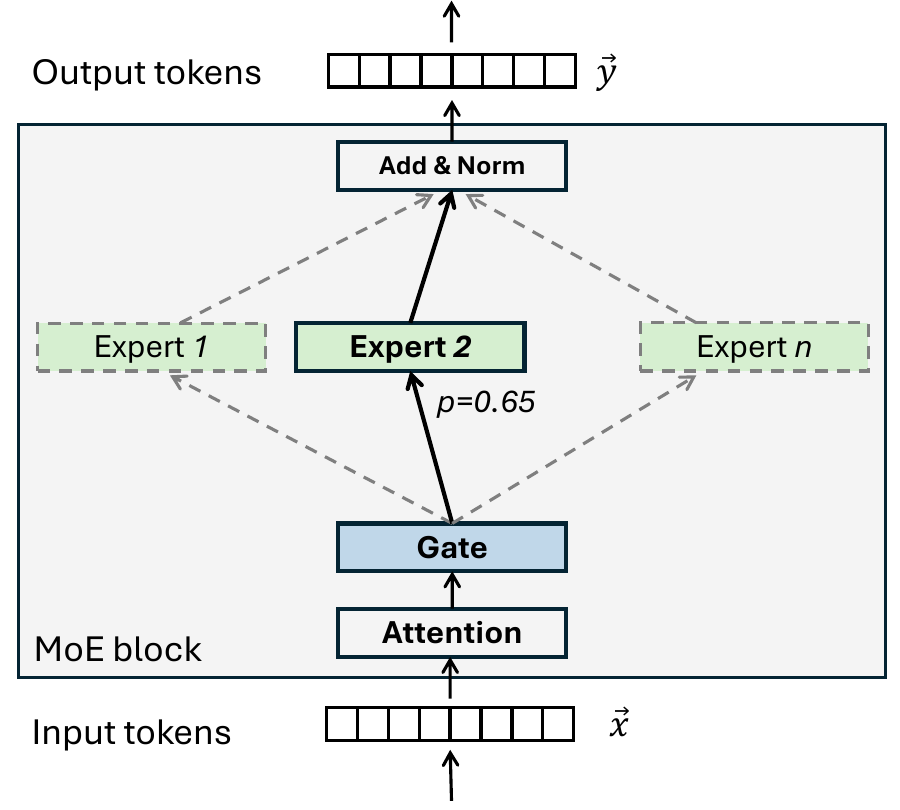}
        \vspace{-0.2em}
        \caption{An MoE block with multiple experts. In this example, the \emph{gate} only activates \emph{Expert 2}.}
        \label{fig:design:layer-forward}
    \end{subfigure}
    \hspace{1em}
    \begin{subfigure}[t]{0.6\linewidth}
        \centering
        \includegraphics[height=11em]{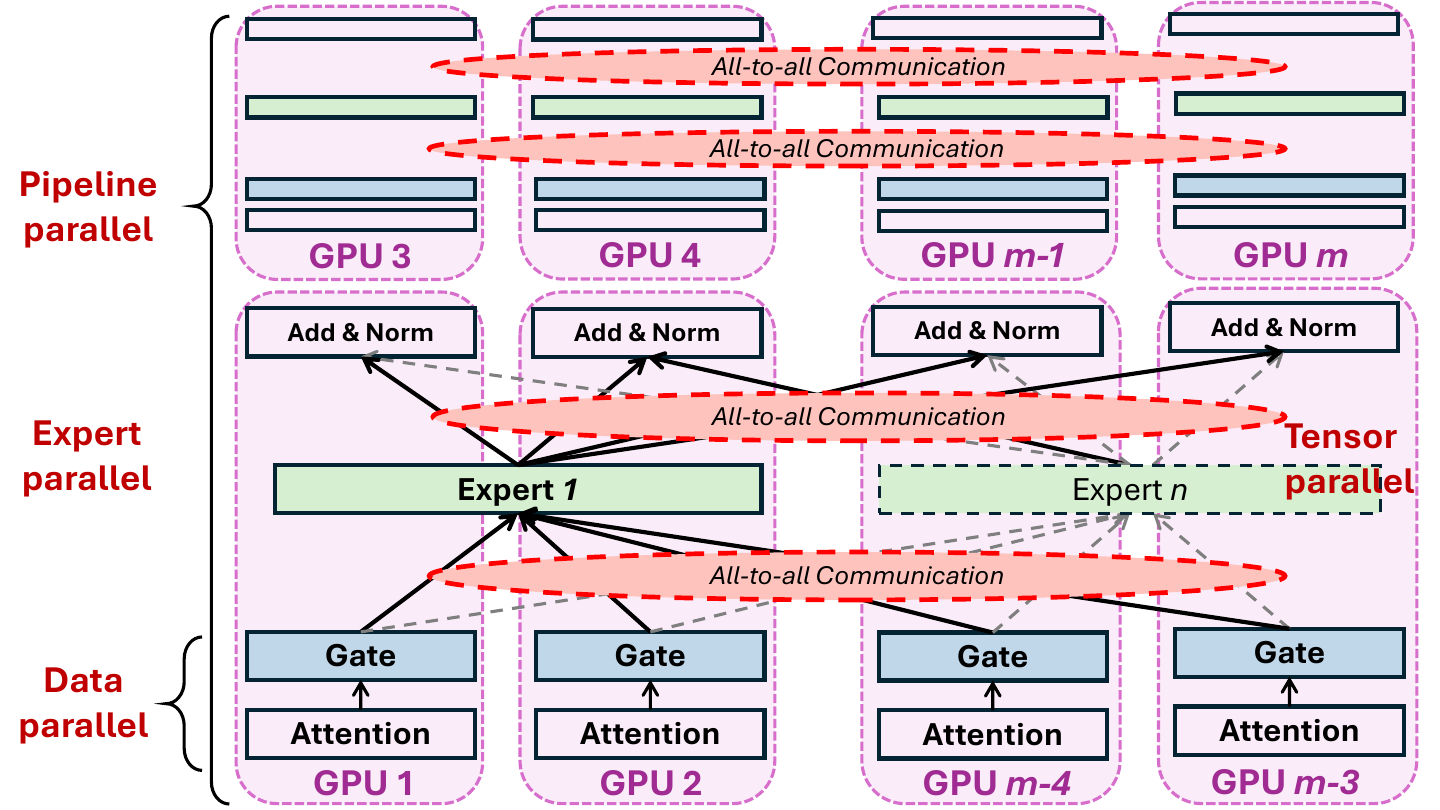}
        \vspace{-0.2em}
        \caption{Example of a hybrid parallelism that combines DP (gate), EP (parallel expert layers), PP (MoE blocks) and TP (a single expert layer).}
        \label{fig:design:moe-distribution}
    \end{subfigure}
    \vspace{-0.8em}
    \caption{Illustration of the MoE's gated expert architecture and its distributed training strategies.}
\end{figure*}

Based on this insight, we introduce \sys{}, a novel system designed to overcome these challenges by enabling efficient topology reconfiguration during distributed MoE training. At the core of \sys{} is a \emph{regionally reconfigurable high-bandwidth domain} 
\sigcomm{based on millisecond-scale reconfigurable OCS that sits at the boundary between scale-up and scale-out networks.}
This design \sigcomm{augments the existing static electrical interconnect with the capability of rapid regional reconfiguration while still preserving its scalability (\S\ref{sec:arch}).}

\sys{} contains the following key components: 1) \sigcomm{\sys{} exploits the partially predictable nature of all-to-all communications to track regional traffic demands at runtime} (\S\ref{sec:design:traffic-demand-prediction}); 2) Based on the obtained demands, \sys{} \sigcomm{uses a greedy algorithm that generates the tailored network topology and reconfigures the OCS to realize it}  (\S\ref{sec:design:topology-generation}); 3) \sigcomm{With the reconfigured topology, \sys{} uses a customized collective communication runtime to orchestrate inter-host DP and EP communications on EPS and regional OCS fabrics (\S\ref{sec:design:traffic-routing}).
}

To demonstrate \sys{}, we build a fully functional prototype with 32 Nvidia A100 GPUs, 16 Mellanox NICs~\cite{100g-nic}, a Polatis millisecond OCS~\cite{polatis}, and an Ethernet switch. 
In addition, we develop a custom collective communication runtime based on Nvidia collective communications library (NCCL)~\cite{nccl} to support in-training topology reconfigurations. \sigcomm{Using this prototype, we successfully demonstrated the benefits of \sys{} on state-of-the-art MoE models (\S\ref{sec:testbed}).}

To evaluate the performance of \sys{} at scale, we perform packet-level simulations using four representative real-world MoE models. Our results reveal that 
\sys{} outperforms the state-of-the-art fabrics, exhibiting a training speed comparable to Rail-optimized~\cite{rail-optimized} and Fat-tree~\cite{fattree} fabrics while significantly improving cost-efficiency.
Specifically, \sys{} improves networking infrastructure cost-efficiency by 1.2$\times$--1.5$\times$ (1.9$\times$--2.3$\times$) compared to Fat-tree and 1.4$\times$--1.5$\times$ (2.3$\times$--2.4$\times$) compared to Rail-optimized for 100Gbps (400Gbps) links. We also observe that \sys{} outperforms TopoOpt~\cite{topoopt} by up to 2.5$\times$ and supports scalability with the cluster size increasing to 30K+ GPUs (\S\ref{sec:sim}). When connected to co-packaged optical ports attached to GPUs, \sys{} augments high-radix scale-up systems like NVL72 by 1.3$\times$ (\S\ref{sec:nvl72}).

For more information, please visit the website: \href{https://mixnet-project.github.io/}{\url{https://mixnet-project.github.io/}}.

\section{Background}\label{sec:motiv}

In this section, we first describe MoE's model architecture and training parallelization strategies (\S\ref{sec:motiv:moe}). Then, we discuss several GPU interconnects for distributed training (\S\ref{sec:motiv:switching}).

\subsection{Distributed MoE Training}\label{sec:motiv:moe}

\parab{MoE model architecture.} MoE models contain several sequential MoE blocks~\cite{shazeer2017outrageously, switch-transformers}. As shown in \figref{fig:design:layer-forward}, each MoE block has an attention layer, a gate unit, and several parallel feed-forward networks (FFNs) called \emph{experts}. The input token $x$ is first fed into the attention layer. After that, the gate unit selects the most relevant experts based on the output of the attention layer. 
This is called computation-based routing and is the key to enabling MoE's sparse architecture that scales model parameters without a linear increase in computation cost. The output token, $y$, is the weighted sum of the outputs of all activated experts. 

\parab{Data Parallelism (DP).} In DP~\cite{pytorch-dist}, the model parameters are replicated across multiple GPUs, and each GPU hosts a different subset of training data. 
Because only gradients are transferred across GPUs, the traffic volume is relatively small compared to other parallelisms. In MoE models, DP is usually applied to gate units and add \& norm layers (\figref{fig:design:moe-distribution}). DP also applies, as we create replicas of the whole model onto several different clusters.

\begin{table}[t!]
    \centering
    \renewcommand{\arraystretch}{1.0}
    \begin{tabular}{c c c c}
        \toprule
        \textbf{Models} & \textbf{Mixtral} & \textbf{LLaMA-MoE} & \textbf{Qwen-MoE} \\
        \textbf{Size} & 8$\times$7B & 6.7B & 14.3B \\
        \midrule
        \# of MoE blocks & 32 & 32 & 24 \\ 
        \# experts & 8 & 16 & 64 \\
        EP degree & 8 & 16 & 16\\
        TP degree & 4 & 1 & 1\\
        PP degree & 4 & 4 & 4\\
        Seq. len. & 4096 & 4096 & 4096 \\
        Micro-batch size & 8 & 8 & 8\\
        \bottomrule
    \end{tabular}
    \caption{State-of-the-art MoE training configurations.}
    \label{tab:measure-setting}
\end{table}

\begin{figure}[t!]
    \vspace{-1em}
    \centering
    \includegraphics[width=0.8\linewidth]{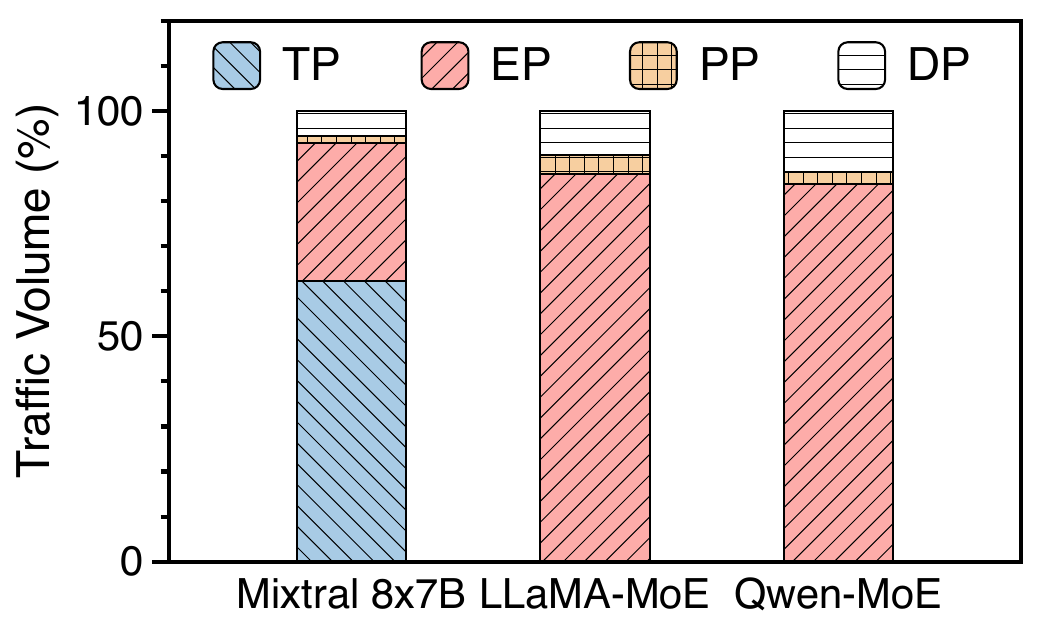}
    \vspace{-0.5em}
    \caption{Traffic volume distribution of different parallelism of three state-of-the-art MoE models. 
    }
    \label{fig:motiv:traffic-dist}
\end{figure}

\parab{Tensor Parallelism (TP).} TP~\cite{megatron-lm} is a technique to partition a layer among multiple GPUs. In distributed MoE training (\figref{fig:design:moe-distribution}), the expert layers are usually partitioned across different GPUs, with intermediate hidden states being transferred through collective communication primitives like broadcast, all-gather and reduce-scatter. 
Therefore, TP is the most communication-intensive operation, 
and its spatial scale is generally limited to a few GPUs in practice~\cite{megatron-lm}.

\parab{Pipeline Parallelism (PP).} In PP~\cite{pipedream, megatron-lm}, multiple sequential stages of the model are distributed to different GPUs (\figref{fig:design:moe-distribution}). Therefore, only hidden activation states are transferred through point-to-point all-reduce collective communication primitives, generating the least amount of communication with deterministic volume. 

\begin{table}[t!]
    \centering
    \fontsize{8}{10}\selectfont
    \renewcommand{\arraystretch}{1.1}
    \begin{tabular}{c c c}
    \toprule
    \textbf{Commodity OCS} & \textbf{Port count} & \textbf{Reconfig. delay}\\
    \midrule
    Robotic (Telescent)~\cite{topoopt} & 1008$\times$1008 & Several minutes\\
    Piezo (Polatis)~\cite{polatis} & 576$\times$576 & 10-25~$ms$\\
    3D MEMS (Calient)~\cite{1296-mems,calient-mems} & 320$\times$320 & 10-15 $ms$ \\
    2D MEMS (Google Palomar)~\cite{liu2023lightwave} & 136$\times$136 & Not reported \\
    RotorNet (InFocus)~\cite{rotornet,realizing-rotornet} & 128$\times$128 & 10~$\mu s$\\
    Silicon Photonics (Lightmatter)~\cite{lightmatter-passage} & 32$\times$32 & 7~$\mu s$\\
    PLZT (EpiPhotonics)~\cite{PLZT} & 16$\times$16 & 10~$ns$\\
    \bottomrule
    \end{tabular}
    \caption{Tradeoff between port count and reconfiguration delay in commodity OCS technologies.}
    \label{tab:ocs}
\end{table}

\begin{figure}[t!]
    \centering
    \vspace{-1.5em}
        \centering
        \includegraphics[width=\linewidth]{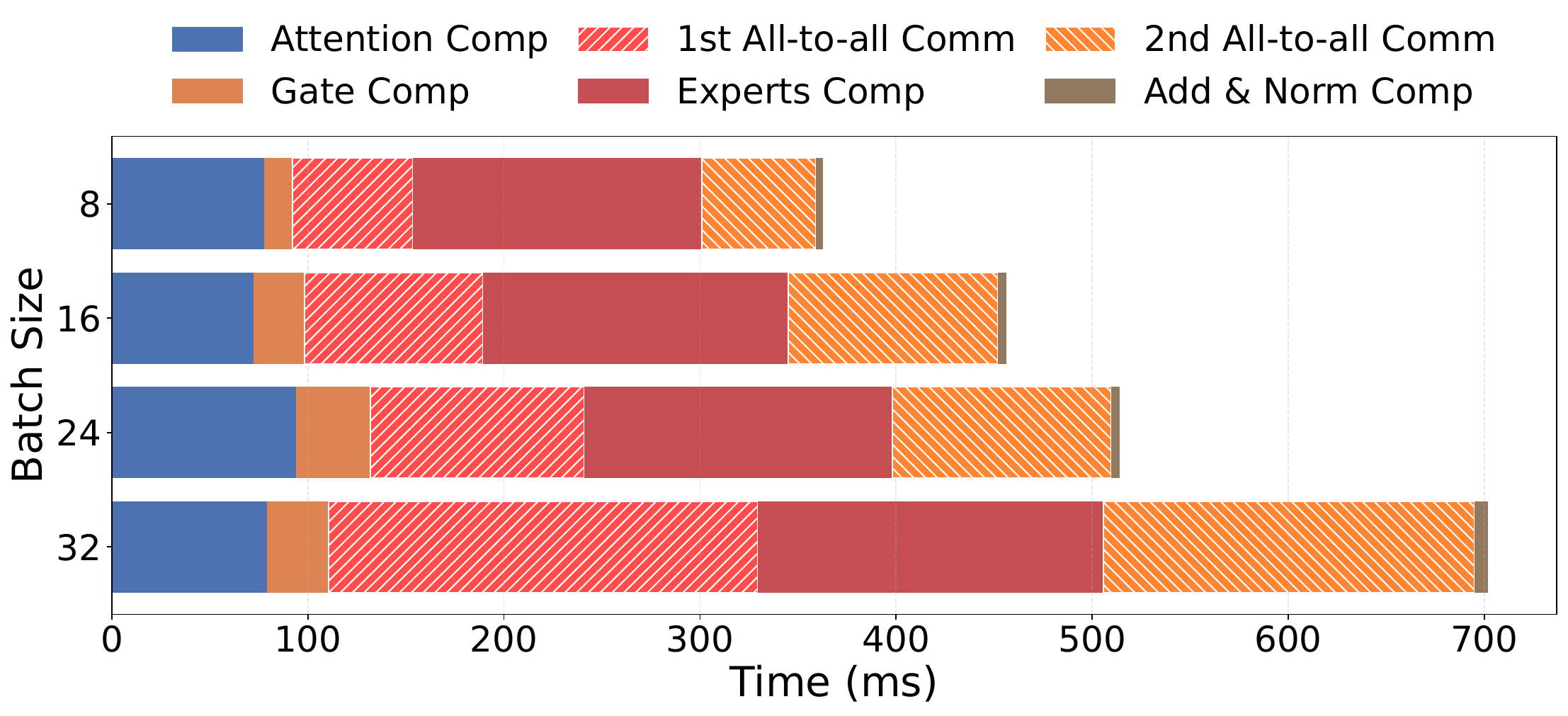}
        \vspace{-2em}
    \caption{[\mixtral{} in production] EP's all-to-all communications occupy 33\% to 55\% of the total training iteration time in 400 Gbps network.}
        \label{fig:motiv:batch-timeline}
\end{figure}

\parab{Expert Parallelism (EP).} In MoE models, different 
experts in an MoE block are allocated to different GPUs~\cite{switch-transformers,gshard,janus,li2024understanding} (\figref{fig:design:moe-distribution}). Since each GPU needs to send its local states to other experts and receive remote states from other GPUs, the dispatching of intermediate hidden states and the collection of expert outputs are performed via two all-to-all communications. 
EP's all-to-all communication is non-uniform and non-deterministic across different training iterations (\S\ref{sec:ideal}). 

\begin{figure*}[t!]
    \centering
    \begin{subfigure}{\textwidth}
        \centering
\includegraphics[width=\linewidth]{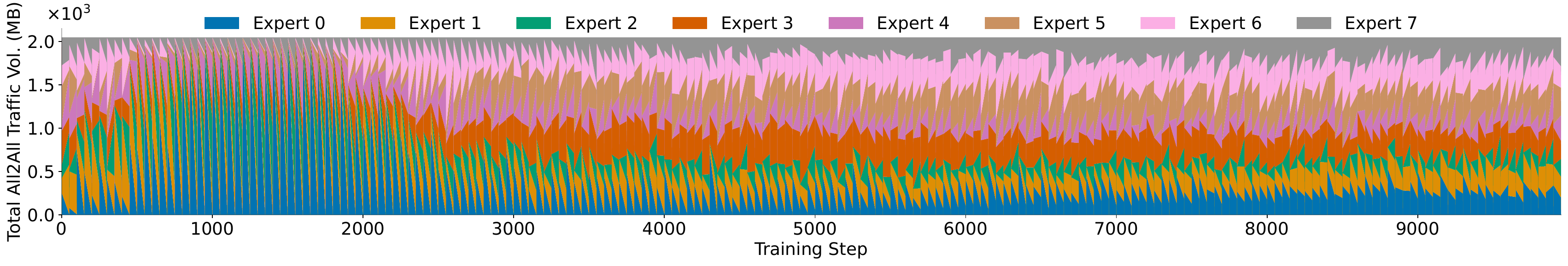}
        \vspace{-2em}
        \caption{In the temporal dimension, the all-to-all traffic volume of each node varies across different training iterations.}
        \label{fig:motiv:traffic:iter}
      \end{subfigure}
      \begin{subfigure}{\textwidth}
        \centering
        \includegraphics[width=\linewidth]{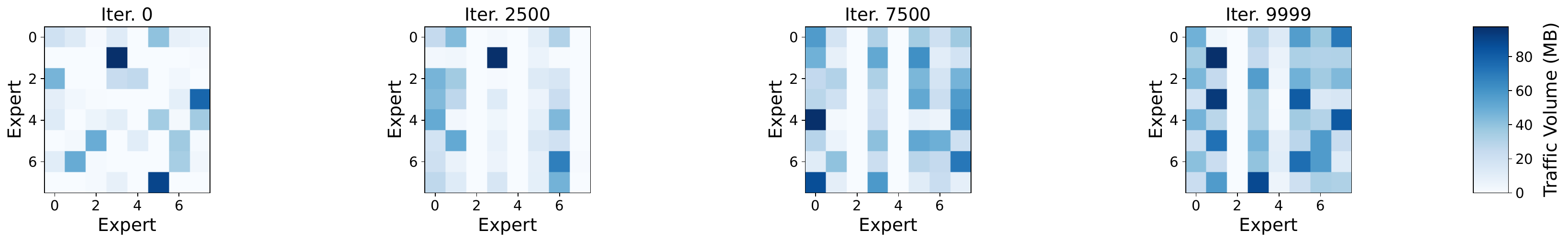}
        \vspace{-1.5em}
        \caption{In spatial dimension, the all-to-all traffic volume is non-uniform across different experts.}
        \label{fig:motiv:traffic:matrix}
      \end{subfigure}
      \vspace{-1.5em}
      \caption{[Mixtral 8$\times$7B in production] All-to-all traffic dynamics during MoE training.}
      \label{fig:motiv:traffic}
\end{figure*}

\parab{Traffic volume of different parallelisms.}
We use Megatron-LM~\cite{megatron-lm} to profile three state-of-the-art MoE models (Mixtral 8$\times$7B MoE~\cite{mixtral}, Llama-MoE~\cite{llama-moe} and Qwen-MoE~\cite{qwen-moe}) and measure the total amount of data transfer. The detailed model configurations are shown in \tabref{tab:measure-setting}.
We plot the distribution of traffic volume in one MoE training iteration in \figref{fig:motiv:traffic-dist}. For \mixtral{}, we observe that TP generates the highest traffic volume  (60\% of the total volume). The EP generates the second-highest traffic volume (30\%), leaving PP and DP contributing less than 6\%. For \llamamoe{} and \qwenmoe{}, we find that EP becomes the most communication intensive (more than 80\%). This is because the size of the largest layer (i.e., expert) fits into a single GPU memory.

\subsection{GPU Interconnects}\label{sec:motiv:switching}

\parab{Scale-up fabrics using NVLink and NVSwitch.}
NVLink and NVSwitch are proprietary technologies provided by Nvidia to support GPU communications within a host server~\cite{nvlink, nvl72}. They offer a higher bandwidth (1.8 TB/s) than PCIe (128 GB/s).

\parab{Scale-out fabrics using electrical packet switching (EPS).} Ethernet- and Infiniband-based EPS has been widely adopted in data center networks with clos-style topologies~\cite{fattree, bcube, dcell,vl2, b4, jupiter-rising}. In such networks, data are encapsulated into packets and switched at layer 2 or above. EPS has the advantage of massive scalability to hundreds of thousands of host servers in modern data centers~\cite{fattree}. However, EPS networks are fixed in topology that cannot be easily reconfigured.

\parab{Scale-out fabrics using optical circuit switching (OCS).} OCS is a layer-1 switching technology that creates dedicated reconfigurable optical circuits between hosts. As depicted in~\tabref{tab:ocs}, today's commodity OCSes have a fundamental tradeoff between the \emph{scalability} (in terms of port counts) and \emph{agility} (in terms of reconfiguration delay). Technologies like the robot optical patch panel~\cite{topoopt} scale up to thousands of ports at the cost of several minutes of reconfiguration delay. At the other end of the spectrum, waveguide-based OCS like silicon photonics~\cite{lightmatter-passage} and PLZT~\cite{PLZT} scores microseconds or nanoseconds latency with limited port counts.

\section{Production Measurements}\label{sec:ideal}

Unlike conventional parallelism (e.g., TP, PP, and DP) with deterministic communication patterns, EP's communications are determined by the gate unit at runtime due to the semantic heterogeneity of the input tokens. To understand the dynamics of EP traffic patterns, we profile Mixtral 8$\times$7B~\cite{mixtral} in a production data center, using a hybrid parallelism that combines an EP degree of 8, TP degree of 4, PP degree of 4 at a sequence length of 4096, and micro-batch size of 8~\cite{llama3-white-paper}.

\parab{The production fabric.} 
We use a Certified Nvidia DGX SuperPOD platform~\cite{nvidia-dgx} with 128 H800 GPUs and 128 ConnectX-7 400Gbps NICs in a production data center. The compute fabric is connected in a rail-optimized topology~\cite{rail-optimized}. We use NCCL~\cite{nccl} to optimize communications on the DGX platform. 

\parab{All-to-all communications within a training iteration.}
We first measure the time it takes for each step in \mixtral{}'s forward pass computation\footnote{The backward pass is a reverse process of the forward pass.} and show the results in \figref{fig:motiv:batch-timeline}.
For the typical micro-batch size used in production (e.g., 8), we observe that the expert computation takes more than 100 ms, which is much larger than the reconfiguration latency of existing optical switches (for example, the MEMS OCS in \tabref{tab:ocs}). Therefore, it provides an opportunity to reconfigure the OCS for the second all-to-all in the expert computation phase. 
For backward propagation, it allows us to hide the reconfiguration latency in the attention computation of its later layer (for the second all-to-all) and expert computation period (for the first all-to-all) as the backward computation often takes more time than the forward. The results of other MoE models are presented in Appendix~\ref{sec:appendix:timeline}.

\sigcomm{
\parab{All-to-all communications are temporally dynamic.} 
\figref{fig:motiv:traffic:iter} plots the total communication volume that each expert receives in all-to-all communication in each MoE layer, which represents the activation intensity of each expert.
We find that the activation intensities of each expert vary significantly across different iterations, which indicates the non-deterministic nature of the EP traffic. 
We also find that as training progresses, the variability of the overall communication volume among experts decreases. The decreasing variability is attributed to the use of \emph{load balancing loss}\footnote{In MoE training, load balancing loss is commonly used to ensure that token loads are evenly distributed across all experts.}. However, even as the overall communication volumes of experts appear to converge, the sparsity of all-to-all traffic matrices persists, as illustrated in \figref{fig:motiv:traffic:matrix}. 
Furthermore, recent advances in the ML community have introduced MoE training techniques that intentionally leave selected experts underutilized at some training stages to achieve improved model performance~\cite{lu2024not, chen2022task}. This further highlights the dynamic communications in distributed MoE training.

\parab{All-to-all communications are spatially non-uniform.} 
\figref{fig:motiv:traffic:matrix} plots the detailed all-to-all communication matrix 
of a selected layer through different iterations. We observe that each traffic matrix of all-to-all communication is non-uniform, with heavy communication only between several GPU pairs. 
Specifically, the state-of-the-art LLM model DeepSeek-V3 reveals that explicitly creating non-uniformity in token distribution across experts while bypassing load-balancing loss\footnote{See Figure 9 and Figure 10 in DeepSeek-V3 report~\cite{deepseekv3} for more details.} improves the MoE training process.
}

\begin{figure}[t!]
    \centering
    \includegraphics[width=0.7\linewidth]{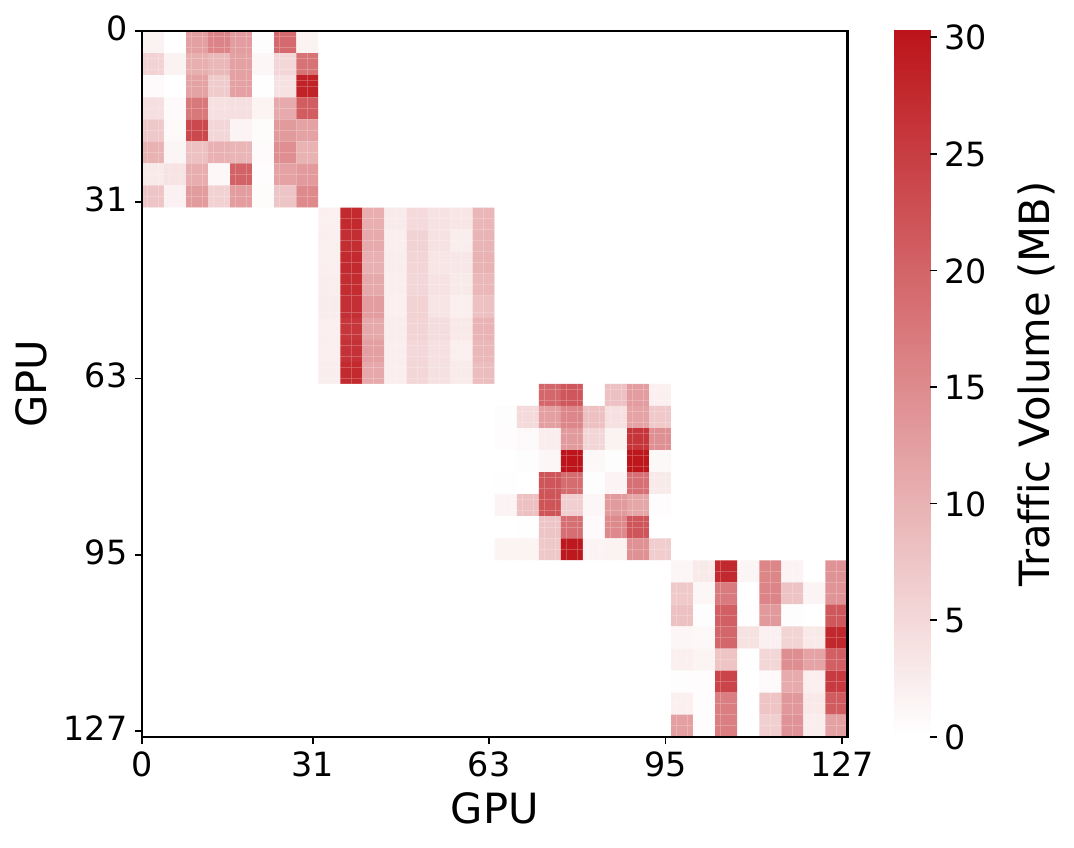}
    \vspace{-1em}
    \caption{[Mixtral 8$\times$7B in production] Traffic matrix of all GPUs showing strong locality.}
    \label{fig:motiv:global-expert-parallel}
\end{figure}

\begin{table*}[t]
    \centering
        \fontsize{8}{10}\selectfont
        \renewcommand{\arraystretch}{1.0}
        \begin{tabular}{c | c c c | c c c | c}
            \toprule
             & \multicolumn{3}{c|}{\textbf{Traffic}} & \multicolumn{3}{c|}{\textbf{Ideal Fabrics}} & \textbf{Best Fit} \\
            \cline{2-7}
             & \textbf{Volume} & \textbf{Temporal Pattern} & \textbf{Spatial Pattern} & \textbf{Bandwidth} & \textbf{Reconfigurability} & \textbf{Scalability} & \textbf{Interconnect Technology}\\
            \midrule
            DP & Low & Deterministic & Global All-Reduce & Low & Slow \& One-Shot & Large & Electrical Packet Switch (Ethernet) \\
            TP & Highest & Deterministic & Local All-Reduce & High & Slow \& One-Shot & Small & Crossbar Switch (NVSwitch) \\
            PP & Low & Deterministic & Global Point-to-Point & Low & Slow \& One-Shot & Large & Electrical Packet Switch (Ethernet)  \\
            EP & High & Non-Deterministic & Regional Sparse All-to-All & High & Fast \& In-Training & Medium & Circuit Switch (Optical) \\
            \bottomrule
        \end{tabular}
        \caption{The quest for a best fit between interconnect fabric and the MoE parallelization strategies.}
        \vspace{-1.2em}
        \label{tab:moe-traffic}
\end{table*}

\parab{All-to-all communications have strong locality.}
\figref{fig:motiv:global-expert-parallel} shows the all-to-all communications among all the 128 GPUs during the training of \mixtral{}. We observe that the EP traffic exhibits \emph{strong locality}. This is because only the expert layers within the same MoE block need all-to-all communications, while expert layers across different MoE blocks at different PP stages do not communicate directly.

\arxiv{The above observations are stemmed from the inherent sparse activation characteristic of MoE layer and the gradual refinement of the gating unit during training. We note that other work in ML community~\cite{deepseekv3, li2023merge} have observed similar behaviors, suggesting that these characteristics are common across different MoE models.}

\section{\sysbf{} Architecture Design}\label{sec:arch}
So far, we have shown that MoE introduces unique traffic patterns that are temporally non-deterministic and spatially non-uniform. Now, an important question is \emph{How to design a network architecture that best serves the requirements of distributed MoE training?} In this section, we first study an ideal yet practical fabric for distributed MoE training (\S\ref{sec:arch:ideal}). Then, we present our key proposal in \sys{}, the regionally reconfigurable high-bandwidth domain with OCS (\S\ref{sec:arch:reconfig}).

\vspace{-0.5em}\subsection{Towards An Ideal yet Practical Fabric}\label{sec:arch:ideal}
We start with a thought experiment from first principles on an ideal yet practical fabric for distributed MoE training.

\parab{The ideal fabric.}
Designing an ideal fabric for distributed MoE training requires the best fit between the traffic patterns generated by different parallelization strategies and the interconnect technologies. In~\tabref{tab:moe-traffic}, we first summarize the requirements in terms of volume, temporal pattern, and spatial pattern for different parallelisms. Then, we review the desired fabrics, considering their bandwidth, reconfigurability, and scalability. In particular, conventional parallelisms like TP, DP, and PP require only a one-shot reconfiguration because their communication patterns are deterministic. While TP requires high bandwidth within a small number of GPUs that host an individual layer, DP and PP span more GPUs with relatively lower communication bandwidth. Notably, EP is significantly different, as its temporally non-deterministic traffic patterns require in-training topology reconfiguration, and the all-to-all communications among experts require a medium fabric radix. Therefore,
the ideal fabric for MoE training should be a reconfigurable network capable of adjusting its topology as soon as traffic patterns vary across training iterations. Moreover, the topology reconfiguration should be completed before each all-to-all communication phase to avoid interrupting the computation. So, the time window left for topology reconfiguration is on the order of tens of milliseconds (based on the measurements in \figref{fig:motiv:batch-timeline}).

\parab{Challenges of the ideal fabric.}
Realizing this ideal fabric presents significant challenges. Recall that in \tabref{tab:ocs} we review a trade-off in commodity OCS technologies\footnote{There are small-scale prototypes that break this tradeoff by using advanced devices (e.g., tunable lasers with arrayed waveguide grating (AWGR)~\cite{sirius}, silicon photonics~\cite{sip-ml}, etc.). They are out of the scope of this paper, which mainly focuses on commodity solutions that are readily deployable at scale.} between reconfiguration delay and port count. Achieving millisecond scale reconfiguration times typically limits the number of ports in today's OCS to a few hundred ports, making it difficult to scale to hundreds of thousands of nodes needed in a large MoE interconnect. In contrast, increasing the port count to support hundreds of thousands of ports results in slower reconfiguration times, failing to meet the rapid reconfiguration required within training iterations.  

\begin{figure}[t!]
    \centering
    \includegraphics[width=\linewidth]{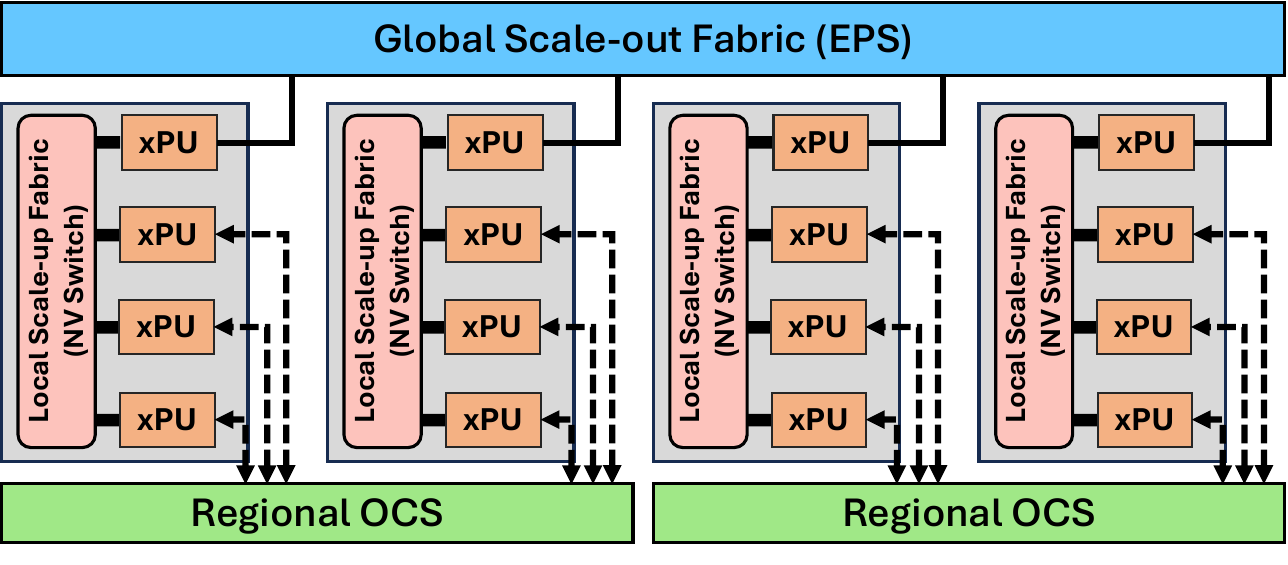}
    \vspace{-1.5em}
    \caption{The \sysbf network architecture. 
    }
    \label{fig:design:network-arch}
\end{figure}

\parab{Landing the ideal fabric in practice.}
To reconcile these challenges, we leverage the key observation that despite the non-deterministic and non-uniform characteristics of MoE all-to-all traffic, there is a strong locality for traffic variations \sigcomm{as MoE blocks are normally placed in a pipeline~\cite{megatron-lm,deepseekv3}}. 
Therefore, instead of building a globally reconfigurable OCS fabric, we propose designing several \emph{regionally reconfigurable OCS} networks (described next in \S\ref{sec:arch:reconfig}). \figref{fig:design:network-arch} depicts the network architecture of \sys{}. By partitioning the network into several domains where the communication locality is strong, each regional OCS rapidly adjusts to  traffic demands of EP without the complexity of global reconfiguration. 
By implementing regionally reconfigurable OCS networks, we leverage this locality to achieve fast reconfiguration within smaller, manageable regions. 
This approach balances the need for reconfigurability with practical hardware limitations, effectively supporting the dynamic communication patterns of MoE training.

\subsection{Regionally Reconfigurable OCS}\label{sec:arch:reconfig}
\sys{}, as the first fabric to support the in-training topology reconfiguration, highlights the core idea of building regionally reconfigurable OCS to offload dynamic EP traffic in the existing electrical fabric.
By leveraging the strong locality inherent in MoE's all-to-all traffic, we partition the network into regions where communication demands are non-deterministic among expert layers. This regional approach allows for rapid reconfiguration within each partition, effectively overcoming the fundamental trade-off between reconfiguration speed and port count in OCS technology. By focusing on regional reconfigurability, \sys{} achieves scalability while maintaining rapid adaptability to dynamic communication patterns of MoE training, alleviating the complexity of global network reconfiguration.

\parab{Where to deploy regionally reconfigurable OCS?}
To best serve the regional sparse all-to-all traffic patterns featured in distributed MoE training, a regional OCS is connected to a cluster of GPU servers where each server splits its NICs between EPS and OCS\footnote{In the near term, \sys{}'s regional OCS leverages the optical transceivers through the NIC for deployment readiness. In the long term, as the co-package optics (CPO) becomes widely adopted, \sys{}'s regional OCS is compatible with the commodity optical I/O solutions (e.g., TeraPHY from Ayar Labs~\cite{ayar-labs}) that is directly connected to the computing chip (e.g., GPU, TPU, etc.).}. Given the fact that today's fast OCSes with millisecond-scale reconfiguration delay support up to 500 ports (\tabref{tab:ocs}) as well as a typical server contains eight NICs, a reconfigurable high-bandwidth domain interconnect through an OCS supports around 80 to 250 servers (each server assigning two to six NICs to OCS).

\parab{When to reconfigure the topology?}
Unlike EPS networks, which are connectionless, OCS networks are connection-oriented and require active control for reconfiguration. During topology reconfiguration, OCS networks are not available to carry packets\footnote{Most commodity optical switches require tens of nanoseconds to several milliseconds or  (\tabref{tab:ocs}) to reconfigure their topology.}. Therefore, we need to choose the right time to start the topology reconfiguration process before the actual data transfer starts to avoid blocking training process.

\parab{How to reconfigure the topology?}
\arxiv{In \sys{}, the OCS fabric is arranged into multiple isolated slices for each reconfigurable high-bandwidth region. Therefore, the regional OCS topology is controlled by its localized topology controller, which frequently collects traffic demands from the host servers. Within each training iteration are four all-to-all communications with the same or transposed traffic pattern. However, these traffic patterns are non-deterministic across different training iterations. Hence, we need to develop a mechanism to reconfigure the topology tailored to traffic patterns. The regional topology reconfiguration means that \sys{} does not require a centralized topology controller, which avoids the scalability concerns of the control plane. 
}

\parab{Towards a mixed optical-electrical fabric.}
\sys{} seeks the best fit between the traffic patterns of the parallelization strategies and the corresponding switching technologies. Therefore, \sys{} uses server-scale NVSwitch for tensor parallelism, regionally reconfigurable OCS for expert parallelism, and large-scale EPS for data parallelism and pipeline parallelism. To distribute data movement tasks on different fabrics, a new collective communication library that supports topology reconfiguration is needed.

\section{\sysbf{} System Implementation}\label{sec:design}

\begin{figure}[t!]
    \centering
    \includegraphics[width=0.9\linewidth]{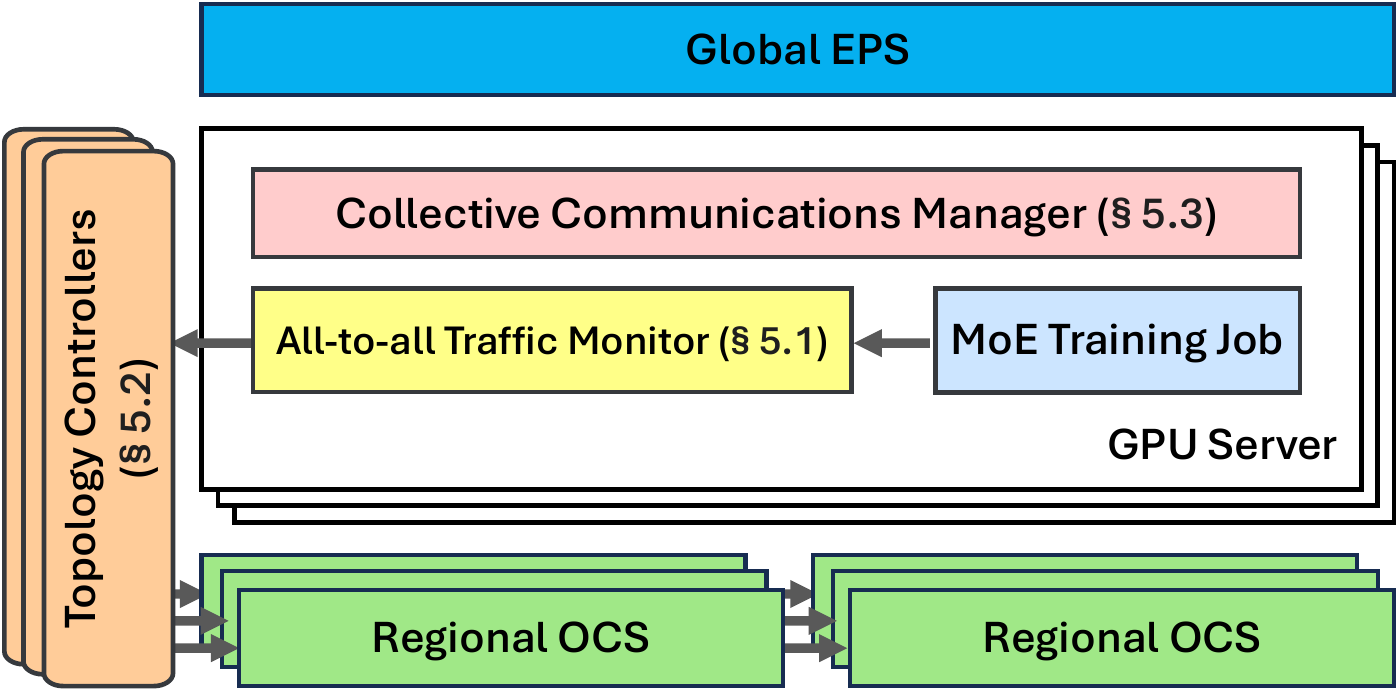}
    \vspace{-0.5em}
    \caption{\sysbf{} system implementation.}
    \label{fig:design:system-arch}
\end{figure}

To enable the aforementioned \sys{} architecture, we design and implement several core components of the control and data planes.
As shown in \figref{fig:design:system-arch}, \sys{}'s implementation contains a traffic monitor that keeps track of the traffic demands in EP to characterize subsequent all-to-all communications (\secref{sec:design:traffic-demand-prediction}). 
Based on the monitored traffic demands, multiple decentralized topology controllers generate and enforce topologies for regional OCSes (\secref{sec:design:topology-generation}). Then, a collective communication manager is responsible to steer the traffic in the \sys{} fabric (\secref{sec:design:traffic-routing}). In addition, we discuss the failure handling mechanism in (\secref{sec:design:failure-handling}).

\subsection{All-to-All Traffic Characterization}\label{sec:design:traffic-demand-prediction}
In each MoE block, there are four all-to-all communication phases during each training iteration: two in the forward pass and two in the backward pass (\figref{fig:design:moe-distribution}). The first all-to-all communication occurs after the gate unit computation. The output from the gate unit\footnote{
The actual output of the gate unit is the dispatching probability distribution for each token. The expert load is derived directly from the probability distribution using the \emph{top-k} parameter.} across the EP groups determines the traffic matrix for this communication phase. \sigcomm{These four all-to-all traffic matrices are strictly the same or transposed due to the symmetry of the token dispatching and collection process. 
}
For communication in the first all-to-all in FP, given that the \sys{} network architecture is provisioned with millisecond-scale reconfigurable OCS, there are two options. \sys{} reconfigures the OCS while blocking the training process, as this reconfiguration time cannot be hidden in the computation. On the other hand, if \sys{} opts to utilize a random topology or reuse topology from previous MoE layers, it cannot benefit from the efficient circuits schedule.

\sigcomm{
Besides, we observe the \emph{partial predicabilty} of the FP's first all-to-all, which offers an opportunity to proactively reconfigure the OCS for it in advance (\eg in attention computation phase). We offload the details of the prediction algorithm for all-to-all traffic demand in \secref{sec:appendix:predict}. Note that \sys{} does not introduce extra demand monitoring overhead as the state-of-the-art MoE training framework already contains a mechanism to collect this information~\cite{megatron-token-collect} to perform on-demand all-to-all transmission.
}

\begin{algorithm}[t!]
    \caption{Reconfigure OCS 
    }
    \small
    \label{alg:reconfigure}
    \begin{algorithmic}[1]
    \Procedure{ReconfigureOCS}{$E$, $\alpha$, $N$, $V$}
    \FuncInput{$E$: all-to-all communication demands of experts}
    \FuncInput{$\alpha$: current optical degree}
    \FuncInput{$N$: number of servers}
    \FuncInput{$V$: server node set}
    \FuncOutput{$S$: NIC level mapping in OCS}
    \State C $\gets$ zero matrix of size $N \times N$
    \State $avail\_ocs[v] \gets \alpha$ for $v \in V$
    \State Initialize finish time $T = \infty$ if $D[i][j] \neq 0$, otherwise $T = 0$
    \LineComment{Step 1: $D$ is translated into an upper triangular matrix.}
    \State D $\gets$ \Call{calculate\_server\_demand}{$E$}
    \While{True}
        \LineComment{Step 2: Find bottleneck links}
        \State (i, j) $\gets$ \Call{findBottleneckLink}{$T$, $C$, $V$}
        \If{$avail\_ocs[i] > 0$ and $avail\_ocs[j] > 0$}
            \LineComment{Step 3: Create a link between $i$ and $j$}
            \State $C[i][j] \gets C[i][j] + 1$ and $C[j][i] \gets C[j][i] + 1$
            \For{$v \in \{i, j\}$}
                \State $avail\_ocs[v] \gets avail\_ocs[v] - 1$
            \EndFor
        \Else
            \State Break
        \EndIf
        \LineComment{Update the time matrix}
        \State $T[i][j] \gets \frac{D[i][j]}{C[i][j]}$, $T[j][i] \gets \frac{D[j][i]}{C[j][i]}$
    \EndWhile
    \LineComment{Step 4: Generate OCS topology}
    \State $S$ $\gets$ \Call{GetNICMapping}{$C$}
    \State $S$ $\gets$ \Call{permuteLinks}{$S$}
    \LineComment{Step 5: Reconfigure the OCS}
    \State \Call{reconfigureOCS}{$S$}
    \State \Return $S$
    \EndProcedure
    \end{algorithmic}
\end{algorithm}

\subsection{Topology Reconfiguration}\label{sec:design:topology-generation}

In \sys{}, finding an optimal topology and deriving an optical schedule is an NP-hard problem~\cite{foerster2018characterizing}. We address this challenge by introducing a lightweight greedy algorithm.
The key insight is that all-to-all communication time is determined by the delay of largest transfers, which implies that the corresponding GPU pair should be allocated with more circuits.
Thus, we identify the communication pairs with the longest transmission time in each iteration and assign them with direct optical links in the OCS topology. 
The detailed OCS reconfiguration algorithm is shown in \algoref{alg:reconfigure}.

\parab{Step 1: Obtain the inter-server demand (line 5).} Given the predicted all-to-all communication demands, the algorithm first maps the traffic matrix to an actual inter-server communication demand with respect to the number of experts per GPU and the number of GPUs per server. Note that we provision the TX and RX bandwidth of each OCS link together, thus making the inter-server demand matrix upper triangular via adding the TX and RX demands together.

\parab{Step 2: Find bottleneck links (line 7).} The algorithm then iteratively finds the bottleneck of the currently allocated links. The bottleneck link is defined as the link with the longest completion time given the demand matrix $D$ and allocated link matrix $C$. We greedily calculate the bottleneck link by calculating the completion time of each link and return the server pairs with the longest completion time.

\parab{Step 3: Allocate OCS circuit (line 9-11).} The algorithm first allocates the OCS link for the found bottleneck server pairs. If the OCS NICs of two servers are not fully allocated, the algorithm will assign the link for them accordingly. 

\parab{Step 4: Generate OCS topology (line 15-16).} The algorithm generates the topology by mapping the TX and RX NICs based on the allocated link matrix $C$. Note that if multiple links exist between two servers, the algorithm permutes the connection to achieve a non-uniform memory access (NUMA) optimized topology to avoid intra-host congestion. 
For example, if there are two links between server A and server B, the algorithm will permute the connection to ensure that the corresponding TX and RX NICs are in two different NUMA nodes for intra-host traffic forwarding (\secref{sec:design:traffic-routing}).

\parab{Step 5: Reconfigure OCS (line 17).} The final step is to reconfigure the OCS cross-link connections accordingly. The topology manager leverages the TX/RX pair mappings to establish the optical path for the aforementioned pairs.

\subsection{Collective Communication Allocation}\label{sec:design:traffic-routing}
With the generated topology, the next step is to allocate network traffic from different parallelisms to \sys{} and generate routing schedules. In the following, we illustrate how \sys{}'s collective communications manager routes network traffic from various parallelisms in the data path.

\parab{TP and PP.} TP traffic is limited to intra-host high bandwidth domain (e.g., the NVSwitch). PP traffic occurs across different PP stages and relies on the high-fanout EPS fabric in the \sys{} architecture. As a result, there are no special configurations for these two types of parallelism.

\parab{DP.} DP traffic typically spans the entire training cluster. Therefore, \sys{} routes it through the EPS fabric. To further improve the communication efficiency of DP transfers, we leverage the hierarchical all-reduce algorithm~\cite{rat, byteps} to reduce the outbound traffic volume from each server. 
First, the GPUs within each server perform an intra-host reduction to aggregate the parameters to a gateway DP GPU connected to the EPS NIC. Next, all servers engage in a global ring all-reduce among the gateway DP GPUs to synchronize the model parameters. Finally, each server broadcasts the synchronized parameters from the gateway DP GPU to all other GPUs. The first and third stages of communication use the high-speed NVSwitch, while the second stage relies on the relatively lower-bandwidth EPS fabric. If multiple EPS NICs are available in the fabric, \sys{} utilizes a multi-ring all-reduce method to fully exploit the bandwidth and reduce communication time.

\begin{figure}[t!]
    \centering
    \includegraphics[width=0.9\linewidth]{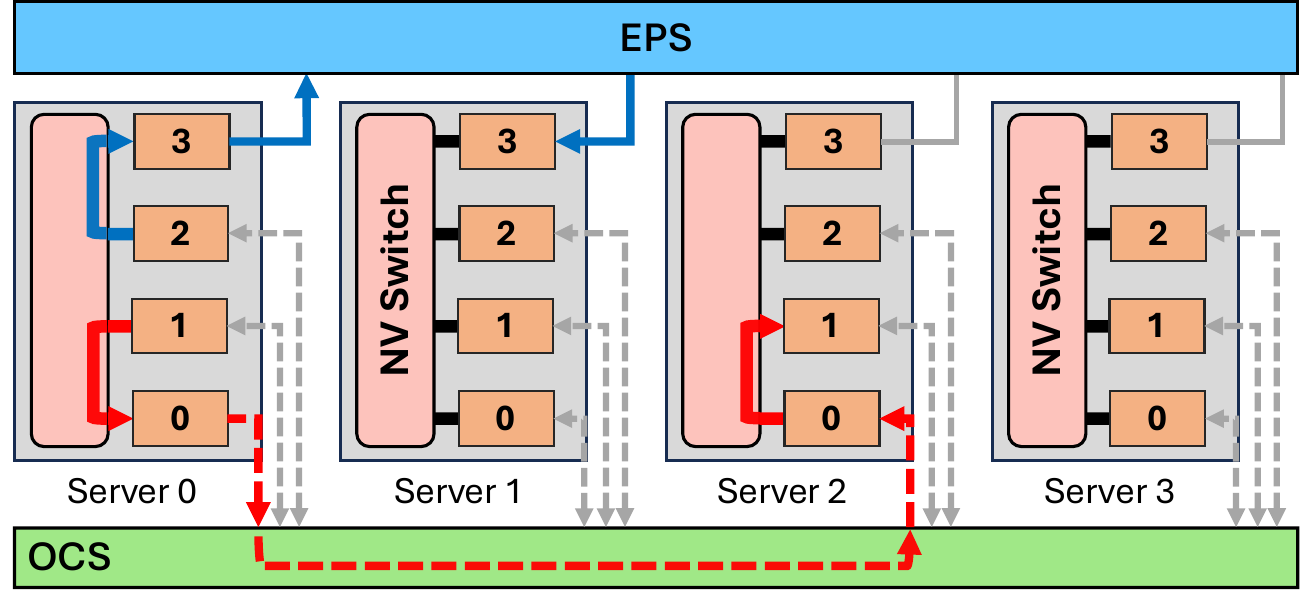}
    \vspace{-0.8em}
    \caption{Routing of all-to-all communications in \sysbf{}. For presentation simplicity, each server only contains 4 GPUs, and the TX/RX links are merged into a single link. In practice, \sysbf{} supports the standard setup of 8 GPUs or more.
     }
    \label{fig:design:all-to-all-routing}
\end{figure}

\parab{Topology-aware EP.} EP traffic is expected to use the regionally reconfigurable high-bandwidth domain. After reconfiguration, \sys{} essentially forms a direct-connect topology for EP transfers. Based on that, \sys{} uses the following steps to route the EP traffic. We illustrate this process in \figref{fig:design:all-to-all-routing}, which depicts a reconfigured topology among four servers with a total of 16 GPUs (EP degree equals 16). 
\vspace{0.5em} 
\begin{enumerate}[leftmargin=*]
    \item Each GPU looks up the topology to identify its intra-server communication delegation GPU for all communication pairs. \sys{} prioritizes directly connected optical circuits over EPS. For example, the delegation GPU from server 0 to server 2 is GPU 2, as they are connected with optical circuits. 
    However, to perform communications between server 0 and server 1, they have to use GPUs that are connected in the EPS.
    \item With the gateway information, each server performs an intra-host \ct{gather}, gathering outbound data to the corresponding delegation GPUs via NVSwitch. Note that in \secref{sec:design:topology-generation}, we balanced the number of NICs across each NUMA node to mitigate intra-host congestion when multiple links are provisioned between a server pair. \sys{} aims to distribute the traffic load across delegation NICs as evenly as possible. 
    \item Each server initiates the inter-host all-to-all communication across all delegation GPUs using NICs in both the EPS and OCS fabrics.
    \item Each server performs an intra-host all-to-all communication among local experts via NVSwitch.
    \item The delegation GPUs in each server \ct{scatter} the received all-to-all data to its final destination. 
\end{enumerate}

As the dataflows in steps (3) and (4) do not interfere with each other, \sys{} overlaps the communication in these two steps to reduce overall completion time.

\subsection{Failure Handling}\label{sec:design:failure-handling}

There are two categories of failures in distributed MoE training: network (NIC/link) failures and GPU failures. \sys{} is tolerant to both transient and permanent failures during distributed MoE training.

\parab{Network fault resilience.}
In practice, each server in \sys{} connects to both the global EPS fabric and a regional OCS domain using multiple NICs. Specifically, the EPS-side typically uses at least two NICs per server (see \secref{sec:arch:reconfig}), providing redundancy for packet-switched communications. Based on prior reports~\cite{alibaba-hpn}, the failure rate of a single NIC or link during a training job is approximately 0.057\%, making dual EPS NICs sufficient to reduce the probability of simultaneous failure to below 0.00003\%.
If both EPS NICs on a server fail—a rare but possible event—\sys{} reroutes traffic through the OCS domain. Specifically, traffic destined for the failed NICs is first routed optically to a healthy peer, and then relayed through that peer's functional EPS interface. Similarly, the EPS can serve as a fallback path if OCS links or ports fail. This dual-path design ensures resilient connectivity, at the cost of some additional intra-cluster forwarding overhead.

\parab{GPU fault tolerance.}
\sys{} also supports failure recovery when one or more GPUs become unavailable. We consider two realistic failure levels:

\begin{itemize}[leftmargin=*]
\item{Single-GPU failure.} If a GPU fails during training, \sys{} remaps the workload to a designated backup GPU. This is aligned with the design of high-availability systems such as NVL72, which reserve spare GPUs per group for fault tolerance. In \sys{}, the backup GPU may be connected via either EPS or OCS: 1) If reachable via EPS, training resumes without additional routing. 2) If reachable only via OCS, \sys{} forwards traffic through a peer GPU with optical connectivity to the backup GPU after topology reconfiguration, maintaining functional interconnect through minor topology adjustments.
\item{Full-node failure.} A complete server failure (e.g., all 8 GPUs) requires a replacement node from the global backup pool. These backup nodes connect via EPS uplinks, ensuring network connectivity without reliance on regional OCS. Upon checkpoint restoration, \sys{} resumes training with minimal disruption.
\end{itemize}

\parab{Runtime reconfiguration.} To maintain topology validity in the presence of failures, \sys{}'s decentralized topology controllers detect communication failures and regenerate the OCS topology accordingly. This involves excluding failed nodes from the candidate set and recomputing optical mappings (see Algorithm 1 in \secref{sec:design:topology-generation}). Because \sys{} relies on regional control, such reconfigurations are localized and incur minimal global disruption. We evaluate the impact of various failure scenarios in \secref{sec:sim:failure-resiliency}. 

\section{\sysbf{} Prototype}\label{sec:testbed}

To evaluate \sys{}, we build a fully functional prototype using commodity hardware\footnote{Due to Nvidia's warranty restrictions, we cannot reconfigure the topology of the DGX SuperPod used in the measurement study (\S\ref{sec:ideal}). Hence, we use commodity servers equipped with Nvidia GPUs for the testbed.} capable of training state-of-the-art MoE models.

\begin{figure}[t!]
    \centering
    \includegraphics[width=0.55\linewidth]{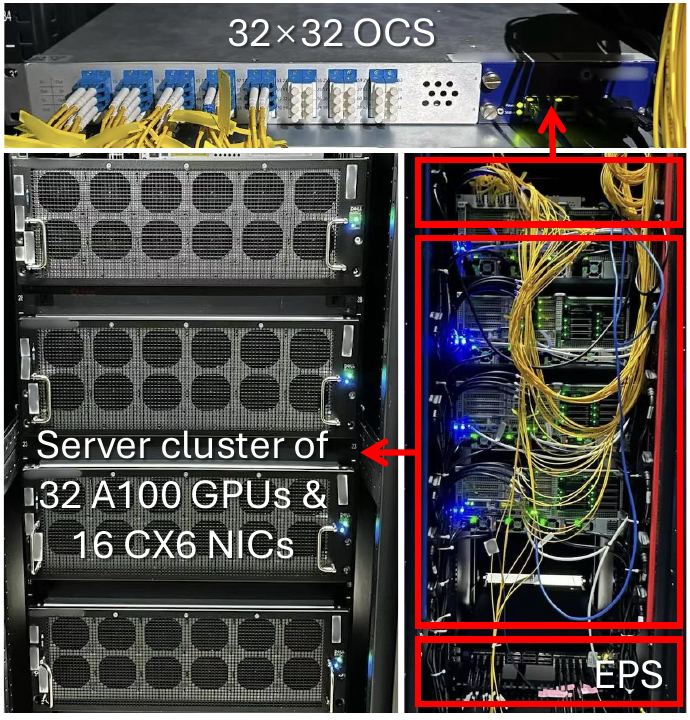}
    \vspace{-1em}
    \caption{\sysbf{} testbed using commodity hardware.}
    \label{fig:testbed:testbed}
\end{figure}

\parab{Hardware setup.} \figref{fig:testbed:testbed} is a picture of our prototype, which contains four commodity servers, each equipped with eight Nvidia A100 GPUs and four Mellanox ConnectX-6 100G NICs. 
For each server, three NICs are connected to a Polatis millisecond-scale OCS~\cite{polatis}, while the remaining one NIC is connected to a Nvidia SN3700 Ethernet switch~\cite{sn3700}. We use 100 Gbps QSFP28 optical transceivers and duplex LC fibers~\cite{duplex-fc-fiber}. All NICs operate in RoCEv2 mode. Each server has a total of four NVLinks that connect two adjacent GPUs. 
Appendix~\ref{sec:appendix:testbed} provides further details of our testbed.

\parab{Software stack.} We implement \sys{}'s software stack using approximately 6K lines of code in C++, including the topology generator, the OCS controller, and the custom collective communication runtime supporting in-training topology reconfigurations. For DP and PP communication over the static EPS fabric, we use NCCL~\cite{nccl} to provide high-speed intra-host and inter-host all-reduce/point-to-point communications. For EP's all-to-all communications that involve both EPS and OCS, our custom collective communication runtime leverages RDMA for high-speed data transfer using the raw \ct{ibverbs} library based on FuseLink~\cite{fuselink}. 
We port the \sys{} runtime to Python to integrate with Megatron-LM~\cite{megatron-lm} for training real-world MoE models.
Specifically, we have implemented communication primitives similar to those in \texttt{torch.dist} and expose them as \texttt{mixnet.all\_to\_all} and \texttt{mixnet.all\_reduce}.

\begin{figure}[t!]
    \centering
    \includegraphics[width=0.65\linewidth]{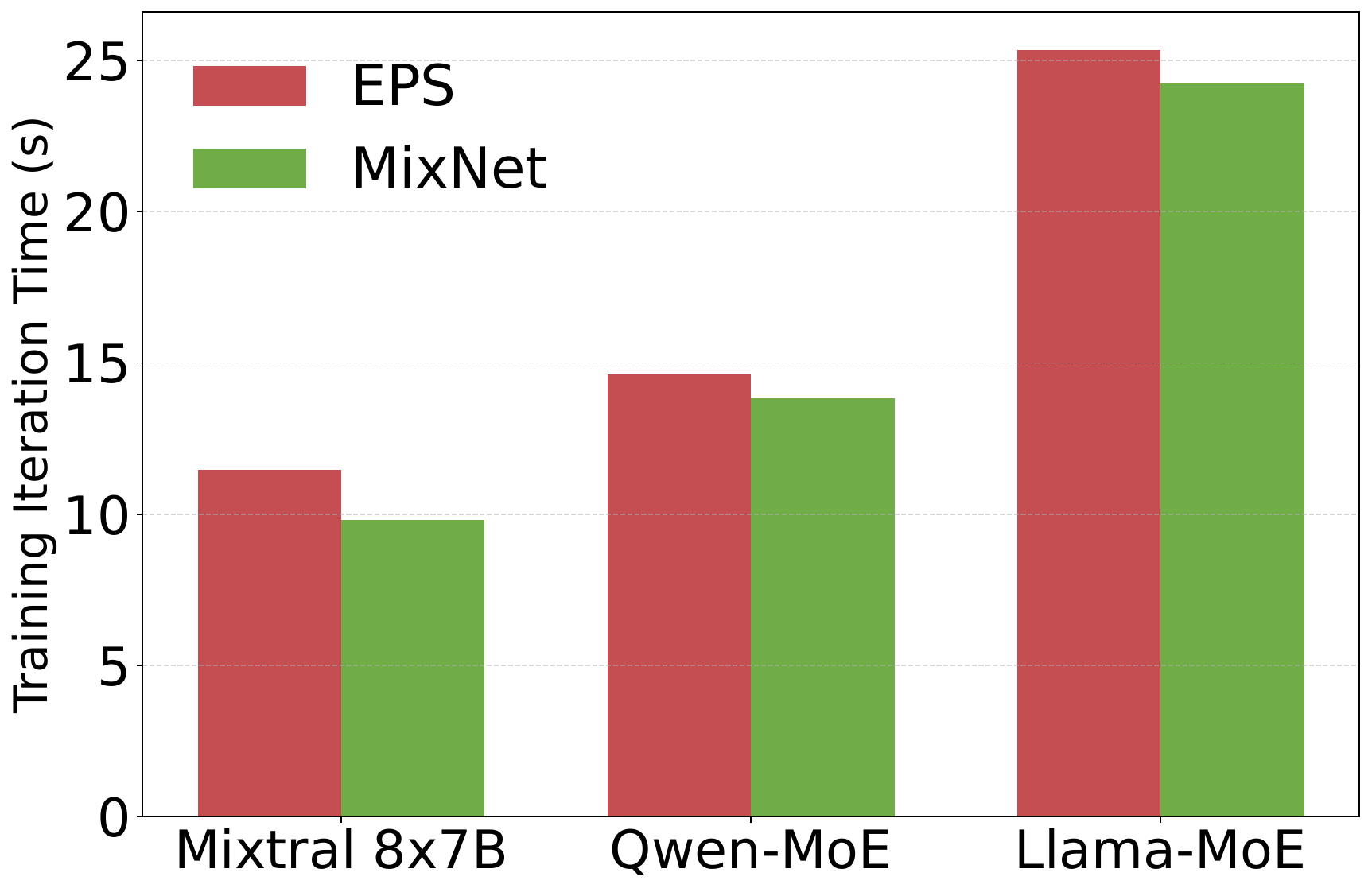}
    \vspace{-1em}
    \caption{[Testbed] End-to-end training iteration time on our 32 GPU prototype.}
    \label{fig:testbed:iteration-time}
\end{figure}

\parab{Training state-of-the-art MoE models.} We use the prototype to train three state-of-the-art MoE models, and compare its performance with a baseline configuration where all the four NICs are connected to an Ethernet switch (the ideal switch baseline). \figref{fig:testbed:iteration-time} shows that \sys{} achieves comparable performance to the 4 $\times$ 100G EPS baseline. \sys{} utilizes one NIC in the EPS fabric and configures the remaining three NICs in an optical circuit fabric (a total of 12 optical ports and 4 electronic ports). In contrast, the EPS baseline uses the four 100 Gbps ConnectX-6 NICs in a non-blocking EPS fabric with 16 electronic ports.
\sys{}'s performance stems from its ability to efficiently provision high-bandwidth optical circuits for communication-intensive pairs in sparsely non-uniform all-to-all traffic, without compromising the transfer speed of DP and TP traffic.
It is important to note that \sys{} \emph{does not alter the parallelization strategies} used in MoE training, but only \emph{optimizes data transfer} through its architectural design and efficient circuit-switching algorithm. As a result, \sys{} does not affect the training accuracy of MoE models.

\section{Large-Scale Simulations}\label{sec:sim}
This section evaluates the performance of \sys{} through large-scale simulation. 
We also present design space explorations on several factors, such as network scalability, EPS link options, and reconfiguration delays in Appendix \secref{sec:appendix:sim-detail}.

\begin{figure*}
    \begin{subfigure}[t]{0.24\linewidth}
        \centering
        \includegraphics[width=\linewidth]{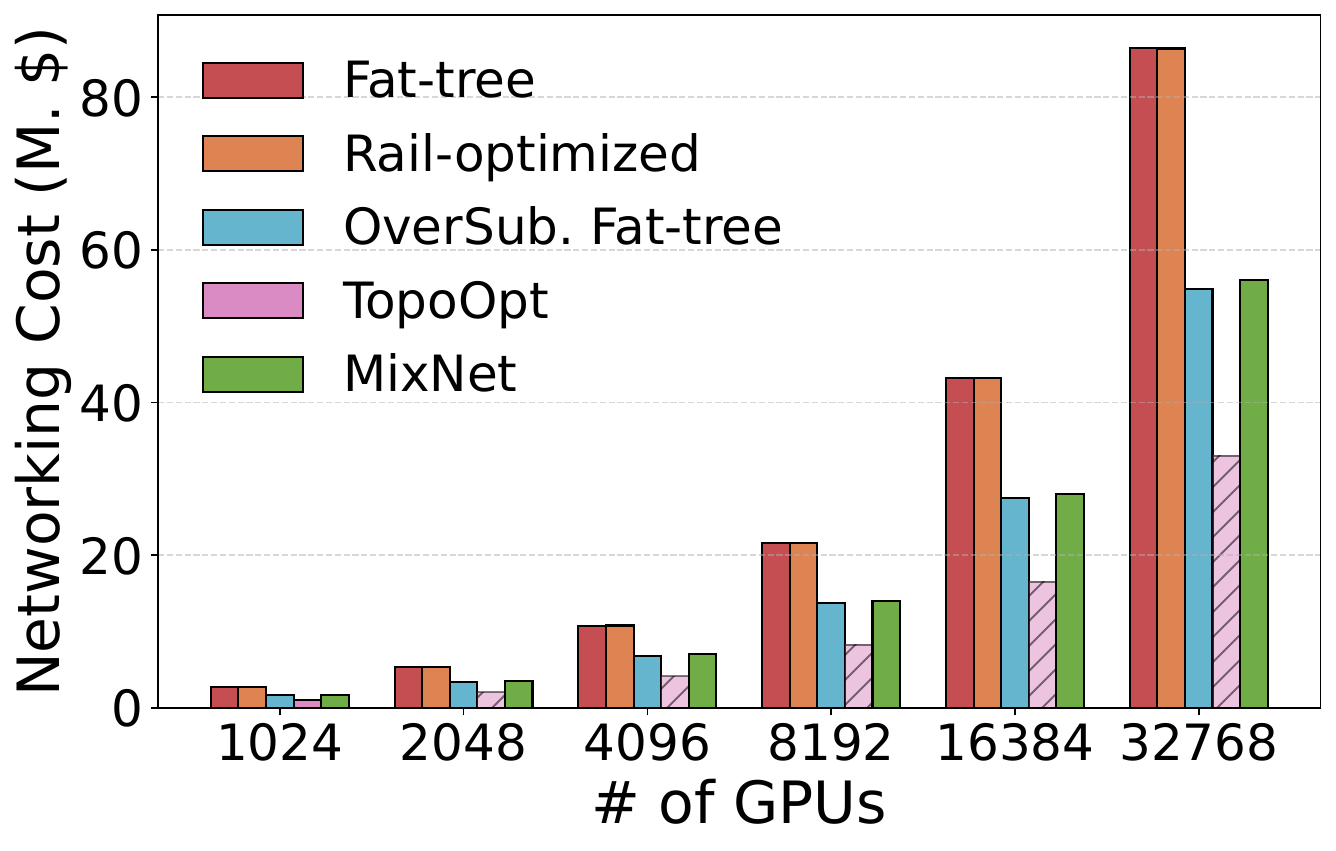}
        \vspace{-1.5em}
        \caption{100Gbps}
        \label{fig:sim:cost-analysis:100g}
    \end{subfigure}
    \hspace{0.2em}
    \begin{subfigure}[t]{0.24\linewidth}
        \centering
        \includegraphics[width=\linewidth]{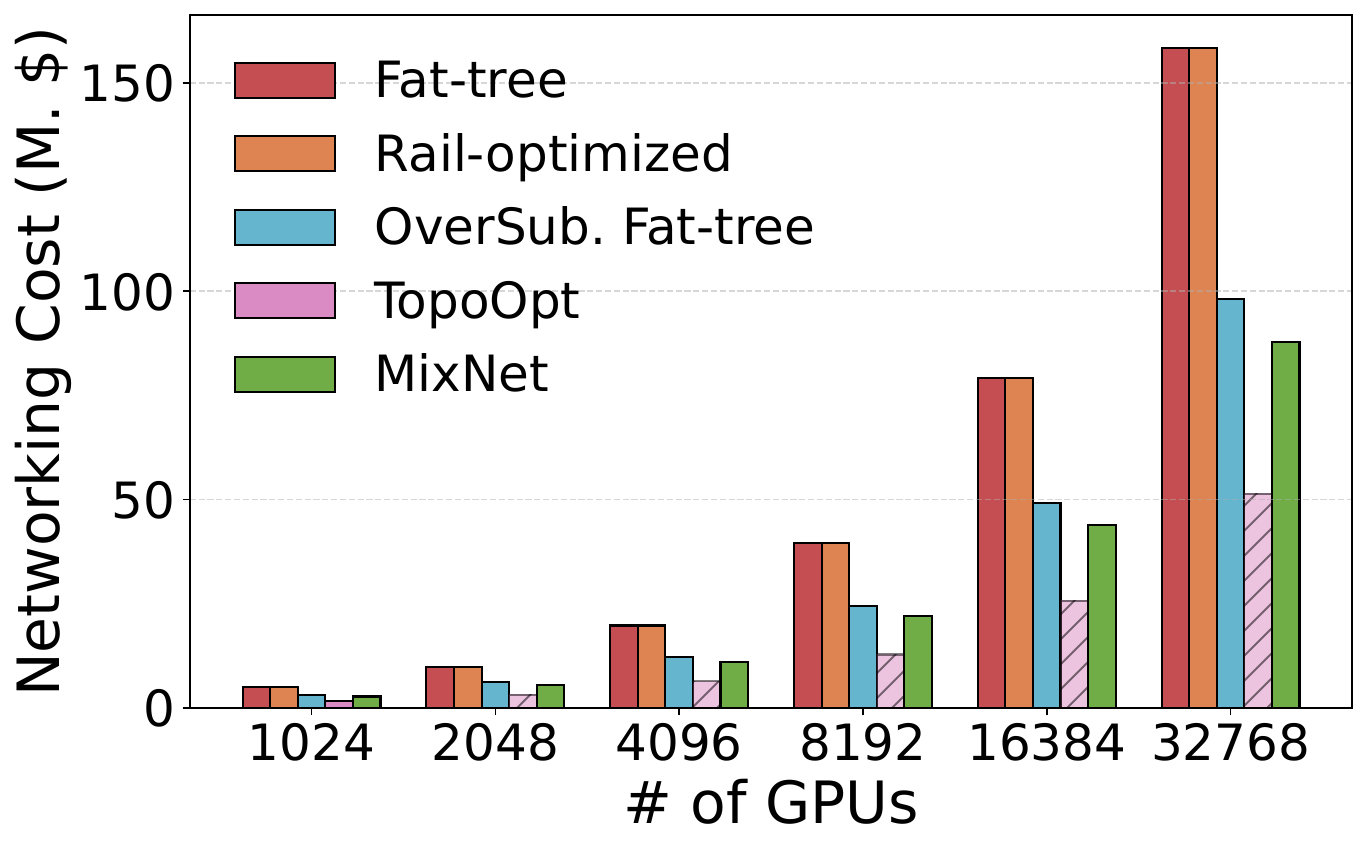}
        \vspace{-1.5em}
        \caption{200Gbps}
        \label{fig:sim:cost-analysis:200g}
    \end{subfigure}
    \hspace{0.2em}
    \begin{subfigure}[t]{0.24\linewidth}
        \centering
        \includegraphics[width=\linewidth]{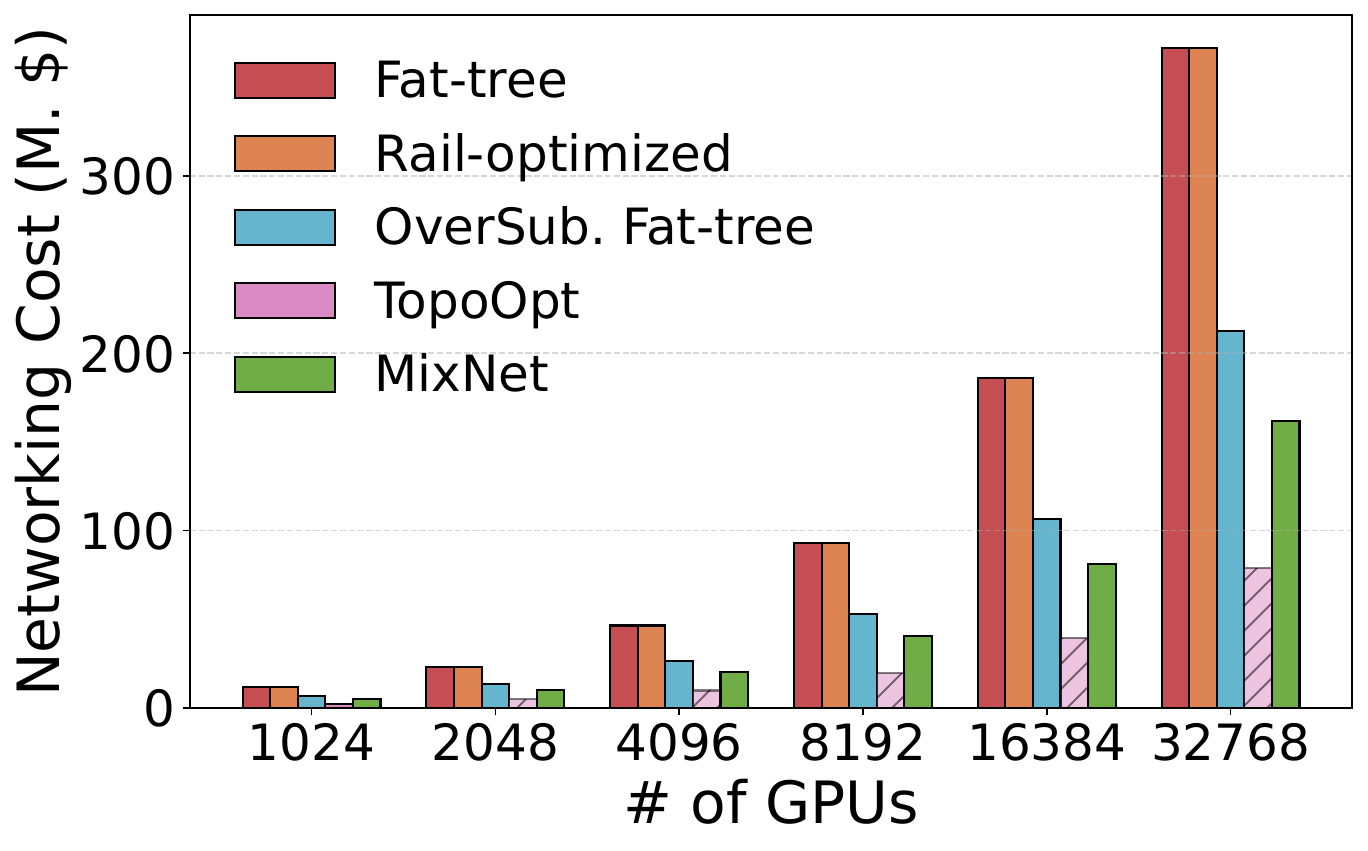}
        \vspace{-1.5em}
        \caption{400Gbps}
        \label{fig:sim:cost-analysis:400g}
    \end{subfigure}
    \hspace{0.2em}
    \begin{subfigure}[t]{0.24\linewidth}
        \centering
        \includegraphics[width=\linewidth]{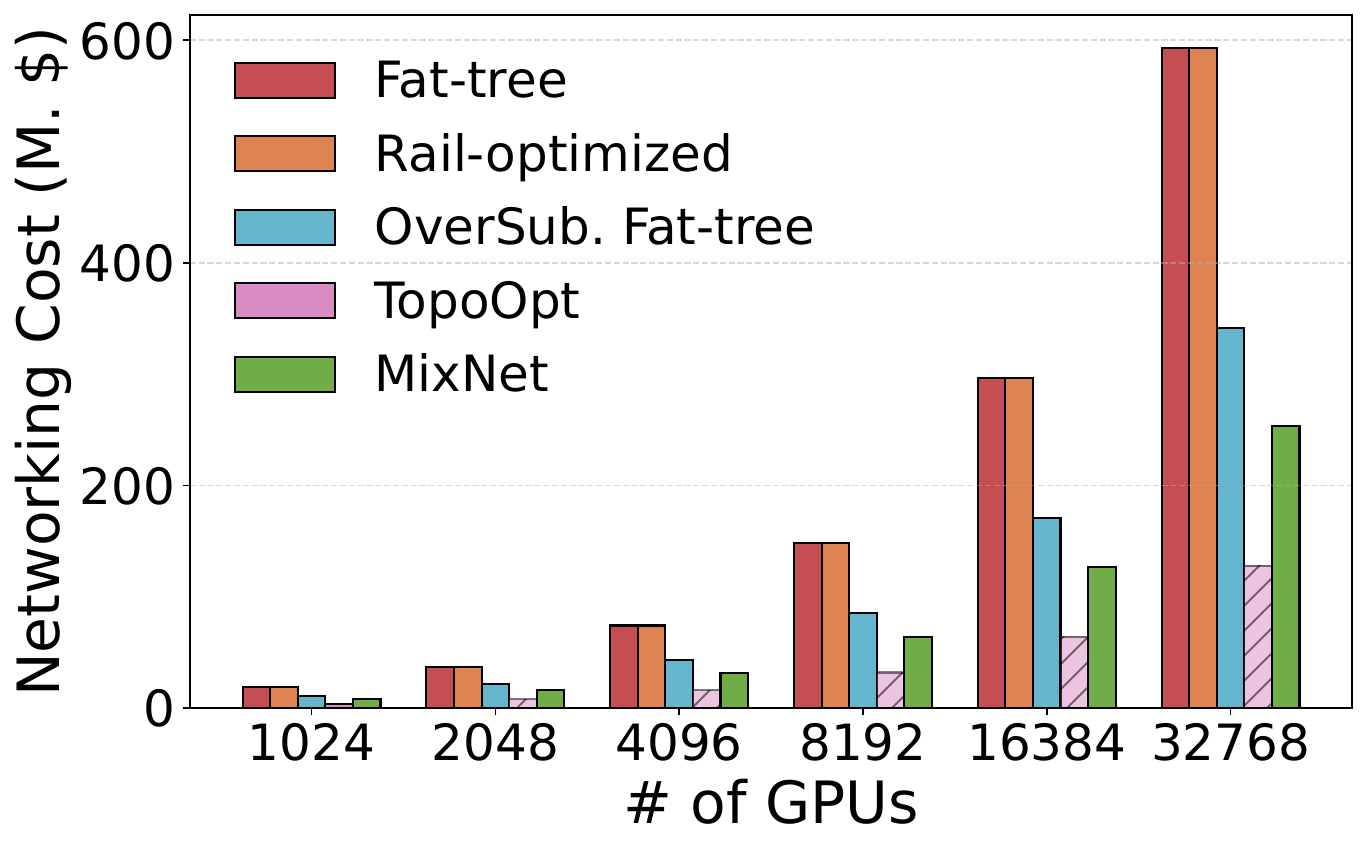}
        \vspace{-1.5em}
        \caption{800Gbps}
        \label{fig:sim:cost-analysis:800g}
    \end{subfigure}
    \vspace{-1.2em}
    \caption{[Simulation] Networking cost analysis.}
    \label{fig:sim:cost-analysis}
    
\end{figure*}

\subsection{Setup}\label{sec:sim:setup}
\parab{Packet-level simulation methodology.} The simulation process is divided into two phases. First, we develop a simulator on top of FlexFlow~\cite{flexflow}. We extend FlexFlow to support pipeline parallelism and rectify its profiler to ensure that the profiled computation time aligns with the actual runtime on the testbed. The simulator is fed with the micro-batch size, an MoE model, and a specified parallelization strategy, and generates a task DAG that describes the computation and communication tasks for the cluster. Using this DAG, we then utilize an event-driven packet-level simulator based on \ct{htsim}
~\cite{htsim}, which simulates packet-based communication between GPUs. The link propagation delay is set to 1 $\mu$s. We set the number of NICs and GPUs per server to 8, with each NIC having a bandwidth of $B$. In our setup, each server has eight GPUs, interconnected via a high-speed NVSwitch (900 GB/s), and eight NICs, reflecting typical configurations used in production environments. The training process for the MoE model is simulated across multiple iterations. The details of used models and parallelization strategies are presented in the Appendix~\ref{sec:appendix:sim:setup}.

\parab{Simulated GPU interconnect fabrics.} We compare the performance of \sys{} with the following interconnects: 
\begin{itemize}[leftmargin=*] 
    \item \textbf{\sysbf{} (this work).} In \sys{}, each server connects two NICs to the EPS fabric using a fat-tree topology and connects the remaining six NICs to the OCS fabric by default. Following the architecture of the regionally reconfigurable high-bandwidth domain in \sys{}, the optical circuit switch only needs to connect the GPUs within a single EP group, which is a maximum of 64 GPUs in our configuration. This can be easily supported by commodity OCS technologies (\tabref{tab:ocs}). \sys{} blocks the network for 25 ms during the reconfiguration of the OCS for the first all-to-all communication in the forward pass and hides the reconfiguration time during computation for subsequent all-to-all communications, as discussed in \secref{sec:design:traffic-demand-prediction}.
    \item \textbf{Fat-tree~\cite{fattree}.} We consider a 1:1 non-blocking \emph{Fat-tree} network. 
    \item \textbf{OverSub. Fat-tree.} We compare \sys{} with a Fat-tree interconnect with the 3:1 over-subscription ratio.
    \item \textbf{Rail-optimized~\cite{rail-optimized}.} It has been the recommended GPU interconnect used by Nvidia. It differs from the fat-tree by connecting GPUs of the same rank to the same ToR switch, providing low latency for GPUs within the same rail.
    \item \textbf{TopoOpt~\cite{topoopt}.} The state-of-the-art optical interconnect that co-optimizes both model parallelization and network topology to minimize communication overhead. For TopoOpt, all NICs are \emph{optimistically} connected via a large and flat optical patch panel.
\end{itemize}

\begin{figure*}[ht!]
    \begin{subfigure}[t]{0.24\linewidth}
        \centering
        \includegraphics[width=\linewidth]{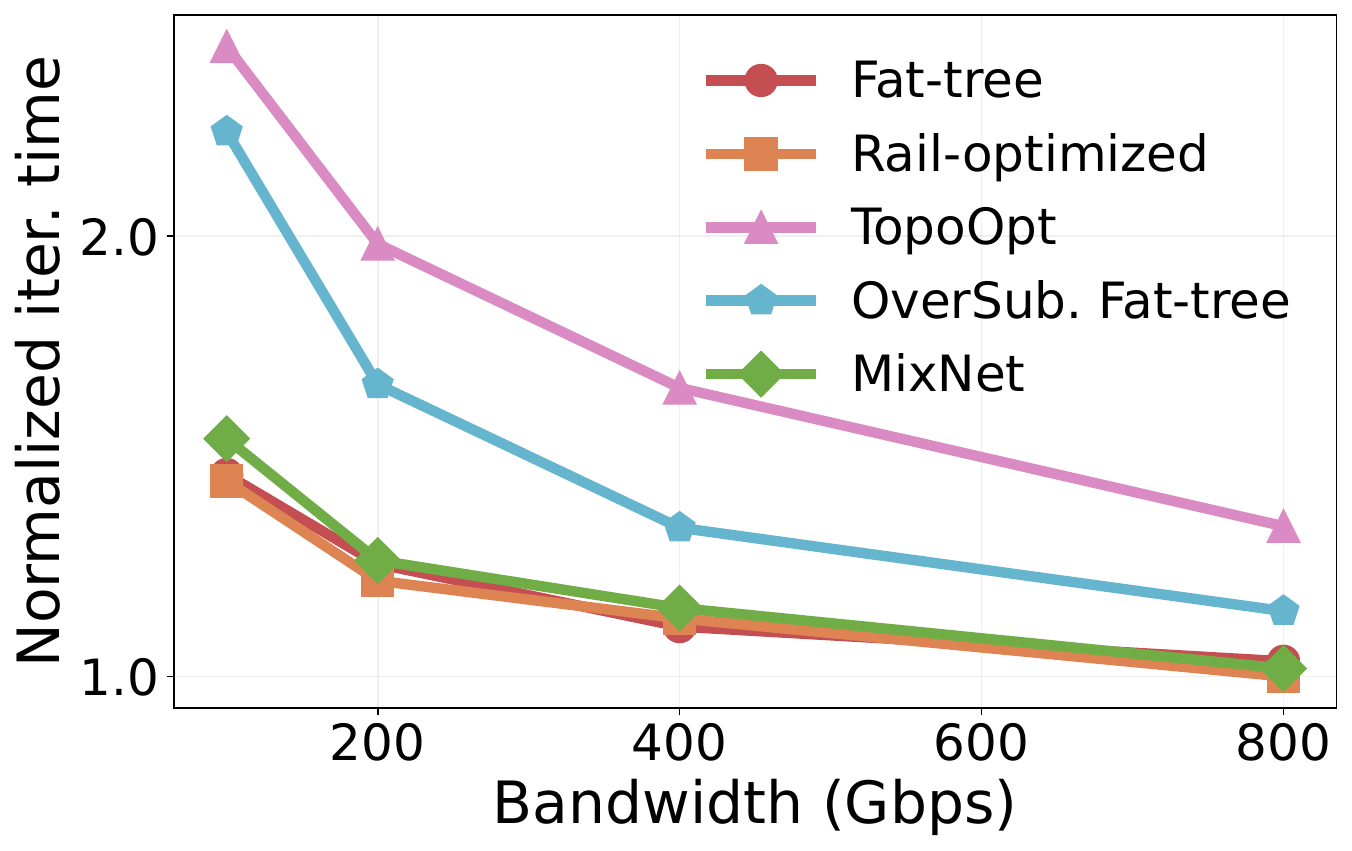}
        \vspace{-1.5em}
        \caption{\mixtrallarge{}}
        \label{fig:sim:speed:mixtral-large}
    \end{subfigure}
    \hspace{0.2em}
    \begin{subfigure}[t]{0.24\linewidth}
        \centering
        \includegraphics[width=\linewidth]{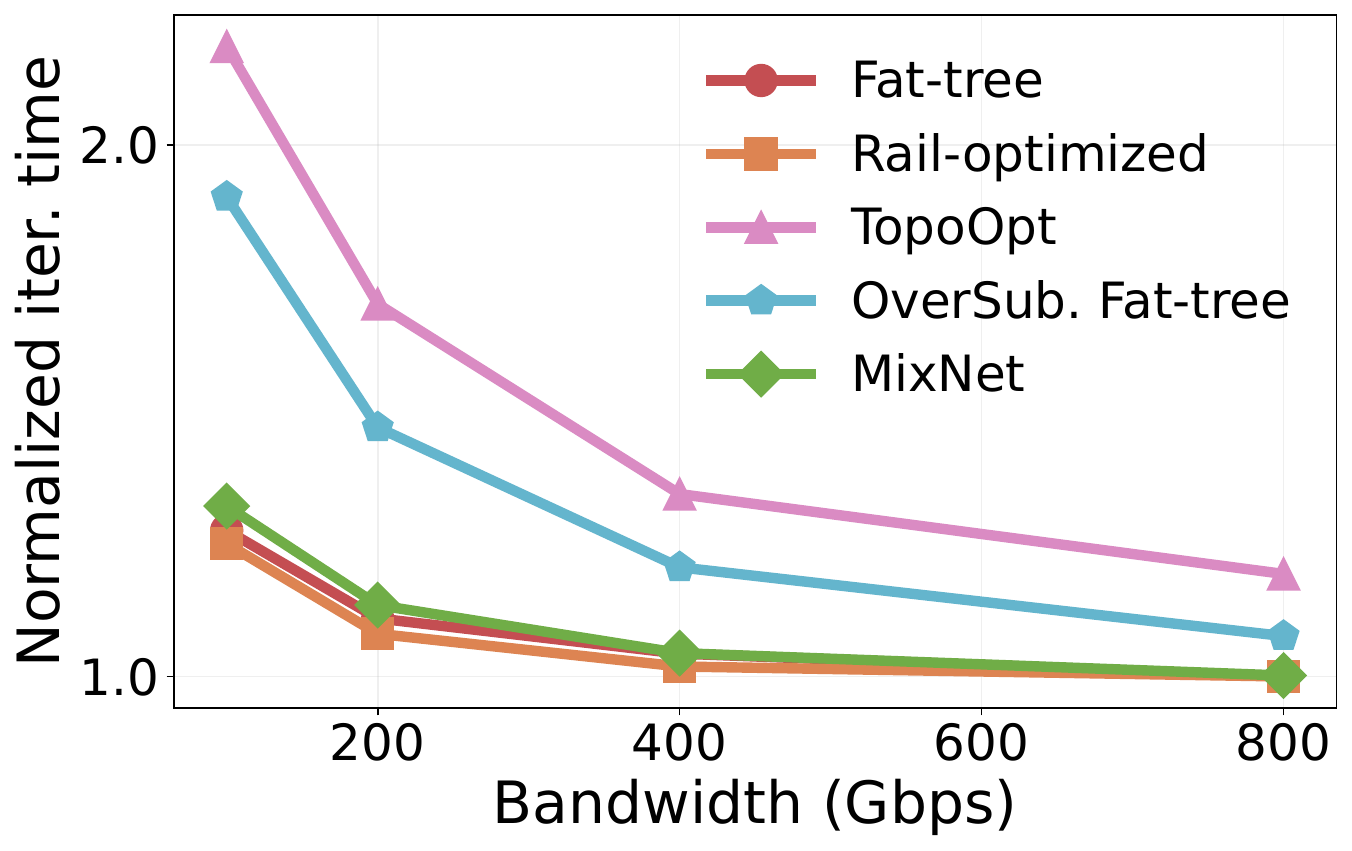}
        \vspace{-1.5em}
        \caption{\mixtral{}}
        \label{fig:sim:speed:mixtral}
    \end{subfigure}
    \hspace{0.2em}
    \begin{subfigure}[t]{0.24\linewidth}
        \centering
        \includegraphics[width=\linewidth]{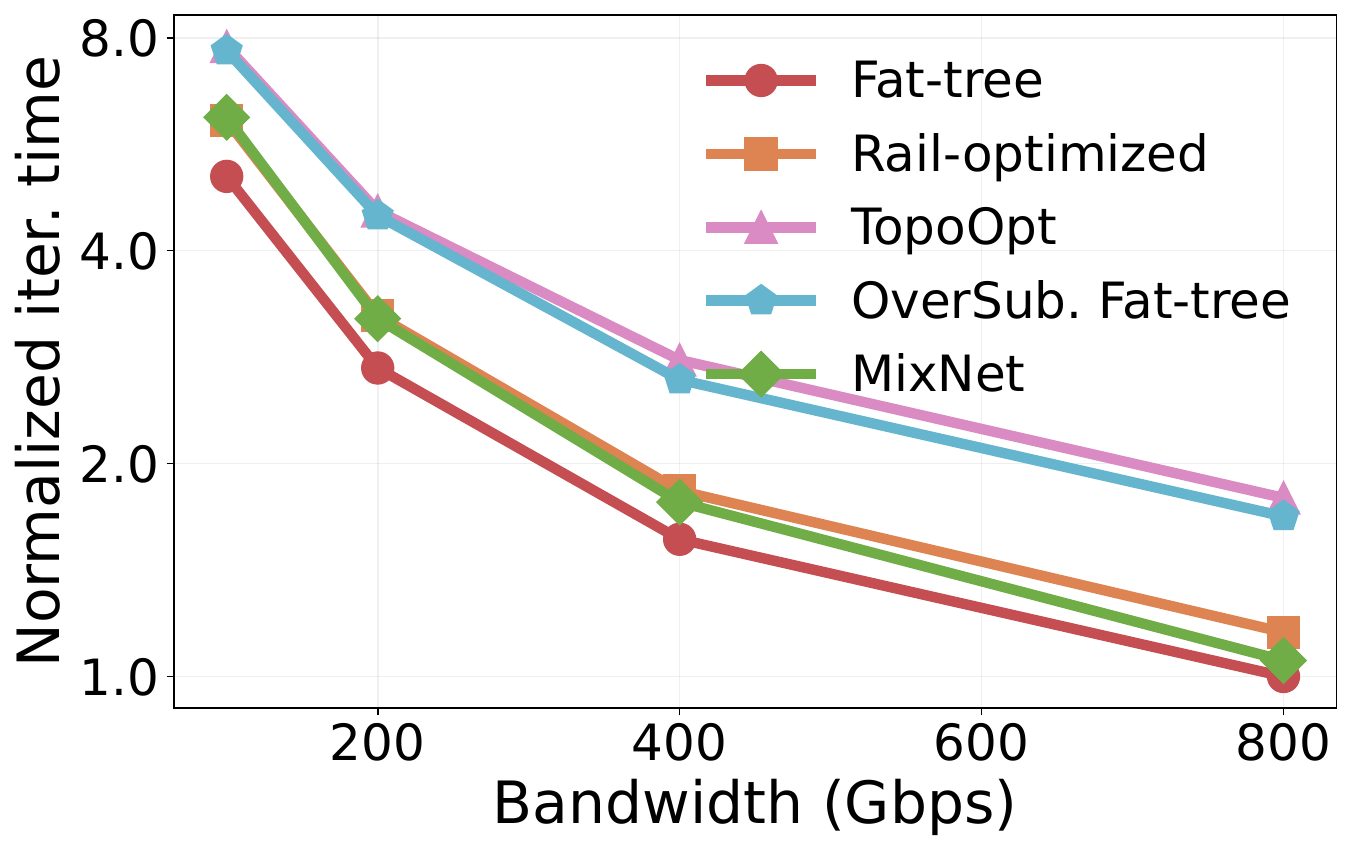}
        \vspace{-1.5em}
        \caption{\qwenmoe{}}
        \label{fig:sim:speed:qwen}
    \end{subfigure}
    \hspace{0.2em}
    \begin{subfigure}[t]{0.24\linewidth}
        \centering
        \includegraphics[width=\linewidth]{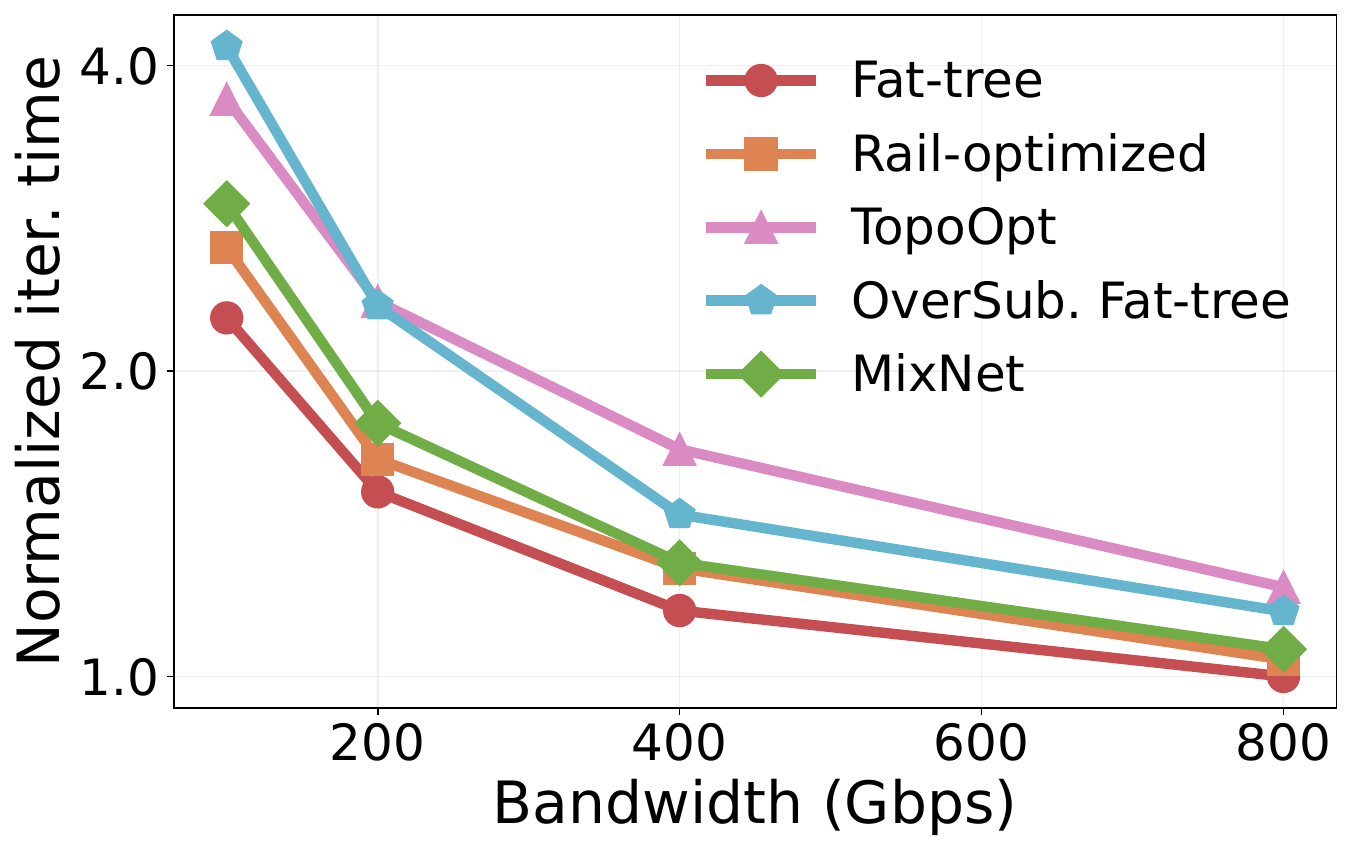}
        \vspace{-1.5em}
        \caption{\deepseek{}}
        \label{fig:sim:speed:deepseek}
    \end{subfigure}
    \vspace{-1em}
    \caption{[Simulation] Training speed-ups in a cluster of 128 servers with 1024 GPUs.}
    \label{fig:sim:speed}
\end{figure*}

\subsection{Networking Cost Analysis}\label{sec:sim:cost}
We present the cost analysis of \sys{} in \figref{fig:sim:cost-analysis}, using a common production setup where each server contains 8 GPUs, following the same methodology as \cite{topoopt}. The networking cost is analyzed with link bandwidths from 100 Gbps to 800 Gbps across different cluster sizes.  It is important to note that we only account for the number of \emph{actually used} switch ports in calculating the cost, as the cluster may not fit perfectly within a fat-tree/rail-optimized topology with a reasonable $K$. More details on the cost of each networking component can be found in Appendix~\ref{sec:appendix:cost}.

First, compared to the non-blocking Fat-tree and Rail-optimized topologies, \sys{} reduces networking costs by an average of 2.0$\times$, as it organizes its high-bandwidth domain using OCS interconnects, which is significantly cheaper than EPS fabrics at high link bandwidth. Specifically, as shown in \figref{fig:sim:cost-analysis:400g}, \sys{}'s OCS fabric incurs 2.3$\times$ lower cost on average than fat-tree topology at 400 Gbps.

Second, we acknowledge that \sys{} incurs slightly higher expenses than TopoOpt at the cluster size of 128 servers (1024 GPUs). This is because: 1) \sys{} requires EPS fabric to maintain global network-wide connectivity, and 2) \sys{}'s high-bandwidth domains assume millisecond-level reconfigurable OCS to adapt to runtime MoE traffic, which is more expensive than TopoOpt's slowly reconfigurable patch panel. However, TopoOpt requires a multi-tier patch panel fabric to form a network of more than 1K GPUs. Achieving this requires extensive patch panel ports and expensive long-reach transceivers to compensate for the insertion loss of optical signals across multiple switching layers. As a result, it remains unclear whether TopoOpt is able to interconnect such large clusters and maintain its cost-efficiency.

\subsection{Performance: Training Speed Ups}\label{sec:sim:speed}
This section compares the end-to-end training iteration time of \sys{} against other interconnects across four MoE models on the cluster with 128 servers and 1024 GPUs.

\figref{fig:sim:speed:mixtral-large} compares the training iteration time of various interconnects for the \mixtrallarge{} model. Due to its efficient bandwidth allocation, we observe that \sys{} achieves performance very close to the ideal Fat-tree and Rail-optimized topologies. 
In particular, with a TP degree of 8, \sys{} provides direct optical circuits for almost all high-traffic server pairs during all-to-all communication (24 optical circuits for 8 EP participants). Compared to TopoOpt, \sys{} reduces the training iteration time by 1.5$\times$ on average, as TopoOpt's static topology cannot adapt to real-time traffic variations. In addition, we observe that \sys{} outperforms the over-subscribed fat-tree by up to 1.6$\times$.
\figref{fig:sim:speed:mixtral} shows a similar trend for the \mixtral{} model, with \sys{} reducing the iteration time by 1.4$\times$ on average compared to TopoOpt.
Both Mixtral models show diminishing returns from increased link bandwidth, as they are computation-bound at a micro-batch size of 8 (\figref{fig:motiv:batch-timeline}). At higher bandwidths, the communication overhead shrinks, narrowing the performance gap between \sys{} and others. The results for larger batch sizes of these Mixtral models are in the Appendix~\ref{sec:appendix:mixtral-large}.

\figref{fig:sim:speed:qwen} and \figref{fig:sim:speed:deepseek} compare \sys{} with other interconnects on \qwenmoe{} (32 experts with 32-way EP) and \deepseek{} (256 experts with 64-way EP), both of which use larger numbers of experts and higher EP degrees than Mixtral models. We observe that \sys{} achieves performance comparable to Rail-optimized and Fat-tree topologies, and outperforms TopoOpt by 1.5$\times$ on average in \qwenmoe{} and 1.3$\times$ in \deepseek{}.
Compared to Fat-tree, \sys{} exhibits slightly larger performance gaps at low bandwidths as the number of experts increases (e.g., 32 in \qwenmoe{} vs. 8 in \mixtrallarge{}). This is because more bandwidth-intensive GPU pairs require dedicated optical circuits, which slightly exceeds \sys{}'s default optical fanout under the 8-GPU setup (6 circuits for 32 EP participants). Nevertheless, owing to the sparsity of EP traffic, \sys{} narrows the performance gap as link bandwidth increases.
We also observe that, unlike \qwenmoe{}, \sys{} shows smaller performance gains on \deepseek{} as bandwidth increases. This is because, although \deepseek{} uses a higher EP degree, its larger model size results in a lower communication-to-computation ratio, thus reducing the benefit of additional bandwidth.

\begin{figure*}[t!]
    \vspace{-0.3em}
    \centering
    \begin{subfigure}[t]{0.9\linewidth}
        \centering
        \vspace{0.3em}
        \includegraphics[width=\linewidth]{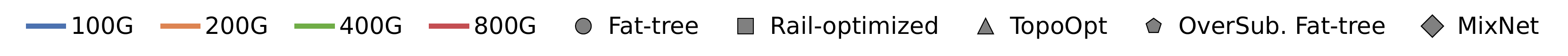}
    \end{subfigure}
    \\
    \vspace{-0.5em}
    \begin{subfigure}[t]{0.24\linewidth}
        \centering
        \includegraphics[width=\linewidth]{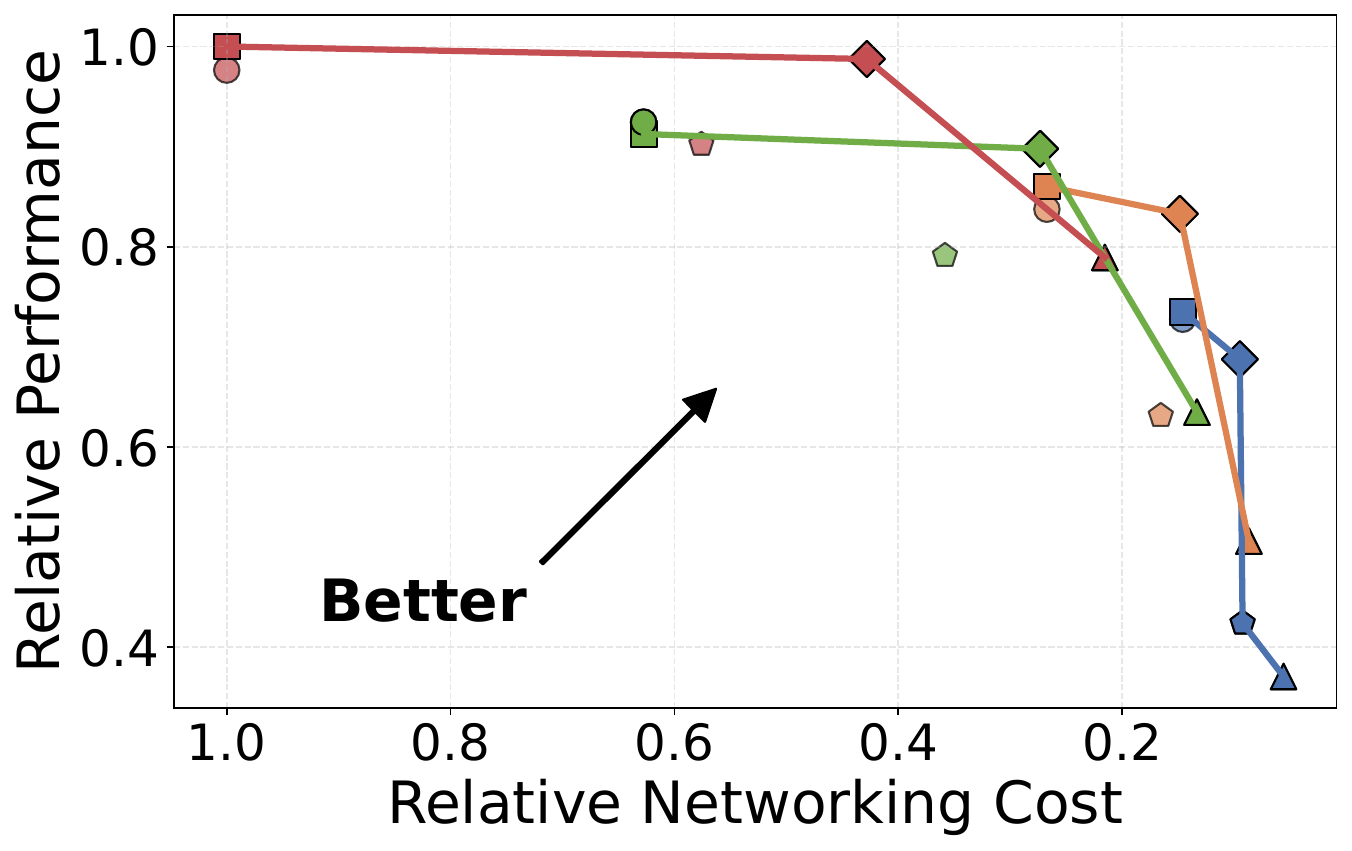}
        \vspace{-1.5em}
        \caption{\mixtrallarge{}}
        \label{fig:sim:perf-cost:100G}
    \end{subfigure} 
    \hspace{0.2em}
    \begin{subfigure}[t]{0.24\linewidth}
        \centering
        \includegraphics[width=\linewidth]{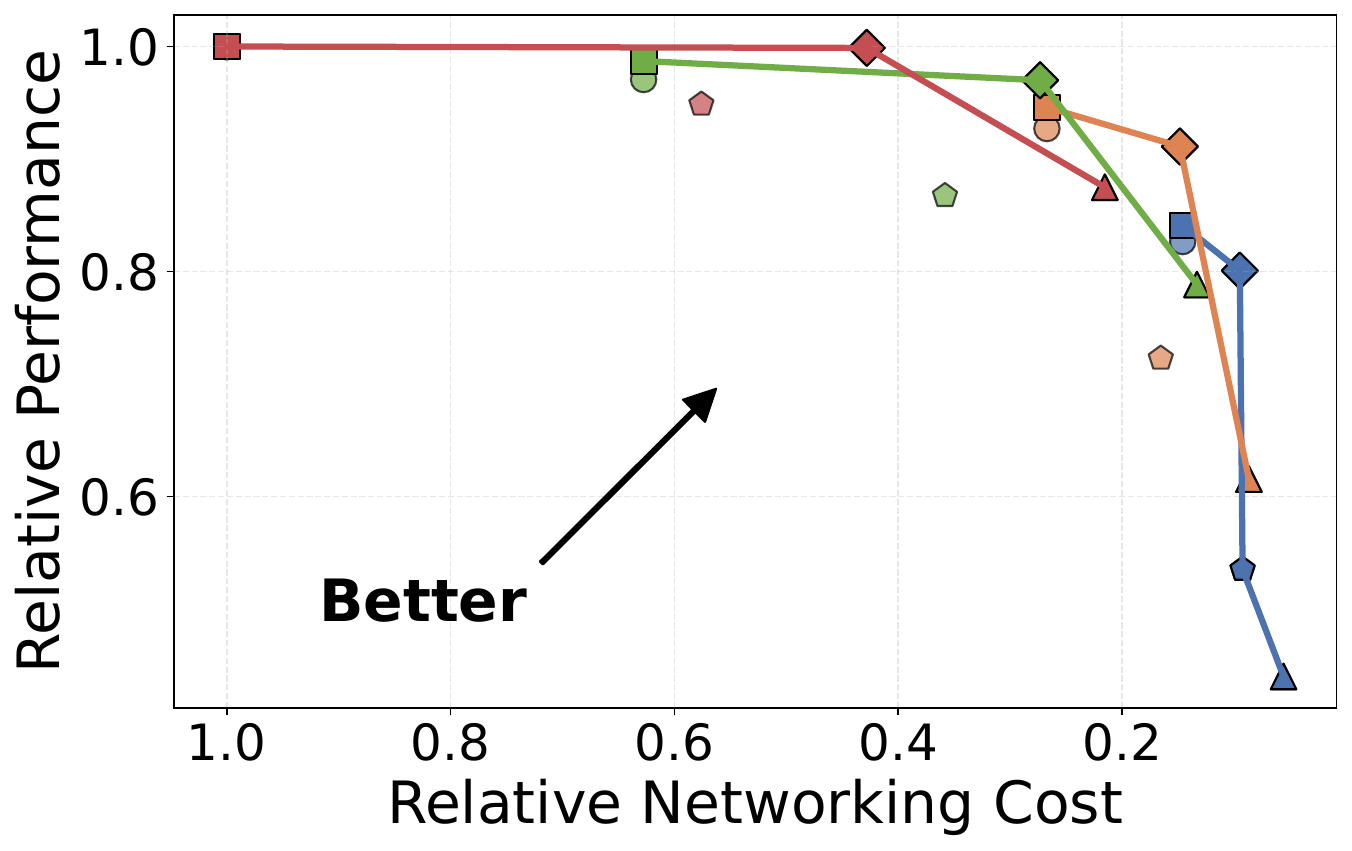}
        \vspace{-1.5em}
        \caption{\mixtral{}}
        \label{fig:sim:perf-cost:200G}
    \end{subfigure}
    \hspace{0.2em}
    \begin{subfigure}[t]{0.24\linewidth}
        \centering
        \includegraphics[width=\linewidth]{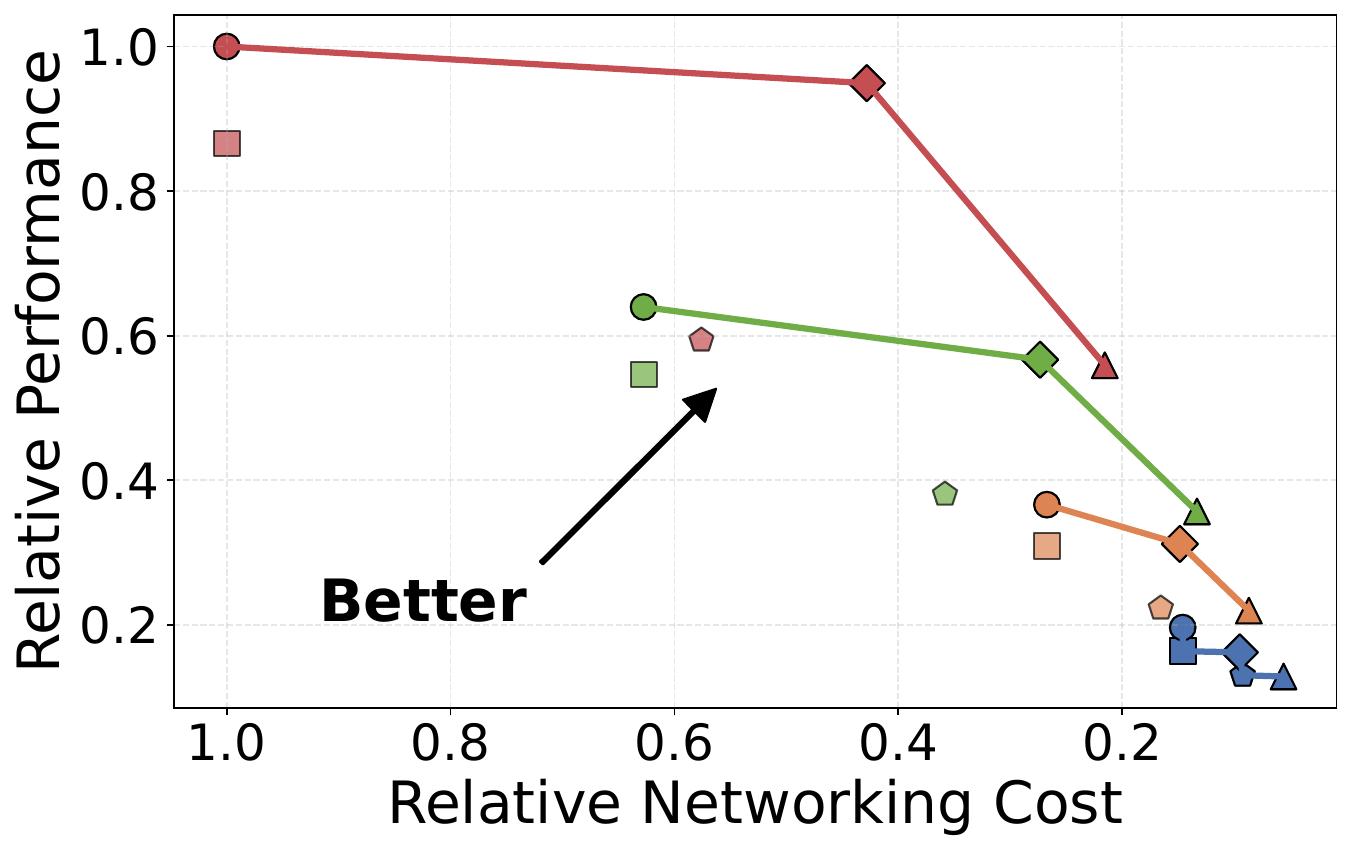}
        \vspace{-1.5em}
        \caption{\qwenmoe{}}
        \label{fig:sim:perf-cost:400G}
    \end{subfigure}
    \hspace{0.2em}
    \begin{subfigure}[t]{0.24\linewidth}
        \centering
        \includegraphics[width=\linewidth]{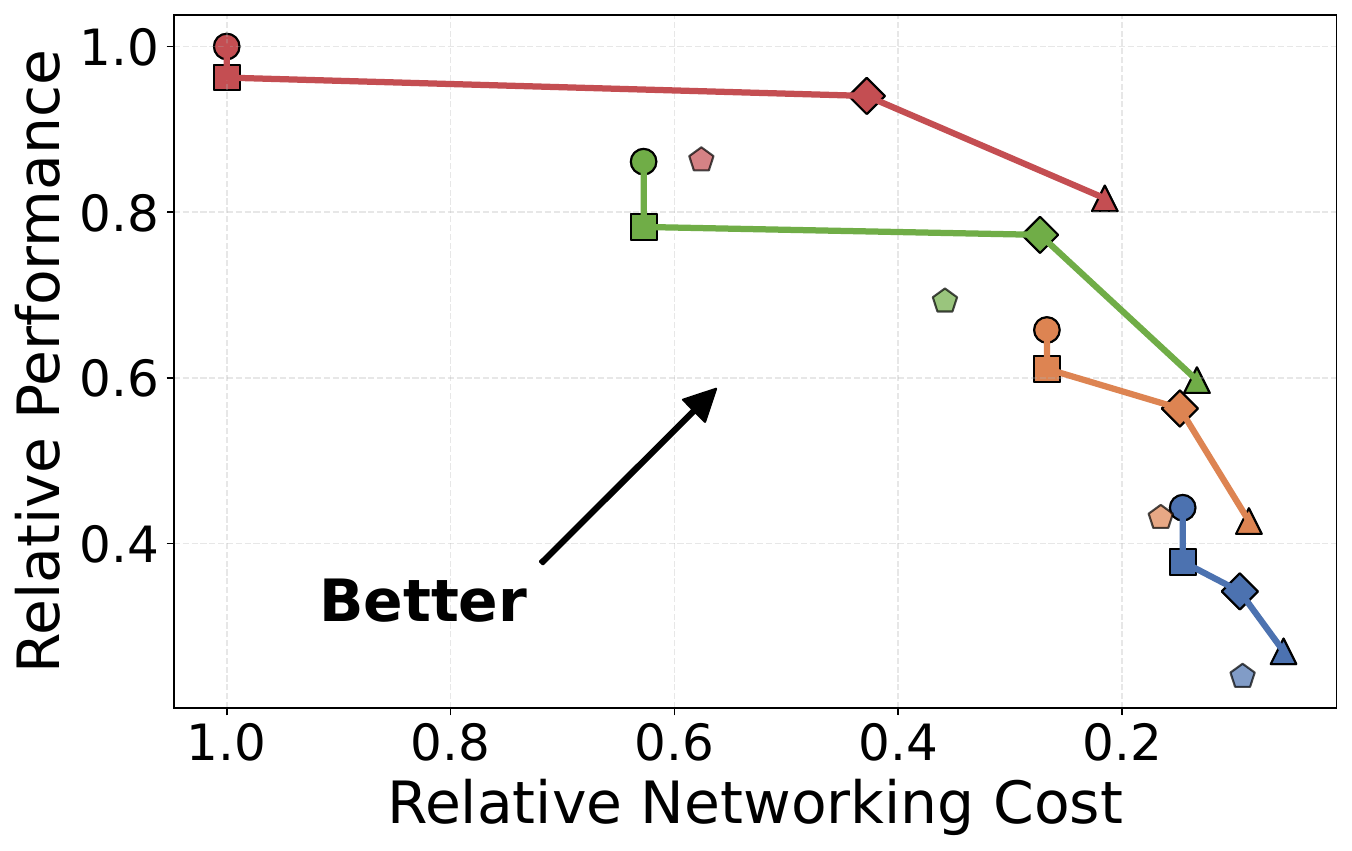}
        \vspace{-1.5em}
        \caption{\deepseek{}}
        \label{fig:sim:perf-cost:800G}
    \end{subfigure}
    \vspace{-1em}
    \caption{[Simulation] Performance-cost comparison of different interconnects on four state-of-the-art MoE models.
    }
    \label{fig:sim:perf-cost}
\end{figure*}

\subsection{Cost Efficiency: Pareto Front Analysis}
To better capture the trade-offs between networking cost and performance, we present the Pareto Front analysis of different interconnects in \figref{fig:sim:perf-cost}. This approach offers a more balanced view that avoids favoring low-cost yet low-performance designs (\eg TopoOpt or over-subscribed topologies), which may not be practically useful despite their low cost.
We observe that \sys{} consistently defines the Pareto Front and significantly outperforms Fat-tree and Rail-optimized across all four evaluated models in terms of cost efficiency, which is quantified as a performance-per-dollar metric (inverse of training iteration time normalized by networking cost).
At 100 Gbps link bandwidth, \sys{} achieves 1.2$\times$ to 1.5$\times$ higher cost-efficiency compared to Fat-tree, with \mixtral{} showing the highest improvement. Moreover, \sys{} outperforms Rail-optimized by 1.4$\times$ to 1.5$\times$.
At 200 Gbps link bandwidth, At 200 Gbps, the advantage grows to 1.4$\times$ to 1.8$\times$ over Fat-tree and 1.7$\times$ to 1.9$\times$ over Rail-optimized.
For 400 Gbps networks,  \sys{} demonstrates even higher cost-efficiency gains: 2.3$\times$ for \mixtral{}, 2.2$\times$ for \mixtrallarge{}. For \deepseek{}, \sys{} improves the training cost efficiency by 2.1$\times$ compared to Fat-tree.

Notably, \sys{} demonstrates strong cost-efficiency across varying link bandwidths, maintaining a 2.0$\times$–2.4$\times$ advantage over Fat-tree and 2.2$\times$–2.6$\times$ over Rail-optimized even at forward-looking 800 Gbps networks. These gains stem from two factors: 1) \sys{} directly connects high-traffic regional GPU pairs with optical circuits, reducing the need for excessive electrical switches and optical transceivers in Fat-tree; 2) unlike Fat-tree's underutilized uniform bisection bandwidth, \sys{} optimizes resource allocation to match MoE's sparse, non-uniform communication patterns, cutting hardware costs while sustaining high performance.

\subsection{Failure Resiliency}\label{sec:sim:failure-resiliency}
\begin{figure}[t!]
    \begin{subfigure}[t]{0.48\linewidth}
        \centering
        \includegraphics[width=\linewidth]{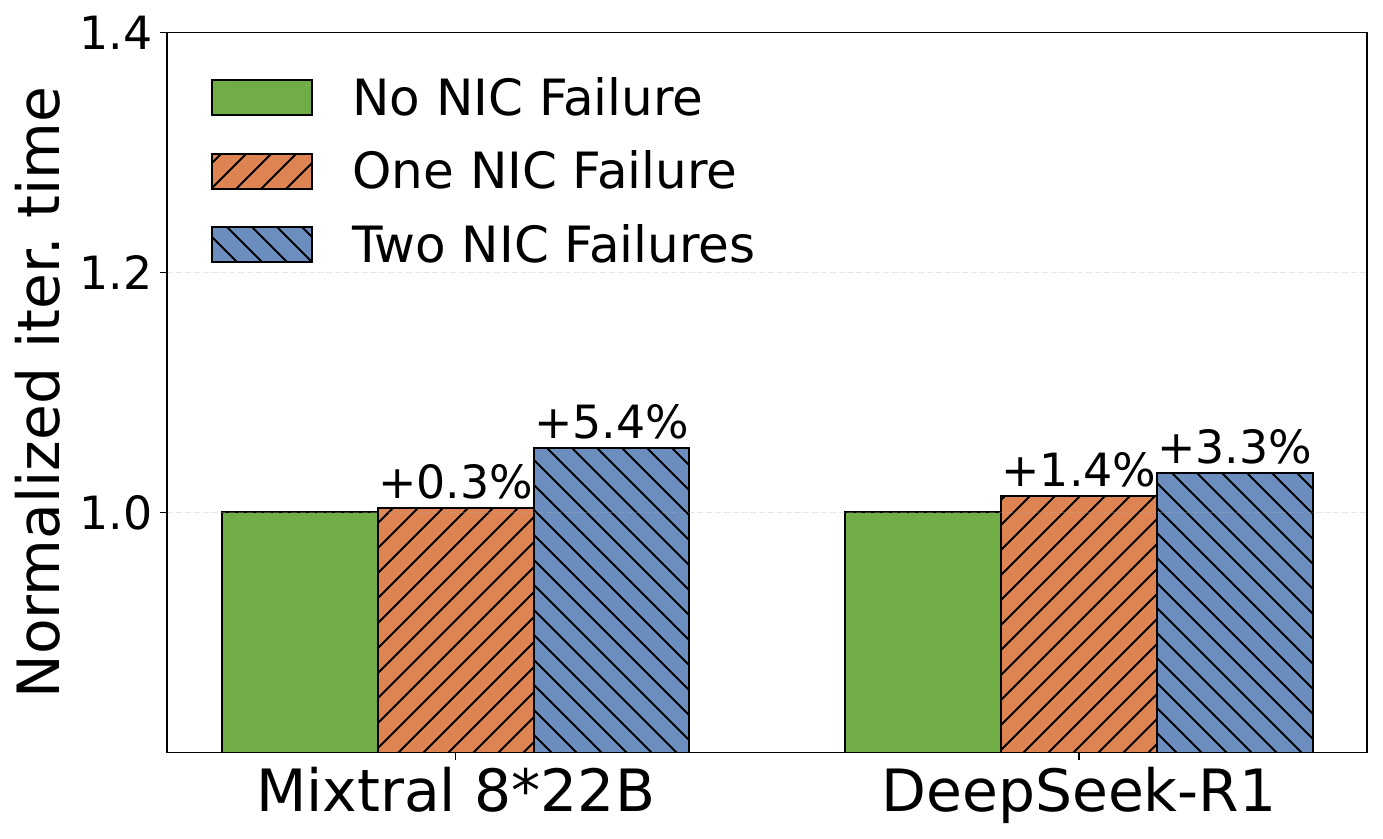}
        \vspace{-1.5em}
        \caption{NIC failure.}
        \label{fig:sim:failure:eps-nic}
    \end{subfigure}
    \hspace{0.2em}
    \begin{subfigure}[t]{0.48\linewidth}
        \centering
        \includegraphics[width=\linewidth]{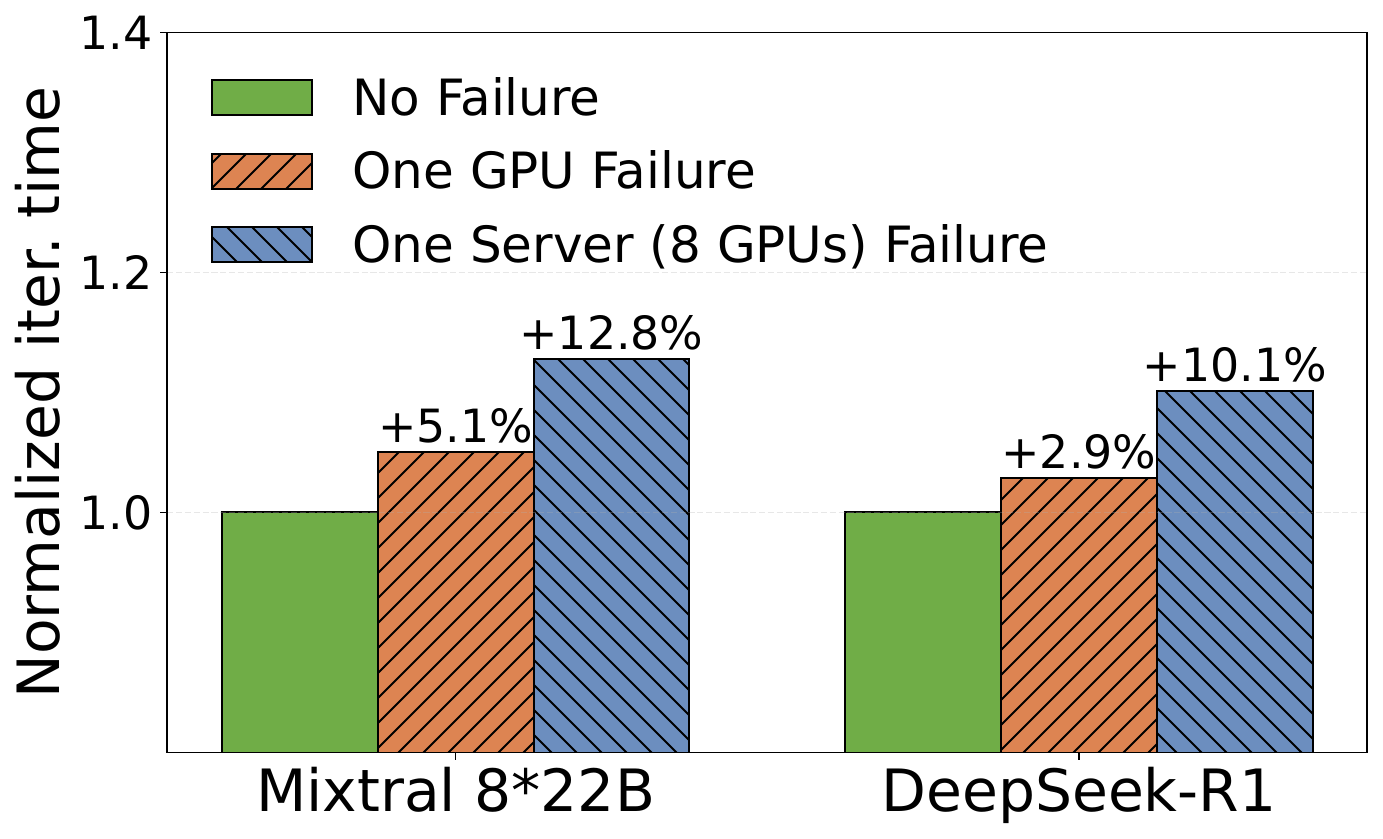}
        \vspace{-1.5em}
        \caption{GPU failure.}
        \label{fig:sim:failure:gpu}
    \end{subfigure}
    \vspace{-1em}
    \caption{[Simulation] Failure resiliency of \sysbf{}.}
    \label{fig:sim:failure}
\end{figure}

Following \secref{sec:design:failure-handling}, we evaluate the impact of different failure cases on training performance using the \mixtrallarge{} and \deepseek{} models on a 1024-GPU cluster with 400 Gbps link bandwidth.

\parab{NIC failures.} \figref{fig:sim:failure:eps-nic} shows the training performance of \sys{} when encountering NIC failures, where indirect forwarding is employed to bypass the failed NIC. We observe that \sys{} maintains acceptable performance, with only a 3.3\% increase in total training time for the \deepseek{} model. This minor overhead is attributed to the inherent network-wide backup policy, where EPS and OCS provide mutual fallback paths, and the intra-host scale-up domain offers sufficient forwarding bandwidth.

\parab{GPU failures.} We also evaluate the impact of GPU failures, as shown in \figref{fig:sim:failure:gpu}. For example, in the \mixtrallarge{} model, working around a single failed GPU via a regional backup GPU with OCS-based indirect forwarding leads to a 5.1\% increase in total training time. This additional overhead arises because the TP communication in \mixtrallarge{} occurs between servers through the low-bandwidth scale-out fabric, rather than the original high-bandwidth intra-host scale-up domain. In the more severe case of a full server failure (all 8 GPUs), the performance degradation is higher, as all EP traffic to and from the backup GPUs has to traverse the two connected EPS NICs, as discussed in \secref{sec:design:failure-handling}. Similarly, replacing a fully failed GPU server in the \deepseek{} model results in a 6.5\% performance degradation due to the constrained network connectivity for EP traffic in this scenario.

In summary, \sys{} exhibits resilience to both network and GPU failures, consistently delivering acceptable performance across the evaluated scenarios.

\section{Look Ahead: High-Radix Scale-Up Domains}
\label{sec:nvl72}
\begin{figure}
        \centering
        \includegraphics[width=\linewidth]{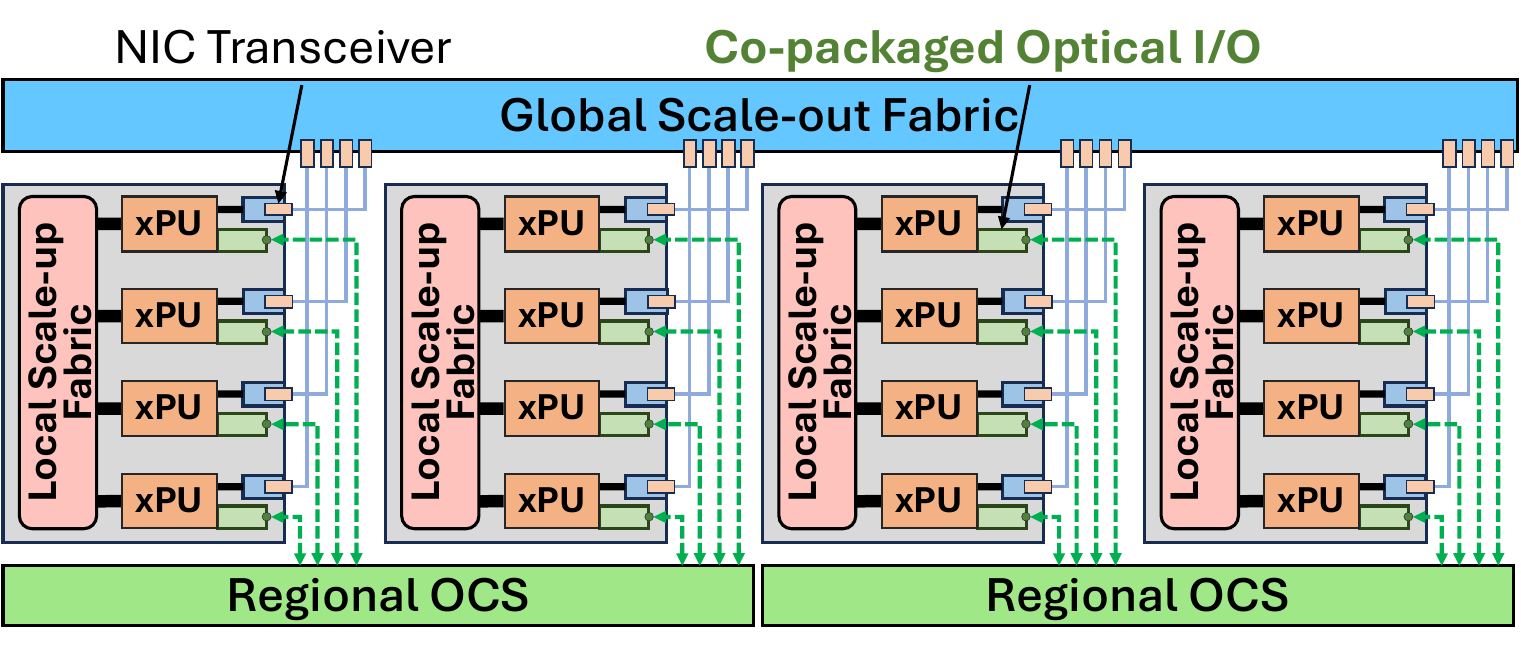}
        \vspace{-2.2em}
        \caption{\sysbf{} with co-packaged optical ports directly attached to xPUs.}
        \label{fig:conceptual:arch-5years}
\end{figure}

So far, we have discussed \sys{} as a production-ready system only using commodity OCS (\tabref{tab:ocs}) and networking equipment (e.g., NICs, transceivers) that are agnostic of local scale-up high-bandwidth domains.
In this section, we present a look-ahead study and extend the concept of \sys{} to better support the emerging trend of high-radix scale-up domains (e.g., Nvidia GB200 NVL72 system interconnects 72 GPUs in a single scale-up high-bandwidth domain)\footnote{\sys{} works with NVL72 by splitting scale-out NICs between OCS and EPS.}. In these systems, EP's all-to-all communications consume more bandwidth in the scale-up high-bandwidth domain than that of scale-out fabrics. Therefore, different from conventional settings, we consider the port of regional OCS to be directly attached to the GPU chip, which is enabled by co-packaged optical I/O~\cite{ayar-labs, lightmatter-passage}. Based on this, we envision \sys{} evolving towards a regional OCS architecture capable of directly receiving optical signals from xPUs (e.g., GPU, TPU, NPU, etc.), as illustrated in \figref{fig:conceptual:arch-5years}. Such an optical switching architecture would enable a long-reach, high-speed (\eg 4 Tbps or more) \emph{regional} fabric that sits at the boundary of scale-up and scale-out domains, supporting larger-scale switching fabric compared to state-of-the-art NVL72 copper-based interconnects.

\begin{figure}[t!]
    \centering
    \includegraphics[width=0.65\linewidth]{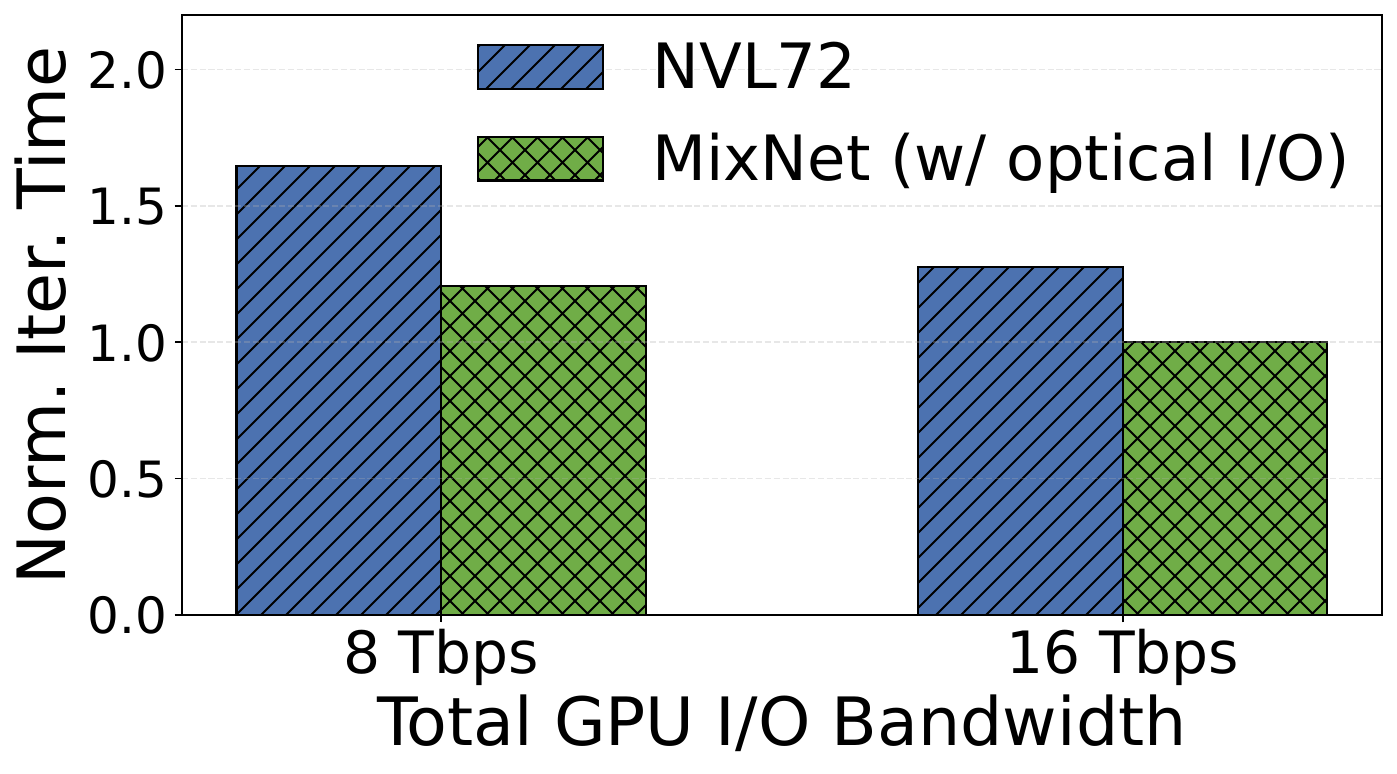}
    \vspace{-1em}
    \caption{[Simulation] Performance comparison with a 2048-GPU cluster of NVL72 systems.}
    \label{fig:sim:nvl72}
\end{figure}

Using packet-level simulations\footnote{Same setup as discussed in Section~\ref{sec:sim:setup}}, we compare the training iteration time of \sys{} (with optical I/O) with Nvidia GB200 NVL72~\cite{nvl72}. We consider a 2048-GPU cluster training the state-of-the-art MoE model, DeepSeek-V3~\cite{deepseekv3} (with an EP degree of 128, a PP degree of 16, and a micro-batch size of 240)\footnote{Same parameters as in \cite{deepseekv3}, with a larger EP size to explore larger training batch sizes and better expert capacities.}.
The NVL72 system is modeled with a per-GPU NVLink bandwidth of 7.2 Tbps~\cite{nvl72,nvlinkspec} in the scale-up domain, alongside 800 Gbps Ethernet for scale-out communication.
We assign 64 GPUs within each NVL72 domain to align with parallelism allocation constraints\footnote{In most production practices, only 64 out of 72 GPUs in NVL72 are used because training parallelisms are on the orders of 2 (e.g., 4, 8, 16, 32, etc.).}.
To ensure a fair comparison, we match \sys{}'s total GPU bandwidth in \sys{} to NVL72's 8 Tbps, allocating 800 Gbps to Ethernet and splitting the remaining bandwidth equally between NVLink and \sys{} (with optical I/O)'s regional OCS.

\figref{fig:sim:nvl72} presents the normalized training iteration time. We observe that \sys{} (with optical I/O) lowers iteration time by 1.3$\times$ compared to NVL72, as it offloads intensive cross-node communication overhead to regional reconfigurable OCS. Yet NVL72 has to use the scale-out networks for cross-node transfers. Additionally, \sys{} continues to deliver performance gains even when GPU total I/O bandwidth scales to 16 Tbps in next-generation systems.

\section{Discussion}\label{sec:discuss}

\parab{Additional training parallelisms.} Other potential parallelisms for LLM training, such as context parallelism~\cite{nvidia-cp}, optimizes the computation of long sequences in the attention module. Notably, its traffic is static and does not overlap with EP's all-to-all communications. \sys{} is ready to accommodate this traffic with its intra-host scale-up fabric or reconfigurable OCS fabric.

\parab{Support for model scaling.} 
Currently, MoE models have two scaling trends. On the one hand, recent models like Mixtral (8$\times$7B and 8$\times$22B) and Grok-2 contain a small number of \emph{huge} experts. This fits into the design of \sys{} that contains multiple scale-up networks within each reconfigurable high-bandwidth domain to support huge expert layers.
On the other hand, some models~\cite{qwen-moe, deepseek2, deepseekv3} choose to involve a large number of \emph{small} experts. Therefore, the radius of the EP all-to-all communications does not grow linearly with the number of experts, as multiple small experts will be packed into one GPU for training efficiency. For example, the state-of-the-art MoE model DeepSeek-V3~\cite{deepseekv3}, which uses an EP degree of 64 and a PP degree of 16, can be accommodated within multiple 64-port OCSes in an \sys{} fabric.

\parab{NIC's optical fanout options.} So far, we consider \sys{}'s NIC to be equipped with a dedicated port for each TX/RX channel connecting to the regional reconfigurable OCS. 
In practice, the \sys{} architecture is also compatible with other optical fanout options. For example, bi-directional transceivers using optical circulators merge TX and RX ports into one fiber port~\cite{jupiter-evolving, liu2023lightwave}, and optical breakout cables separate multiple fiber channels within a single MPO/MTP port into individual LC connections~\cite{breakout-cable}.

\parab{Supporting models beyond MoE.}
While \sys{} is primarily optimized for large-scale MoE training, it is also applicable to other non-MoE LLMs, such as GPT-3~\cite{gpt3} and LLaMA~\cite{llama}. The core benefit of \sys{} is the ability to reconfigure the network topology for non-uniform traffic patterns at a regional radius between the local scale-up networks and global scale-out networks. By blurring the boundaries of scale-up and scale-out networks using reconfigurable OCS, \sys{} supports cases
where non-MoE LLMs' non-uniform traffic patterns require optimized network topologies. For example, a ring-based topology serves DP all-reduce traffic in a more cost-efficient way than other topologies. Meanwhile, emerging LLM architectures in the ML community propose to build unconventional MoE-like models on top of dense LLMs that exhibit non-uniform traffic patterns at regional scales~\cite{lin2024moe, sukhbaatar2024branch}, highlighting the need of \sys{}'s reconfigurable topologies with for distributed training.

\parab{Support for Multi-Tenant Training} 
Different from conventional small-scale training jobs, running multiple concurrent MoE training jobs on the same set of GPUs is not preferred in industry practice. For example, Alibaba HPN~\cite{alibaba-hpn} highlights that a single training job already occupies 3K GPUs. Meta shares the experience of training Llama-3 405B model with 15K GPUs~\cite{llama3-white-paper}. DeepSeek-V3 was trained with 2K GPUs~\cite{deepseekv3}. Moreover, xAI discloses that they train the Grok model with 100K GPUs~\cite{grok-training-cluster}. Nevertheless, if needed, \sys{} supports cluster-wide multi-tenancy because its regional OCS high-bandwidth domains can be reconfigured as isolated sub-networks for each small-scale tenant job.

\parab{Comparability with other OCS technologies.} 
\sys{}'s regional reconfiguration design is orthogonal to the selection of OCS technologies. 
We note the emergence of new OCS technologies, such as iPronics~\cite{ipronics}, which offer microsecond-scale reconfiguration latencies and low per-port costs comparable to MEMS-based OCS. These advancements present a promising opportunity for \sys{} to reconfigure the topology for every all-to-all communication phase in MoE training, further improving training performance while maintaining cost efficiency.
Furthermore, in the main text, we mainly discuss OCS technologies that can actively switch optical signals at the network core. Other emerging OCS techniques, such as tunable laser at endpoints (e.g., transceivers) with arrayed waveguide grating router (AWGR)-based optical switch that pushes reconfigurability into the network endpoints while leaving the network core to be passive~\cite{sirius,zhang2024fast,ye2012awgr}, are also compatible with \sys{}. 
For example, assigning dedicated wavelengths to critical expert pairs eliminates bandwidth contention for all-to-all EP communication while organizing ring-based topologies for all-reduced communication of TP. However, these AWGR and endpoint-based OCS are primarily in the research stage and not commercially ready for mass production yet, leaving a longer way to be integrated into today's computer systems.

\section{Related work}\label{sec:related-work}
\parab{Network architecture for distributed training.} 
There has been a series of proposals on network architecture for large-scale distributed training in both industry and academia. Notably, ByteDance MegaScale~\cite{megascale} uses a Clos-based topology interconnecting more than 10,000 GPUs. Meta~\cite{meta-llm-network} shared its insights on the tuning of routing strategies, optimizing collective operations, and strengthening network resilience to design large-scale RoCE networks for AI training. Alibaba HPN~\cite{alibaba-hpn} introduced a dual-plane network to enhance resilience to failure. Nvidia developed a rail-optimized network~\cite{rail-optimized} to fully leverage the heterogeneous networking capabilities of different fabrics, which has been widely adopted in its computing clusters. Besides these industry proposals, academic researchers further proposed a rail-only design that removes the core switching layer for inter-rail GPUs, albeit at the cost of degrading cross-rail traffic performance~\cite{rail-only}. 

\parab{Reconfigurable networks for distributed training.} SiP-ML~\cite{sip-ml} explores silicon photonics for high-bandwidth optical interconnects for optimizing static communications from traditional DP and model-parallel communications. TopoOpt~\cite{topoopt} proposed optimizing the network topology for distributed training jobs using one-shot optically reconfigurable networks. To the best of our knowledge, \sys{} is the first to propose an optically reconfigurable network using commodity hardware for MoE training.

\parab{Reconfigurable data center networks.} There has been a decades-long research agenda focused on designing reconfigurable networks for data centers~\cite{helios,c-through,osa,mordia,reactor,sirius,xia2015enabling,liu2015scheduling,megaswitch,rotornet,opera,firefly,topoopt,shale,mellette2016scalable,farrington2013multiport,saran2024semi,wilson2024breaking,amir2023poster,wilson2023extending,amir2022optimal,raja2021ultrafast,gerard2021fast,zhang2021gemini,yu2017fast}. These proposals target generic data center networks, which are not optimized to provide cost-efficient solutions for large-scale MoE training. In particular, traffic-oblivious solutions, such as RotorNet~\cite{rotornet, realizing-rotornet} and Opera~\cite{opera}, result in suboptimal performance for MoE training, as they cannot deliver timely transfers for bandwidth-intensive all-to-all traffic. Meanwhile, the hardware innovations like Sirius~\cite{sirius} feature faster optical switching latency, hence allowing \sys{} to achieve much faster topology reconfiguration delay. Shoal~\cite{shoal} proposes  reconfigurable electronic circuit switches in rack-scale networks and is not unsuitable for large-scale MoE training.

\parab{OCS deployments in data centers.} Google has pioneered the deployments of OCS technology in production data centers. Over the years, they have transitioned from electronic packet switching (EPS) to hybrid optical-electrical solutions~\cite{fattree, b4,jupiter-rising,gibson2022aquila,jupiter-evolving,liu2023lightwave,zu2024resiliency}. Early systems like Jupiter~\cite{jupiter-rising} used Clos topologies with EPS to achieve scalability and high bandwidth but faced limitations in power efficiency and adaptability to dynamic workloads. Subsequent innovations in Jupiter Evolving~\cite{jupiter-evolving} introduced OCS to complement EPS in a hybrid architecture that improved cost and power efficiency. 
Most recently, Lightwave Fabrics from Google~\cite{liu2023lightwave} used reconfigurable MEMS-based OCS to establish topologies for TPU supercomputers~\cite{tpuv4}.
In particular, it performs one-shot topology reconfiguration prior to training such that the topology remains fixed throughout the training process. In contrast, \sys{} proposes runtime topology reconfiguration during training to accommodate dynamic MoE traffic patterns.
Furthermore, TPU's optical interconnect only reconfigures links between 4$\times$4$\times$4 cubes, while the intra-cube topology remains static (3D Torus) during OCS reconfiguration. This fixed intra-cube structure is ill-suited for dynamic all-to-all communication, which requires multi-hop forwarding in response to changing and sparse traffic demands~\cite{tpuv4}.

\parab{OCS for scale-up interconnects.} 
The regionally reconfigurable OCS design in \sys{} is applicable to expand the high-bandwidth circuit-switched connectivity enabled by NVLink and NVSwitch. 
For example, forward-looking techniques like Lightmatter passage optical interconnect~\cite{lightmatter-passage} benefit from \sys{} by reconfiguring all GPUs within the same EP group at the chip level. 
This would enable high-radix on-chip photonic communication with massive bandwidth to handle the communication-intensive demands of both TP and EP in MoE training. 

\parab{Emerging OCS hardware devices and systems.}
There are recent proposals on designing novel OCS hardware at server scale for chip-to-chip interconnects~\cite{lightmatter-passage,seok2016large, zhang2022beam,bunandar2024optical}. We would like to highlight that these emerging devices require a system-level design to be practical. Similar to other system-level work~\cite{kumar2024case}, \sys{}'s regional OCS and its algorithmic designs are compatible with this vibrant line of exploration on novel OCS hardware.

\section{Conclusion}
This paper presented \sys{}, a novel reconfigurable fabric using commodity hardware for large-scale MoE training. At the core of \sys{} is the design and implementation of the regionally reconfigurable high-bandwidth domain based on distributed OCS. 
Through proof-of-concept prototype and large-scale packet simulations, we show that \sys{} delivers distributed training performance comparable to state-of-the-art electrical and optical interconnects while significantly reducing networking cost. 

\parab{Ethics:} This work does not raise any ethical issues.\label{bodypage}

\section*{Acknowledgments}
We sincerely thank the anonymous SIGCOMM reviewers and our shepherd, Alex C. Snoeren, for their insightful feedback. We also thank Haiyang Chen, Decang Sun, Jipeng Zhang, and the Polatis team for their support in building the testbed. This work is supported in part by the Hong Kong RGC TRS T41-603/20R, ITC ACCESS, TACC~\cite{tacc-asplos}, EmbedWay research project, Beijing Municipal Science and Technology Project No. Z241100004224023, NSFC Excellent Young Scientists Fund Program (Overseas). Zhizhen Zhong and Kai Chen are the corresponding authors.

\bibliographystyle{ACM-Reference-Format}
\bibliography{main}


\begin{thebibliography}{118}


\ifx \showCODEN    \undefined \def \showCODEN     #1{\unskip}     \fi
\ifx \showDOI      \undefined \def \showDOI       #1{#1}\fi
\ifx \showISBNx    \undefined \def \showISBNx     #1{\unskip}     \fi
\ifx \showISBNxiii \undefined \def \showISBNxiii  #1{\unskip}     \fi
\ifx \showISSN     \undefined \def \showISSN      #1{\unskip}     \fi
\ifx \showLCCN     \undefined \def \showLCCN      #1{\unskip}     \fi
\ifx \shownote     \undefined \def \shownote      #1{#1}          \fi
\ifx \showarticletitle \undefined \def \showarticletitle #1{#1}   \fi
\ifx \showURL      \undefined \def \showURL       {\relax}        \fi
\providecommand\bibfield[2]{#2}
\providecommand\bibinfo[2]{#2}
\providecommand\natexlab[1]{#1}
\providecommand\showeprint[2][]{arXiv:#2}

\bibitem[\protect\citeauthoryear{??}{100}{[n. d.]a}]%
        {100g-transceiver}
 \bibinfo{year}{[n. d.]}\natexlab{a}.
\newblock \bibinfo{title}{{100GBASE-SR4 850nm 100m DOM MPO-12/UPC MMF Optical
  Transceiver Module}}.
\newblock \bibinfo{howpublished}{\url{https://www.fs.com/products/48354.html}}.
    (\bibinfo{year}{[n. d.]}).
\newblock


\bibitem[\protect\citeauthoryear{??}{200}{[n. d.]a}]%
        {200g-transceiver}
 \bibinfo{year}{[n. d.]}\natexlab{a}.
\newblock \bibinfo{title}{{200GBASE-SR4 850nm 100m DOM MPO-12/UPC MMF Optical
  Transceiver Module}}.
\newblock
  \bibinfo{howpublished}{\url{https://www.fs.com/products/139696.html}}.
  (\bibinfo{year}{[n. d.]}).
\newblock


\bibitem[\protect\citeauthoryear{??}{400}{[n. d.]a}]%
        {400g-transceiver}
 \bibinfo{year}{[n. d.]}\natexlab{a}.
\newblock \bibinfo{title}{{400GBASE-SR4 PAM4 850nm 100m DOM MPO-12/APC MMF
  Optical Transceiver Module}}.
\newblock
  \bibinfo{howpublished}{\url{https://www.fs.com/products/226577.html}}.
  (\bibinfo{year}{[n. d.]}).
\newblock


\bibitem[\protect\citeauthoryear{??}{meg}{[n. d.]}]%
        {megatron-token-collect}
 \bibinfo{year}{[n. d.]}\natexlab{}.
\newblock \bibinfo{title}{{All-to-all traffic demand collection in
  Megatron-LM}}.
\newblock
  \bibinfo{howpublished}{\url{https://github.com/NVIDIA/Megatron-LM/blob/461b06cd6d1fb4a625cebdbca499dac9484087fc/megatron/core/transformer/moe/token\_dispatcher.py\#L432}}.
    (\bibinfo{year}{[n. d.]}).
\newblock


\bibitem[\protect\citeauthoryear{??}{bre}{[n. d.]}]%
        {breakout-cable}
 \bibinfo{year}{[n. d.]}\natexlab{}.
\newblock \bibinfo{title}{{Breakout optical cables}}.
\newblock
  \bibinfo{howpublished}{\url{https://arubanetworking.hpe.com/techdocs/Switches/xcvrs/xcvr_guide/Content/Chp_overview/spl-opt-cab.htm}}.
    (\bibinfo{year}{[n. d.]}).
\newblock


\bibitem[\protect\citeauthoryear{??}{cal}{[n. d.]}]%
        {calient-mems}
 \bibinfo{year}{[n. d.]}\natexlab{}.
\newblock \bibinfo{title}{{Calient Optical Circuit Switch}}.
\newblock \bibinfo{howpublished}{\url{www.calient.net}}.   (\bibinfo{year}{[n.
  d.]}).
\newblock


\bibitem[\protect\citeauthoryear{??}{bur}{[n. d.]a}]%
        {burst-mode-transceiver1}
 \bibinfo{year}{[n. d.]}\natexlab{a}.
\newblock \bibinfo{title}{{Calix® 100-04482 Compatible TAA 10GBs XGS-PON OLT
  XFP Transceiver with Burst Mode (SMF, 1577nmTx/1270nmRx, SC, N1, DOM)}}.
\newblock
  \bibinfo{howpublished}{\url{https://www.addonnetworks.com/products/transceivers/calix/xfp/10gbase/100-04482-ao}}.
    (\bibinfo{year}{[n. d.]}).
\newblock


\bibitem[\protect\citeauthoryear{??}{nvi}{[n. d.]a}]%
        {nvidia-cp}
 \bibinfo{year}{[n. d.]}\natexlab{a}.
\newblock \bibinfo{title}{Context parallelism overview}.
\newblock
  \bibinfo{howpublished}{\url{https://docs.nvidia.com/megatron-core/developer-guide/latest/api-guide/context_parallel.html}}.
    (\bibinfo{year}{[n. d.]}).
\newblock


\bibitem[\protect\citeauthoryear{??}{dee}{[n. d.]}]%
        {deepseekv3}
 \bibinfo{year}{[n. d.]}\natexlab{}.
\newblock \bibinfo{title}{{DeepSeek-V3}}.
\newblock
  \bibinfo{howpublished}{\url{https://github.com/deepseek-ai/DeepSeek-V3/blob/main/DeepSeek_V3.pdf}}.
    (\bibinfo{year}{[n. d.]}).
\newblock


\bibitem[\protect\citeauthoryear{??}{gpt}{[n. d.]}]%
        {gpt3}
 \bibinfo{year}{[n. d.]}\natexlab{}.
\newblock \bibinfo{title}{DeepSpeed}.
\newblock \bibinfo{howpublished}{\url{https://openai.com/index/gpt-3-apps/}}.
  (\bibinfo{year}{[n. d.]}).
\newblock


\bibitem[\protect\citeauthoryear{??}{rai}{[n. d.]}]%
        {rail-optimized}
 \bibinfo{year}{[n. d.]}\natexlab{}.
\newblock \bibinfo{title}{Doubling all2all Performance with NVIDIA Collective
  Communication Library 2.12}.
\newblock
  \bibinfo{howpublished}{\url{https://developer.nvidia.com/blog/doubling-all2all-performance-with-nvidia-collective-communication-library-2-12/}}.
    (\bibinfo{year}{[n. d.]}).
\newblock


\bibitem[\protect\citeauthoryear{??}{fle}{[n. d.]}]%
        {flexflow}
 \bibinfo{year}{[n. d.]}\natexlab{}.
\newblock \bibinfo{title}{FlexFlow}.
\newblock \bibinfo{howpublished}{\url{https://github.com/flexflow/FlexFlow}}.
  (\bibinfo{year}{[n. d.]}).
\newblock


\bibitem[\protect\citeauthoryear{??}{800}{[n. d.]a}]%
        {800g-transceiver}
 \bibinfo{year}{[n. d.]}\natexlab{a}.
\newblock \bibinfo{title}{{Generic Compatible 800GBASE-SR8 QSFP-DD PAM4 850nm
  50m DOM MPO-16/APC MMF Optical Transceiver Module}}.
\newblock
  \bibinfo{howpublished}{\url{https://www.fs.com/products/200921.html?attribute=93760&id=3569240}}.
    (\bibinfo{year}{[n. d.]}).
\newblock


\bibitem[\protect\citeauthoryear{??}{gro}{[n. d.]a}]%
        {grok}
 \bibinfo{year}{[n. d.]}\natexlab{a}.
\newblock \bibinfo{title}{Grok}.
\newblock \bibinfo{howpublished}{\url{https://x.ai/blog/grok-1.5v}}.
  (\bibinfo{year}{[n. d.]}).
\newblock


\bibitem[\protect\citeauthoryear{??}{gro}{[n. d.]b}]%
        {grok-training-cluster}
 \bibinfo{year}{[n. d.]}\natexlab{b}.
\newblock \bibinfo{title}{{Grok training cluster}}.
\newblock
  \bibinfo{howpublished}{\url{https://www.windowscentral.com/software-apps/elon-musk-flaunts-the-most-powerful-training-cluster-in-the-world-that-will-transform-grok-into-the-most-powerful-ai-by-december-to-take-on-microsoft-and-openai}}.
    (\bibinfo{year}{[n. d.]}).
\newblock


\bibitem[\protect\citeauthoryear{??}{hts}{[n. d.]}]%
        {htsim}
 \bibinfo{year}{[n. d.]}\natexlab{}.
\newblock \bibinfo{title}{htsim simulator}.
\newblock
  \bibinfo{howpublished}{\url{https://github.com/nets-cs-pub-ro/NDP/wiki/NDP-Simulator}}.
    (\bibinfo{year}{[n. d.]}).
\newblock


\bibitem[\protect\citeauthoryear{??}{ipr}{[n. d.]}]%
        {ipronics}
 \bibinfo{year}{[n. d.]}\natexlab{}.
\newblock \bibinfo{title}{{IPRONICS One}}.
\newblock
  \bibinfo{howpublished}{\url{https://ipronics.com/ipronics-optical-networking-engine/}}.
    (\bibinfo{year}{[n. d.]}).
\newblock


\bibitem[\protect\citeauthoryear{??}{dup}{[n. d.]}]%
        {duplex-fc-fiber}
 \bibinfo{year}{[n. d.]}\natexlab{}.
\newblock \bibinfo{title}{{LC UPC to LC UPC, Duplex, 2 Fibers}}.
\newblock \bibinfo{howpublished}{\url{https://www.fs.com/products/40191.html}}.
    (\bibinfo{year}{[n. d.]}).
\newblock


\bibitem[\protect\citeauthoryear{??}{lig}{[n. d.]}]%
        {lightmatter-passage}
 \bibinfo{year}{[n. d.]}\natexlab{}.
\newblock \bibinfo{title}{{LightMatter}}.
\newblock
  \bibinfo{howpublished}{\url{https://lightmatter.co/products/passage/}}.
  (\bibinfo{year}{[n. d.]}).
\newblock


\bibitem[\protect\citeauthoryear{??}{lla}{[n. d.]}]%
        {llama}
 \bibinfo{year}{[n. d.]}\natexlab{}.
\newblock \bibinfo{title}{LLaMA}.
\newblock \bibinfo{howpublished}{\url{https://llama.meta.com}}.
  (\bibinfo{year}{[n. d.]}).
\newblock


\bibitem[\protect\citeauthoryear{??}{mix}{[n. d.]a}]%
        {mixtral22b}
 \bibinfo{year}{[n. d.]}\natexlab{a}.
\newblock \bibinfo{title}{{Mixtral 8x22B}}.
\newblock
  \bibinfo{howpublished}{\url{https://huggingface.co/mistralai/Mixtral-8x22B-Instruct-v0.1}}.
    (\bibinfo{year}{[n. d.]}).
\newblock


\bibitem[\protect\citeauthoryear{??}{mix}{[n. d.]b}]%
        {mixtral-huggingface}
 \bibinfo{year}{[n. d.]}\natexlab{b}.
\newblock \bibinfo{title}{{Mixtral-8x7B-Instruct-v0.1}}.
\newblock
  \bibinfo{howpublished}{\url{https://huggingface.co/mistralai/Mixtral-8x7B-Instruct-v0.1}}.
    (\bibinfo{year}{[n. d.]}).
\newblock


\bibitem[\protect\citeauthoryear{??}{mix}{[n. d.]c}]%
        {mixtral}
 \bibinfo{year}{[n. d.]}\natexlab{c}.
\newblock \bibinfo{title}{{Mixtral of experts}}.
\newblock
  \bibinfo{howpublished}{\url{https://mistral.ai/news/mixtral-of-experts/}}.
  (\bibinfo{year}{[n. d.]}).
\newblock


\bibitem[\protect\citeauthoryear{??}{400}{[n. d.]b}]%
        {400g-switch}
 \bibinfo{year}{[n. d.]}\natexlab{b}.
\newblock \bibinfo{title}{{MSN4700-WS2FC, NVIDIA® Mellanox Spectrum-3 Based
  32-Port Ethernet L3 Data Center Switch, 32 x 400Gb QSFP-DD}}.
\newblock
  \bibinfo{howpublished}{\url{https://www.colfaxdirect.com/store/pc/viewPrd.asp?idproduct=4135}}.
    (\bibinfo{year}{[n. d.]}).
\newblock


\bibitem[\protect\citeauthoryear{??}{PLZ}{[n. d.]}]%
        {PLZT}
 \bibinfo{year}{[n. d.]}\natexlab{}.
\newblock \bibinfo{title}{{Nano-Second Speed PLZT Photonics}}.
\newblock \bibinfo{howpublished}{\url{http://epiphotonics.com/products.html}}.
   (\bibinfo{year}{[n. d.]}).
\newblock


\bibitem[\protect\citeauthoryear{??}{ncc}{[n. d.]}]%
        {nccl}
 \bibinfo{year}{[n. d.]}\natexlab{}.
\newblock \bibinfo{title}{NVIDIA Collective Communications Library (NCCL)}.
\newblock \bibinfo{howpublished}{\url{https://developer.nvidia.com/nccl}}.
  (\bibinfo{year}{[n. d.]}).
\newblock


\bibitem[\protect\citeauthoryear{??}{nvi}{[n. d.]b}]%
        {nvidia-dgx}
 \bibinfo{year}{[n. d.]}\natexlab{b}.
\newblock \bibinfo{title}{{NVIDIA DGX SuperPOD}}.
\newblock
  \bibinfo{howpublished}{\url{https://www.nvidia.com/en-us/data-center/dgx-superpod/}}.
    (\bibinfo{year}{[n. d.]}).
\newblock


\bibitem[\protect\citeauthoryear{??}{nvl}{[n. d.]a}]%
        {nvl72}
 \bibinfo{year}{[n. d.]}\natexlab{a}.
\newblock \bibinfo{title}{NVIDIA GB200 NVL72 Delivers Trillion-Parameter LLM
  Training and Real-Time Inference}.
\newblock
  \bibinfo{howpublished}{\url{https://developer.nvidia.com/blog/nvidia-gb200-nvl72-delivers-trillion-parameter-llm-training-and-real-time-inference/}}.
    (\bibinfo{year}{[n. d.]}).
\newblock


\bibitem[\protect\citeauthoryear{??}{100}{[n. d.]b}]%
        {100g-nic}
 \bibinfo{year}{[n. d.]}\natexlab{b}.
\newblock \bibinfo{title}{{NVIDIA Mellanox MCX515A-CCAT ConnectX®-5 EN Network
  Interface Card, 100GbE Single-Port QSFP28}}.
\newblock
  \bibinfo{howpublished}{\url{https://www.fs.com/products/119648.html}}.
  (\bibinfo{year}{[n. d.]}).
\newblock


\bibitem[\protect\citeauthoryear{??}{200}{[n. d.]b}]%
        {200g-nic}
 \bibinfo{year}{[n. d.]}\natexlab{b}.
\newblock \bibinfo{title}{{NVIDIA Mellanox MCX653105A-HDAT ConnectX®-6
  InfiniBand/VPI Adapter Card 200GbE/HDR, Single-Port QSFP56}}.
\newblock
  \bibinfo{howpublished}{\url{https://www.fs.com/products/168437.html}}.
  (\bibinfo{year}{[n. d.]}).
\newblock


\bibitem[\protect\citeauthoryear{??}{400}{[n. d.]c}]%
        {400g-nic}
 \bibinfo{year}{[n. d.]}\natexlab{c}.
\newblock \bibinfo{title}{{NVIDIA Mellanox MCX75310AAS-NEAT ConnectX®-7
  InfiniBand/VPI Adapter Card 400GbE/NDR, Single-Port OSFP}}.
\newblock
  \bibinfo{howpublished}{\url{https://www.fs.com/products/212161.html}}.
  (\bibinfo{year}{[n. d.]}).
\newblock


\bibitem[\protect\citeauthoryear{??}{800}{[n. d.]b}]%
        {800g-switch}
 \bibinfo{year}{[n. d.]}\natexlab{b}.
\newblock \bibinfo{title}{NVIDIA Mellanox Spectrum-4 SN5600 800G 64-Port
  51.2Tb/s 2U Data Center Switch}.
\newblock
  \bibinfo{howpublished}{\url{https://fireowls.com/products/920-9n42f-00ri-7c0-nvidia-spectrum-sn5600-ethernet-switch?srsltid=AfmBOoqC5-OQefz1MpAi_2QkzW2tnaCTXA_xGtgoj9b3NphWUXw8dBrl}}.
    (\bibinfo{year}{[n. d.]}).
\newblock


\bibitem[\protect\citeauthoryear{??}{nvl}{[n. d.]b}]%
        {nvlink}
 \bibinfo{year}{[n. d.]}\natexlab{b}.
\newblock \bibinfo{title}{NVIDIA NVLink and NVLink Switch}.
\newblock
  \bibinfo{howpublished}{\url{https://www.nvidia.com/en-us/data-center/nvlink/}}.
    (\bibinfo{year}{[n. d.]}).
\newblock


\bibitem[\protect\citeauthoryear{??}{nvs}{[n. d.]}]%
        {nvswitch}
 \bibinfo{year}{[n. d.]}\natexlab{}.
\newblock \bibinfo{title}{NVIDIA NVSwitch Technical Overview}.
\newblock
  \bibinfo{howpublished}{\url{https://images.nvidia.com/content/pdf/nvswitch-technical-overview.pdf}}.
    (\bibinfo{year}{[n. d.]}).
\newblock


\bibitem[\protect\citeauthoryear{??}{sn3}{[n. d.]}]%
        {sn3700}
 \bibinfo{year}{[n. d.]}\natexlab{}.
\newblock \bibinfo{title}{{NVIDIA Spectrum SN3700}}.
\newblock
  \bibinfo{howpublished}{\url{https://marketplace.nvidia.com/en-us/enterprise/networking/sn3700/}}.
    (\bibinfo{year}{[n. d.]}).
\newblock


\bibitem[\protect\citeauthoryear{??}{nvl}{[n. d.]c}]%
        {nvlinkspec}
 \bibinfo{year}{[n. d.]}\natexlab{c}.
\newblock \bibinfo{title}{{NVLink Switch Specifications}}.
\newblock
  \bibinfo{howpublished}{\url{https://www.nvidia.com/en-us/data-center/nvlink/?ncid=no-ncid\#nvlink-switch-specifications}}.
    (\bibinfo{year}{[n. d.]}).
\newblock


\bibitem[\protect\citeauthoryear{??}{ocs}{[n. d.]}]%
        {ocs-price}
 \bibinfo{year}{[n. d.]}\natexlab{}.
\newblock \bibinfo{title}{{Polatis Optical Circuit Switch}}.
\newblock \bibinfo{howpublished}{\url{https:
  //www.polatis.com/series-7000-384x384-port-software-controlled-optical-circuitswitch-sdn-enabled.asp}}.
    (\bibinfo{year}{[n. d.]}).
\newblock


\bibitem[\protect\citeauthoryear{??}{pol}{[n. d.]}]%
        {polatis}
 \bibinfo{year}{[n. d.]}\natexlab{}.
\newblock \bibinfo{title}{Polatis Optical Switches}.
\newblock
  \bibinfo{howpublished}{\url{http://www.polatis.com/series-7000-384x384-port-software-controlled-optical-circuit-switch-sdn-enabled.asp}}.
    (\bibinfo{year}{[n. d.]}).
\newblock


\bibitem[\protect\citeauthoryear{??}{bur}{[n. d.]b}]%
        {burst-mode-transceiver2}
 \bibinfo{year}{[n. d.]}\natexlab{b}.
\newblock \bibinfo{title}{{PON/Burst-mode Transceivers}}.
\newblock
  \bibinfo{howpublished}{\url{https://oesolutions.com/product-type/burst-mode-transceivers/}}.
    (\bibinfo{year}{[n. d.]}).
\newblock


\bibitem[\protect\citeauthoryear{??}{qwe}{[n. d.]}]%
        {qwen-moe}
 \bibinfo{year}{[n. d.]}\natexlab{}.
\newblock \bibinfo{title}{Qwen1.5-MoE-A2.7B}.
\newblock
  \bibinfo{howpublished}{\url{https://huggingface.co/Qwen/Qwen1.5-MoE-A2.7B}}.
   (\bibinfo{year}{[n. d.]}).
\newblock


\bibitem[\protect\citeauthoryear{??}{aya}{[n. d.]}]%
        {ayar-labs}
 \bibinfo{year}{[n. d.]}\natexlab{}.
\newblock \bibinfo{title}{{Rethinking AI Architectures with Optical I/O}}.
\newblock
  \bibinfo{howpublished}{\url{https://www.windowscentral.com/software-apps/elon-musk-flaunts-the-most-powerful-training-cluster-in-the-world-that-will-transform-grok-into-the-most-powerful-ai-by-december-to-take-on-microsoft-and-openai}}.
    (\bibinfo{year}{[n. d.]}).
\newblock


\bibitem[\protect\citeauthoryear{??}{pat}{[n. d.]}]%
        {patch-panel-price}
 \bibinfo{year}{[n. d.]}\natexlab{}.
\newblock \bibinfo{title}{{Telescent G4 Network Topology Manager}}.
\newblock \bibinfo{howpublished}{\url{https://www.telescent.com/products}}.
  (\bibinfo{year}{[n. d.]}).
\newblock


\bibitem[\protect\citeauthoryear{??}{xve}{[n. d.]}]%
        {xverse-moe}
 \bibinfo{year}{[n. d.]}\natexlab{}.
\newblock \bibinfo{title}{{XVERSE-MoE-A36B MoE base model}}.
\newblock
  \bibinfo{howpublished}{\url{https://huggingface.co/xverse/XVERSE-MoE-A36B}}.
   (\bibinfo{year}{[n. d.]}).
\newblock


\bibitem[\protect\citeauthoryear{Al-Fares, Loukissas, and Vahdat}{Al-Fares
  et~al\mbox{.}}{2008}]%
        {fattree}
\bibfield{author}{\bibinfo{person}{Mohammad Al-Fares},
  \bibinfo{person}{Alexander Loukissas}, {and} \bibinfo{person}{Amin Vahdat}.}
  \bibinfo{year}{2008}\natexlab{}.
\newblock \showarticletitle{A scalable, commodity data center network
  architecture}. In \bibinfo{booktitle}{{\em ACM SIGCOMM Computer Communication
  Review}}, Vol.~\bibinfo{volume}{38}. \bibinfo{publisher}{ACM New York, NY,
  USA}, \bibinfo{pages}{63--74}.
\newblock


\bibitem[\protect\citeauthoryear{Amir, Saran, Wilson, Kleinberg, Shrivastav,
  and Weatherspoon}{Amir et~al\mbox{.}}{2024}]%
        {shale}
\bibfield{author}{\bibinfo{person}{Daniel Amir}, \bibinfo{person}{Nitika
  Saran}, \bibinfo{person}{Tegan Wilson}, \bibinfo{person}{Robert Kleinberg},
  \bibinfo{person}{Vishal Shrivastav}, {and} \bibinfo{person}{Hakim
  Weatherspoon}.} \bibinfo{year}{2024}\natexlab{}.
\newblock \showarticletitle{Shale: A Practical, Scalable Oblivious
  Reconfigurable Network}. In \bibinfo{booktitle}{{\em Proceedings of the ACM
  SIGCOMM 2024 Conference}} {\em (\bibinfo{series}{ACM SIGCOMM '24})}.
\newblock
\showURL{%
\url{https://doi.org/10.1145/3651890.3672248}}


\bibitem[\protect\citeauthoryear{Amir, Wilson, Shrivastav, Weatherspoon, and
  Kleinberg}{Amir et~al\mbox{.}}{2023}]%
        {amir2023poster}
\bibfield{author}{\bibinfo{person}{Daniel Amir}, \bibinfo{person}{Tegan
  Wilson}, \bibinfo{person}{Vishal Shrivastav}, \bibinfo{person}{Hakim
  Weatherspoon}, {and} \bibinfo{person}{Robert Kleinberg}.}
  \bibinfo{year}{2023}\natexlab{}.
\newblock \showarticletitle{Poster: Scalability and congestion control in
  oblivious reconfigurable networks}. In \bibinfo{booktitle}{{\em Proceedings
  of the ACM SIGCOMM 2023 Conference}}. \bibinfo{pages}{1138--1140}.
\newblock


\bibitem[\protect\citeauthoryear{Amir, Wilson, Shrivastav, Weatherspoon,
  Kleinberg, and Agarwal}{Amir et~al\mbox{.}}{2022}]%
        {amir2022optimal}
\bibfield{author}{\bibinfo{person}{Daniel Amir}, \bibinfo{person}{Tegan
  Wilson}, \bibinfo{person}{Vishal Shrivastav}, \bibinfo{person}{Hakim
  Weatherspoon}, \bibinfo{person}{Robert Kleinberg}, {and}
  \bibinfo{person}{Rachit Agarwal}.} \bibinfo{year}{2022}\natexlab{}.
\newblock \showarticletitle{Optimal oblivious reconfigurable networks}. In
  \bibinfo{booktitle}{{\em Proceedings of the 54th Annual ACM SIGACT Symposium
  on Theory of Computing}}. \bibinfo{pages}{1339--1352}.
\newblock


\bibitem[\protect\citeauthoryear{Ballani, Costa, Behrendt, Cletheroe, Haller,
  Jozwik, Karinou, Lange, Shi, Thomsen, et~al\mbox{.}}{Ballani
  et~al\mbox{.}}{2020}]%
        {sirius}
\bibfield{author}{\bibinfo{person}{Hitesh Ballani}, \bibinfo{person}{Paolo
  Costa}, \bibinfo{person}{Raphael Behrendt}, \bibinfo{person}{Daniel
  Cletheroe}, \bibinfo{person}{Istvan Haller}, \bibinfo{person}{Krzysztof
  Jozwik}, \bibinfo{person}{Fotini Karinou}, \bibinfo{person}{Sophie Lange},
  \bibinfo{person}{Kai Shi}, \bibinfo{person}{Benn Thomsen}, {et~al\mbox{.}}}
  \bibinfo{year}{2020}\natexlab{}.
\newblock \showarticletitle{Sirius: A flat datacenter network with nanosecond
  optical switching}. In \bibinfo{booktitle}{{\em Proceedings of the Annual
  conference of the ACM Special Interest Group on Data Communication on the
  applications, technologies, architectures, and protocols for computer
  communication}}. \bibinfo{pages}{782--797}.
\newblock


\bibitem[\protect\citeauthoryear{Bunandar, Gupta, Rosenberg, Chao, Liu, and
  Harris}{Bunandar et~al\mbox{.}}{2024}]%
        {bunandar2024optical}
\bibfield{author}{\bibinfo{person}{Darius Bunandar}, \bibinfo{person}{Shashank
  Gupta}, \bibinfo{person}{Jessie Rosenberg}, \bibinfo{person}{Clifford Chao},
  \bibinfo{person}{Kuang Liu}, {and} \bibinfo{person}{Nicholas~C Harris}.}
  \bibinfo{year}{2024}\natexlab{}.
\newblock \bibinfo{title}{Optical communication substrate using glass
  interposer}.
\newblock   (\bibinfo{date}{Oct.~24} \bibinfo{year}{2024}).
\newblock
\newblock
\shownote{US Patent App. 18/638,820.}


\bibitem[\protect\citeauthoryear{Chen, Singla, Singh, Ramachandran, Xu, Zhang,
  Wen, and Chen}{Chen et~al\mbox{.}}{2012}]%
        {osa}
\bibfield{author}{\bibinfo{person}{Kai Chen}, \bibinfo{person}{Ankit Singla},
  \bibinfo{person}{Atul Singh}, \bibinfo{person}{Kishore Ramachandran},
  \bibinfo{person}{Lei Xu}, \bibinfo{person}{Yueping Zhang},
  \bibinfo{person}{Xitao Wen}, {and} \bibinfo{person}{Yan Chen}.}
  \bibinfo{year}{2012}\natexlab{}.
\newblock \showarticletitle{{OSA}: An Optical Switching Architecture for Data
  Center Networks with Unprecedented Flexibility}. In \bibinfo{booktitle}{{\em
  9th USENIX Symposium on Networked Systems Design and Implementation (NSDI
  12)}}. \bibinfo{publisher}{USENIX Association}.
\newblock


\bibitem[\protect\citeauthoryear{Chen, Chen, Zhu, Yu, Porter, Qiao, and
  Zhong}{Chen et~al\mbox{.}}{2017}]%
        {megaswitch}
\bibfield{author}{\bibinfo{person}{Li Chen}, \bibinfo{person}{Kai Chen},
  \bibinfo{person}{Zhonghua Zhu}, \bibinfo{person}{Minlan Yu},
  \bibinfo{person}{George Porter}, \bibinfo{person}{Chunming Qiao}, {and}
  \bibinfo{person}{Shan Zhong}.} \bibinfo{year}{2017}\natexlab{}.
\newblock \showarticletitle{Enabling {Wide-Spread} Communications on Optical
  Fabric with {MegaSwitch}}. In \bibinfo{booktitle}{{\em 14th USENIX Symposium
  on Networked Systems Design and Implementation (NSDI 17)}}.
  \bibinfo{publisher}{USENIX Association}, \bibinfo{address}{Boston, MA},
  \bibinfo{pages}{577--593}.
\newblock
\showISBNx{978-1-931971-37-9}
\showURL{%
\url{https://www.usenix.org/conference/nsdi17/technical-sessions/presentation/chen}}


\bibitem[\protect\citeauthoryear{Chen, Huang, Xie, Jiao, Jiang, Zhou, Li, and
  Wei}{Chen et~al\mbox{.}}{2022}]%
        {chen2022task}
\bibfield{author}{\bibinfo{person}{Tianyu Chen}, \bibinfo{person}{Shaohan
  Huang}, \bibinfo{person}{Yuan Xie}, \bibinfo{person}{Binxing Jiao},
  \bibinfo{person}{Daxin Jiang}, \bibinfo{person}{Haoyi Zhou},
  \bibinfo{person}{Jianxin Li}, {and} \bibinfo{person}{Furu Wei}.}
  \bibinfo{year}{2022}\natexlab{}.
\newblock \showarticletitle{Task-specific expert pruning for sparse
  mixture-of-experts}.
\newblock \bibinfo{journal}{{\em arXiv preprint arXiv:2206.00277\/}}
  (\bibinfo{year}{2022}).
\newblock


\bibitem[\protect\citeauthoryear{Clark, Ballani, Bayvel, Cletheroe, Gerard,
  Haller, Jozwik, Shi, Thomsen, Watts, et~al\mbox{.}}{Clark
  et~al\mbox{.}}{2018}]%
        {clark2018sub}
\bibfield{author}{\bibinfo{person}{Kari Clark}, \bibinfo{person}{Hitesh
  Ballani}, \bibinfo{person}{Polina Bayvel}, \bibinfo{person}{Daniel
  Cletheroe}, \bibinfo{person}{Thomas Gerard}, \bibinfo{person}{Istvan Haller},
  \bibinfo{person}{Krzysztof Jozwik}, \bibinfo{person}{Kai Shi},
  \bibinfo{person}{Benn Thomsen}, \bibinfo{person}{Philip Watts},
  {et~al\mbox{.}}} \bibinfo{year}{2018}\natexlab{}.
\newblock \showarticletitle{Sub-nanosecond clock and data recovery in an
  optically-switched data centre network}. In \bibinfo{booktitle}{{\em 2018
  European Conference on Optical Communication (ECOC)}}. IEEE,
  \bibinfo{pages}{1--3}.
\newblock


\bibitem[\protect\citeauthoryear{DeepSeek-AI, Guo, Yang, Zhang, Song, Zhang,
  Xu, Zhu, Ma, Wang, Bi, Zhang, Yu, Wu, Wu, Gou, Shao, Li, Gao, Liu, Xue, Wang,
  Wu, Feng, Lu, Zhao, Deng, Zhang, Ruan, Dai, Chen, Ji, Li, Lin, Dai, Luo, Hao,
  Chen, Li, Zhang, Bao, Xu, Wang, Ding, Xin, Gao, Qu, Li, Guo, Li, Wang, Chen,
  Yuan, Qiu, Li, Cai, Ni, Liang, Chen, Dong, Hu, Gao, Guan, Huang, Yu, Wang,
  Zhang, Zhao, Wang, Zhang, Xu, Xia, Zhang, Zhang, Tang, Li, Wang, Li, Tian,
  Huang, Zhang, Wang, Chen, Du, Ge, Zhang, Pan, Wang, Chen, Jin, Chen, Lu,
  Zhou, Chen, Ye, Wang, Yu, Zhou, Pan, Li, Zhou, Wu, Ye, Yun, Pei, Sun, Wang,
  Zeng, Zhao, Liu, Liang, Gao, Yu, Zhang, Xiao, An, Liu, Wang, Chen, Nie,
  Cheng, Liu, Xie, Liu, Yang, Li, Su, Lin, Li, Jin, Shen, Chen, Sun, Wang,
  Song, Zhou, Wang, Shan, Li, Wang, Wei, Zhang, Xu, Li, Zhao, Sun, Wang, Yu,
  Zhang, Shi, Xiong, He, Piao, Wang, Tan, Ma, Liu, Guo, Ou, Wang, Gong, Zou,
  He, Xiong, Luo, You, Liu, Zhou, Zhu, Xu, Huang, Li, Zheng, Zhu, Ma, Tang,
  Zha, Yan, Ren, Ren, Sha, Fu, Xu, Xie, Zhang, Hao, Ma, Yan, Wu, Gu, Zhu, Liu,
  Li, Xie, Song, Pan, Huang, Xu, Zhang, and Zhang}{DeepSeek-AI
  et~al\mbox{.}}{2025}]%
        {deepseek-r1}
\bibfield{author}{\bibinfo{person}{DeepSeek-AI}, \bibinfo{person}{Daya Guo},
  \bibinfo{person}{Dejian Yang}, \bibinfo{person}{Haowei Zhang},
  \bibinfo{person}{Junxiao Song}, \bibinfo{person}{Ruoyu Zhang},
  \bibinfo{person}{Runxin Xu}, \bibinfo{person}{Qihao Zhu},
  \bibinfo{person}{Shirong Ma}, \bibinfo{person}{Peiyi Wang},
  \bibinfo{person}{Xiao Bi}, \bibinfo{person}{Xiaokang Zhang},
  \bibinfo{person}{Xingkai Yu}, \bibinfo{person}{Yu Wu}, \bibinfo{person}{Z.~F.
  Wu}, \bibinfo{person}{Zhibin Gou}, \bibinfo{person}{Zhihong Shao},
  \bibinfo{person}{Zhuoshu Li}, \bibinfo{person}{Ziyi Gao},
  \bibinfo{person}{Aixin Liu}, \bibinfo{person}{Bing Xue},
  \bibinfo{person}{Bingxuan Wang}, \bibinfo{person}{Bochao Wu},
  \bibinfo{person}{Bei Feng}, \bibinfo{person}{Chengda Lu},
  \bibinfo{person}{Chenggang Zhao}, \bibinfo{person}{Chengqi Deng},
  \bibinfo{person}{Chenyu Zhang}, \bibinfo{person}{Chong Ruan},
  \bibinfo{person}{Damai Dai}, \bibinfo{person}{Deli Chen},
  \bibinfo{person}{Dongjie Ji}, \bibinfo{person}{Erhang Li},
  \bibinfo{person}{Fangyun Lin}, \bibinfo{person}{Fucong Dai},
  \bibinfo{person}{Fuli Luo}, \bibinfo{person}{Guangbo Hao},
  \bibinfo{person}{Guanting Chen}, \bibinfo{person}{Guowei Li},
  \bibinfo{person}{H. Zhang}, \bibinfo{person}{Han Bao},
  \bibinfo{person}{Hanwei Xu}, \bibinfo{person}{Haocheng Wang},
  \bibinfo{person}{Honghui Ding}, \bibinfo{person}{Huajian Xin},
  \bibinfo{person}{Huazuo Gao}, \bibinfo{person}{Hui Qu}, \bibinfo{person}{Hui
  Li}, \bibinfo{person}{Jianzhong Guo}, \bibinfo{person}{Jiashi Li},
  \bibinfo{person}{Jiawei Wang}, \bibinfo{person}{Jingchang Chen},
  \bibinfo{person}{Jingyang Yuan}, \bibinfo{person}{Junjie Qiu},
  \bibinfo{person}{Junlong Li}, \bibinfo{person}{J.~L. Cai},
  \bibinfo{person}{Jiaqi Ni}, \bibinfo{person}{Jian Liang},
  \bibinfo{person}{Jin Chen}, \bibinfo{person}{Kai Dong}, \bibinfo{person}{Kai
  Hu}, \bibinfo{person}{Kaige Gao}, \bibinfo{person}{Kang Guan},
  \bibinfo{person}{Kexin Huang}, \bibinfo{person}{Kuai Yu},
  \bibinfo{person}{Lean Wang}, \bibinfo{person}{Lecong Zhang},
  \bibinfo{person}{Liang Zhao}, \bibinfo{person}{Litong Wang},
  \bibinfo{person}{Liyue Zhang}, \bibinfo{person}{Lei Xu},
  \bibinfo{person}{Leyi Xia}, \bibinfo{person}{Mingchuan Zhang},
  \bibinfo{person}{Minghua Zhang}, \bibinfo{person}{Minghui Tang},
  \bibinfo{person}{Meng Li}, \bibinfo{person}{Miaojun Wang},
  \bibinfo{person}{Mingming Li}, \bibinfo{person}{Ning Tian},
  \bibinfo{person}{Panpan Huang}, \bibinfo{person}{Peng Zhang},
  \bibinfo{person}{Qiancheng Wang}, \bibinfo{person}{Qinyu Chen},
  \bibinfo{person}{Qiushi Du}, \bibinfo{person}{Ruiqi Ge},
  \bibinfo{person}{Ruisong Zhang}, \bibinfo{person}{Ruizhe Pan},
  \bibinfo{person}{Runji Wang}, \bibinfo{person}{R.~J. Chen},
  \bibinfo{person}{R.~L. Jin}, \bibinfo{person}{Ruyi Chen},
  \bibinfo{person}{Shanghao Lu}, \bibinfo{person}{Shangyan Zhou},
  \bibinfo{person}{Shanhuang Chen}, \bibinfo{person}{Shengfeng Ye},
  \bibinfo{person}{Shiyu Wang}, \bibinfo{person}{Shuiping Yu},
  \bibinfo{person}{Shunfeng Zhou}, \bibinfo{person}{Shuting Pan},
  \bibinfo{person}{S.~S. Li}, \bibinfo{person}{Shuang Zhou},
  \bibinfo{person}{Shaoqing Wu}, \bibinfo{person}{Shengfeng Ye},
  \bibinfo{person}{Tao Yun}, \bibinfo{person}{Tian Pei},
  \bibinfo{person}{Tianyu Sun}, \bibinfo{person}{T. Wang},
  \bibinfo{person}{Wangding Zeng}, \bibinfo{person}{Wanjia Zhao},
  \bibinfo{person}{Wen Liu}, \bibinfo{person}{Wenfeng Liang},
  \bibinfo{person}{Wenjun Gao}, \bibinfo{person}{Wenqin Yu},
  \bibinfo{person}{Wentao Zhang}, \bibinfo{person}{W.~L. Xiao},
  \bibinfo{person}{Wei An}, \bibinfo{person}{Xiaodong Liu},
  \bibinfo{person}{Xiaohan Wang}, \bibinfo{person}{Xiaokang Chen},
  \bibinfo{person}{Xiaotao Nie}, \bibinfo{person}{Xin Cheng},
  \bibinfo{person}{Xin Liu}, \bibinfo{person}{Xin Xie},
  \bibinfo{person}{Xingchao Liu}, \bibinfo{person}{Xinyu Yang},
  \bibinfo{person}{Xinyuan Li}, \bibinfo{person}{Xuecheng Su},
  \bibinfo{person}{Xuheng Lin}, \bibinfo{person}{X.~Q. Li},
  \bibinfo{person}{Xiangyue Jin}, \bibinfo{person}{Xiaojin Shen},
  \bibinfo{person}{Xiaosha Chen}, \bibinfo{person}{Xiaowen Sun},
  \bibinfo{person}{Xiaoxiang Wang}, \bibinfo{person}{Xinnan Song},
  \bibinfo{person}{Xinyi Zhou}, \bibinfo{person}{Xianzu Wang},
  \bibinfo{person}{Xinxia Shan}, \bibinfo{person}{Y.~K. Li},
  \bibinfo{person}{Y.~Q. Wang}, \bibinfo{person}{Y.~X. Wei},
  \bibinfo{person}{Yang Zhang}, \bibinfo{person}{Yanhong Xu},
  \bibinfo{person}{Yao Li}, \bibinfo{person}{Yao Zhao},
  \bibinfo{person}{Yaofeng Sun}, \bibinfo{person}{Yaohui Wang},
  \bibinfo{person}{Yi Yu}, \bibinfo{person}{Yichao Zhang},
  \bibinfo{person}{Yifan Shi}, \bibinfo{person}{Yiliang Xiong},
  \bibinfo{person}{Ying He}, \bibinfo{person}{Yishi Piao},
  \bibinfo{person}{Yisong Wang}, \bibinfo{person}{Yixuan Tan},
  \bibinfo{person}{Yiyang Ma}, \bibinfo{person}{Yiyuan Liu},
  \bibinfo{person}{Yongqiang Guo}, \bibinfo{person}{Yuan Ou},
  \bibinfo{person}{Yuduan Wang}, \bibinfo{person}{Yue Gong},
  \bibinfo{person}{Yuheng Zou}, \bibinfo{person}{Yujia He},
  \bibinfo{person}{Yunfan Xiong}, \bibinfo{person}{Yuxiang Luo},
  \bibinfo{person}{Yuxiang You}, \bibinfo{person}{Yuxuan Liu},
  \bibinfo{person}{Yuyang Zhou}, \bibinfo{person}{Y.~X. Zhu},
  \bibinfo{person}{Yanhong Xu}, \bibinfo{person}{Yanping Huang},
  \bibinfo{person}{Yaohui Li}, \bibinfo{person}{Yi Zheng},
  \bibinfo{person}{Yuchen Zhu}, \bibinfo{person}{Yunxian Ma},
  \bibinfo{person}{Ying Tang}, \bibinfo{person}{Yukun Zha},
  \bibinfo{person}{Yuting Yan}, \bibinfo{person}{Z.~Z. Ren},
  \bibinfo{person}{Zehui Ren}, \bibinfo{person}{Zhangli Sha},
  \bibinfo{person}{Zhe Fu}, \bibinfo{person}{Zhean Xu}, \bibinfo{person}{Zhenda
  Xie}, \bibinfo{person}{Zhengyan Zhang}, \bibinfo{person}{Zhewen Hao},
  \bibinfo{person}{Zhicheng Ma}, \bibinfo{person}{Zhigang Yan},
  \bibinfo{person}{Zhiyu Wu}, \bibinfo{person}{Zihui Gu},
  \bibinfo{person}{Zijia Zhu}, \bibinfo{person}{Zijun Liu},
  \bibinfo{person}{Zilin Li}, \bibinfo{person}{Ziwei Xie},
  \bibinfo{person}{Ziyang Song}, \bibinfo{person}{Zizheng Pan},
  \bibinfo{person}{Zhen Huang}, \bibinfo{person}{Zhipeng Xu},
  \bibinfo{person}{Zhongyu Zhang}, {and} \bibinfo{person}{Zhen Zhang}.}
  \bibinfo{year}{2025}\natexlab{}.
\newblock \bibinfo{title}{DeepSeek-R1: Incentivizing Reasoning Capability in
  LLMs via Reinforcement Learning}.
\newblock   (\bibinfo{year}{2025}).
\newblock
\showeprint[arxiv]{cs.CL/2501.12948}
\showURL{%
\url{https://arxiv.org/abs/2501.12948}}


\bibitem[\protect\citeauthoryear{DeepSeek-AI, Liu, Feng, Wang, Wang, Liu, Zhao,
  Dengr, Ruan, Dai, Guo, Yang, Chen, Ji, Li, Lin, Luo, Hao, Chen, Li, Zhang,
  Xu, Yang, Zhang, Ding, Xin, Gao, Li, Qu, Cai, Liang, Guo, Ni, Li, Chen, Yuan,
  Qiu, Song, Dong, Gao, Guan, Wang, Zhang, Xu, Xia, Zhao, Zhang, Li, Wang,
  Zhang, Zhang, Tang, Li, Tian, Huang, Wang, Zhang, Zhu, Chen, Du, Chen, Jin,
  Ge, Pan, Xu, Chen, Li, Lu, Zhou, Chen, Wu, Ye, Ma, Wang, Zhou, Yu, Zhou,
  Zheng, Wang, Pei, Yuan, Sun, Xiao, Zeng, An, Liu, Liang, Gao, Zhang, Li, Jin,
  Wang, Bi, Liu, Wang, Shen, Chen, Chen, Nie, Sun, Wang, Liu, Xie, Yu, Song,
  Zhou, Yang, Lu, Su, Wu, Li, Wei, Zhu, Xu, Huang, Li, Zhao, Sun, Li, Wang,
  Zheng, Zhang, Xiong, Zhao, He, Tang, Piao, Dong, Tan, Liu, Wang, Guo, Zhu,
  Wang, Zou, Zha, Ma, Yan, You, Liu, Ren, Ren, Sha, Fu, Huang, Zhang, Xie, Hao,
  Shao, Wen, Xu, Zhang, Li, Wang, Gu, Li, and Xie}{DeepSeek-AI
  et~al\mbox{.}}{2024}]%
        {deepseek2}
\bibfield{author}{\bibinfo{person}{DeepSeek-AI}, \bibinfo{person}{Aixin Liu},
  \bibinfo{person}{Bei Feng}, \bibinfo{person}{Bin Wang},
  \bibinfo{person}{Bingxuan Wang}, \bibinfo{person}{Bo Liu},
  \bibinfo{person}{Chenggang Zhao}, \bibinfo{person}{Chengqi Dengr},
  \bibinfo{person}{Chong Ruan}, \bibinfo{person}{Damai Dai},
  \bibinfo{person}{Daya Guo}, \bibinfo{person}{Dejian Yang},
  \bibinfo{person}{Deli Chen}, \bibinfo{person}{Dongjie Ji},
  \bibinfo{person}{Erhang Li}, \bibinfo{person}{Fangyun Lin},
  \bibinfo{person}{Fuli Luo}, \bibinfo{person}{Guangbo Hao},
  \bibinfo{person}{Guanting Chen}, \bibinfo{person}{Guowei Li},
  \bibinfo{person}{H. Zhang}, \bibinfo{person}{Hanwei Xu}, \bibinfo{person}{Hao
  Yang}, \bibinfo{person}{Haowei Zhang}, \bibinfo{person}{Honghui Ding},
  \bibinfo{person}{Huajian Xin}, \bibinfo{person}{Huazuo Gao},
  \bibinfo{person}{Hui Li}, \bibinfo{person}{Hui Qu}, \bibinfo{person}{J.~L.
  Cai}, \bibinfo{person}{Jian Liang}, \bibinfo{person}{Jianzhong Guo},
  \bibinfo{person}{Jiaqi Ni}, \bibinfo{person}{Jiashi Li}, \bibinfo{person}{Jin
  Chen}, \bibinfo{person}{Jingyang Yuan}, \bibinfo{person}{Junjie Qiu},
  \bibinfo{person}{Junxiao Song}, \bibinfo{person}{Kai Dong},
  \bibinfo{person}{Kaige Gao}, \bibinfo{person}{Kang Guan},
  \bibinfo{person}{Lean Wang}, \bibinfo{person}{Lecong Zhang},
  \bibinfo{person}{Lei Xu}, \bibinfo{person}{Leyi Xia}, \bibinfo{person}{Liang
  Zhao}, \bibinfo{person}{Liyue Zhang}, \bibinfo{person}{Meng Li},
  \bibinfo{person}{Miaojun Wang}, \bibinfo{person}{Mingchuan Zhang},
  \bibinfo{person}{Minghua Zhang}, \bibinfo{person}{Minghui Tang},
  \bibinfo{person}{Mingming Li}, \bibinfo{person}{Ning Tian},
  \bibinfo{person}{Panpan Huang}, \bibinfo{person}{Peiyi Wang},
  \bibinfo{person}{Peng Zhang}, \bibinfo{person}{Qihao Zhu},
  \bibinfo{person}{Qinyu Chen}, \bibinfo{person}{Qiushi Du},
  \bibinfo{person}{R.~J. Chen}, \bibinfo{person}{R.~L. Jin},
  \bibinfo{person}{Ruiqi Ge}, \bibinfo{person}{Ruizhe Pan},
  \bibinfo{person}{Runxin Xu}, \bibinfo{person}{Ruyi Chen},
  \bibinfo{person}{S.~S. Li}, \bibinfo{person}{Shanghao Lu},
  \bibinfo{person}{Shangyan Zhou}, \bibinfo{person}{Shanhuang Chen},
  \bibinfo{person}{Shaoqing Wu}, \bibinfo{person}{Shengfeng Ye},
  \bibinfo{person}{Shirong Ma}, \bibinfo{person}{Shiyu Wang},
  \bibinfo{person}{Shuang Zhou}, \bibinfo{person}{Shuiping Yu},
  \bibinfo{person}{Shunfeng Zhou}, \bibinfo{person}{Size Zheng},
  \bibinfo{person}{T. Wang}, \bibinfo{person}{Tian Pei}, \bibinfo{person}{Tian
  Yuan}, \bibinfo{person}{Tianyu Sun}, \bibinfo{person}{W.~L. Xiao},
  \bibinfo{person}{Wangding Zeng}, \bibinfo{person}{Wei An},
  \bibinfo{person}{Wen Liu}, \bibinfo{person}{Wenfeng Liang},
  \bibinfo{person}{Wenjun Gao}, \bibinfo{person}{Wentao Zhang},
  \bibinfo{person}{X.~Q. Li}, \bibinfo{person}{Xiangyue Jin},
  \bibinfo{person}{Xianzu Wang}, \bibinfo{person}{Xiao Bi},
  \bibinfo{person}{Xiaodong Liu}, \bibinfo{person}{Xiaohan Wang},
  \bibinfo{person}{Xiaojin Shen}, \bibinfo{person}{Xiaokang Chen},
  \bibinfo{person}{Xiaosha Chen}, \bibinfo{person}{Xiaotao Nie},
  \bibinfo{person}{Xiaowen Sun}, \bibinfo{person}{Xiaoxiang Wang},
  \bibinfo{person}{Xin Liu}, \bibinfo{person}{Xin Xie},
  \bibinfo{person}{Xingkai Yu}, \bibinfo{person}{Xinnan Song},
  \bibinfo{person}{Xinyi Zhou}, \bibinfo{person}{Xinyu Yang},
  \bibinfo{person}{Xuan Lu}, \bibinfo{person}{Xuecheng Su}, \bibinfo{person}{Y.
  Wu}, \bibinfo{person}{Y.~K. Li}, \bibinfo{person}{Y.~X. Wei},
  \bibinfo{person}{Y.~X. Zhu}, \bibinfo{person}{Yanhong Xu},
  \bibinfo{person}{Yanping Huang}, \bibinfo{person}{Yao Li},
  \bibinfo{person}{Yao Zhao}, \bibinfo{person}{Yaofeng Sun},
  \bibinfo{person}{Yaohui Li}, \bibinfo{person}{Yaohui Wang},
  \bibinfo{person}{Yi Zheng}, \bibinfo{person}{Yichao Zhang},
  \bibinfo{person}{Yiliang Xiong}, \bibinfo{person}{Yilong Zhao},
  \bibinfo{person}{Ying He}, \bibinfo{person}{Ying Tang},
  \bibinfo{person}{Yishi Piao}, \bibinfo{person}{Yixin Dong},
  \bibinfo{person}{Yixuan Tan}, \bibinfo{person}{Yiyuan Liu},
  \bibinfo{person}{Yongji Wang}, \bibinfo{person}{Yongqiang Guo},
  \bibinfo{person}{Yuchen Zhu}, \bibinfo{person}{Yuduan Wang},
  \bibinfo{person}{Yuheng Zou}, \bibinfo{person}{Yukun Zha},
  \bibinfo{person}{Yunxian Ma}, \bibinfo{person}{Yuting Yan},
  \bibinfo{person}{Yuxiang You}, \bibinfo{person}{Yuxuan Liu},
  \bibinfo{person}{Z.~Z. Ren}, \bibinfo{person}{Zehui Ren},
  \bibinfo{person}{Zhangli Sha}, \bibinfo{person}{Zhe Fu},
  \bibinfo{person}{Zhen Huang}, \bibinfo{person}{Zhen Zhang},
  \bibinfo{person}{Zhenda Xie}, \bibinfo{person}{Zhewen Hao},
  \bibinfo{person}{Zhihong Shao}, \bibinfo{person}{Zhiniu Wen},
  \bibinfo{person}{Zhipeng Xu}, \bibinfo{person}{Zhongyu Zhang},
  \bibinfo{person}{Zhuoshu Li}, \bibinfo{person}{Zihan Wang},
  \bibinfo{person}{Zihui Gu}, \bibinfo{person}{Zilin Li}, {and}
  \bibinfo{person}{Ziwei Xie}.} \bibinfo{year}{2024}\natexlab{}.
\newblock \bibinfo{title}{DeepSeek-V2: A Strong, Economical, and Efficient
  Mixture-of-Experts Language Model}.
\newblock   (\bibinfo{year}{2024}).
\newblock
\showeprint[arxiv]{cs.CL/2405.04434}
\showURL{%
\url{https://arxiv.org/abs/2405.04434}}


\bibitem[\protect\citeauthoryear{Dubey, Jauhri, Pandey, Kadian, Al-Dahle,
  Letman, Mathur, Schelten, Yang, Fan, et~al\mbox{.}}{Dubey
  et~al\mbox{.}}{2024}]%
        {llama3-white-paper}
\bibfield{author}{\bibinfo{person}{Abhimanyu Dubey}, \bibinfo{person}{Abhinav
  Jauhri}, \bibinfo{person}{Abhinav Pandey}, \bibinfo{person}{Abhishek Kadian},
  \bibinfo{person}{Ahmad Al-Dahle}, \bibinfo{person}{Aiesha Letman},
  \bibinfo{person}{Akhil Mathur}, \bibinfo{person}{Alan Schelten},
  \bibinfo{person}{Amy Yang}, \bibinfo{person}{Angela Fan}, {et~al\mbox{.}}}
  \bibinfo{year}{2024}\natexlab{}.
\newblock \showarticletitle{The Llama 3 Herd of Models}.
\newblock \bibinfo{journal}{{\em arXiv preprint arXiv:2407.21783\/}}
  (\bibinfo{year}{2024}).
\newblock


\bibitem[\protect\citeauthoryear{Farrington, Forencich, Porter, Sun, Ford,
  Fainman, Papen, and Vahdat}{Farrington et~al\mbox{.}}{2013}]%
        {farrington2013multiport}
\bibfield{author}{\bibinfo{person}{Nathan Farrington}, \bibinfo{person}{Alex
  Forencich}, \bibinfo{person}{George Porter}, \bibinfo{person}{P-C Sun},
  \bibinfo{person}{Joseph~E Ford}, \bibinfo{person}{Yeshaiahu Fainman},
  \bibinfo{person}{George~C Papen}, {and} \bibinfo{person}{Amin Vahdat}.}
  \bibinfo{year}{2013}\natexlab{}.
\newblock \showarticletitle{A multiport microsecond optical circuit switch for
  data center networking}.
\newblock \bibinfo{journal}{{\em IEEE Photonics Technology Letters\/}}
  \bibinfo{volume}{25}, \bibinfo{number}{16} (\bibinfo{year}{2013}),
  \bibinfo{pages}{1589--1592}.
\newblock


\bibitem[\protect\citeauthoryear{Farrington, Porter, Radhakrishnan, Bazzaz,
  Subramanya, Fainman, Papen, and Vahdat}{Farrington et~al\mbox{.}}{2010}]%
        {helios}
\bibfield{author}{\bibinfo{person}{Nathan Farrington}, \bibinfo{person}{George
  Porter}, \bibinfo{person}{Sivasankar Radhakrishnan},
  \bibinfo{person}{Hamid~Hajabdolali Bazzaz}, \bibinfo{person}{Vikram
  Subramanya}, \bibinfo{person}{Yeshaiahu Fainman}, \bibinfo{person}{George
  Papen}, {and} \bibinfo{person}{Amin Vahdat}.}
  \bibinfo{year}{2010}\natexlab{}.
\newblock \showarticletitle{Helios: a hybrid electrical/optical switch
  architecture for modular data centers}. In \bibinfo{booktitle}{{\em
  Proceedings of the ACM SIGCOMM 2010 Conference}} {\em
  (\bibinfo{series}{SIGCOMM '10})}. \bibinfo{publisher}{Association for
  Computing Machinery}.
\newblock
\showISBNx{9781450302012}
\showDOI{%
\url{https://doi.org/10.1145/1851182.1851223}}


\bibitem[\protect\citeauthoryear{Fedus, Zoph, and Shazeer}{Fedus
  et~al\mbox{.}}{2022}]%
        {switch-transformers}
\bibfield{author}{\bibinfo{person}{William Fedus}, \bibinfo{person}{Barret
  Zoph}, {and} \bibinfo{person}{Noam Shazeer}.}
  \bibinfo{year}{2022}\natexlab{}.
\newblock \showarticletitle{Switch transformers: Scaling to trillion parameter
  models with simple and efficient sparsity}.
\newblock \bibinfo{journal}{{\em Journal of Machine Learning Research\/}}
  \bibinfo{volume}{23}, \bibinfo{number}{120} (\bibinfo{year}{2022}),
  \bibinfo{pages}{1--39}.
\newblock


\bibitem[\protect\citeauthoryear{Foerster, Ghobadi, and Schmid}{Foerster
  et~al\mbox{.}}{2018}]%
        {foerster2018characterizing}
\bibfield{author}{\bibinfo{person}{Klaus-Tycho Foerster},
  \bibinfo{person}{Manya Ghobadi}, {and} \bibinfo{person}{Stefan Schmid}.}
  \bibinfo{year}{2018}\natexlab{}.
\newblock \showarticletitle{Characterizing the algorithmic complexity of
  reconfigurable data center architectures}. In \bibinfo{booktitle}{{\em
  Proceedings of the 2018 Symposium on Architectures for Networking and
  Communications Systems}}. \bibinfo{pages}{89--96}.
\newblock


\bibitem[\protect\citeauthoryear{Gangidi, Miao, Zheng, Bondu, Goes, Morsy,
  Puri, Riftadi, Shetty, Yang, et~al\mbox{.}}{Gangidi et~al\mbox{.}}{2024}]%
        {meta-llm-network}
\bibfield{author}{\bibinfo{person}{Adithya Gangidi}, \bibinfo{person}{Rui
  Miao}, \bibinfo{person}{Shengbao Zheng}, \bibinfo{person}{Sai~Jayesh Bondu},
  \bibinfo{person}{Guilherme Goes}, \bibinfo{person}{Hany Morsy},
  \bibinfo{person}{Rohit Puri}, \bibinfo{person}{Mohammad Riftadi},
  \bibinfo{person}{Ashmitha~Jeevaraj Shetty}, \bibinfo{person}{Jingyi Yang},
  {et~al\mbox{.}}} \bibinfo{year}{2024}\natexlab{}.
\newblock \showarticletitle{RDMA over Ethernet for Distributed Training at Meta
  Scale}. In \bibinfo{booktitle}{{\em Proceedings of the ACM SIGCOMM 2024
  Conference}}. \bibinfo{pages}{57--70}.
\newblock


\bibitem[\protect\citeauthoryear{Gerard, Clark, Funnell, Shi, Thomsen, Watts,
  Jozwik, Haller, Williams, Costa, et~al\mbox{.}}{Gerard et~al\mbox{.}}{2021}]%
        {gerard2021fast}
\bibfield{author}{\bibinfo{person}{Thomas Gerard}, \bibinfo{person}{Kari
  Clark}, \bibinfo{person}{Adam Funnell}, \bibinfo{person}{Kai Shi},
  \bibinfo{person}{Benn Thomsen}, \bibinfo{person}{Philip Watts},
  \bibinfo{person}{Krzysztof Jozwik}, \bibinfo{person}{Istvan Haller},
  \bibinfo{person}{Hugh Williams}, \bibinfo{person}{Paolo Costa},
  {et~al\mbox{.}}} \bibinfo{year}{2021}\natexlab{}.
\newblock \showarticletitle{Fast and uniform optically-switched data centre
  networks enabled by amplitude caching}. In \bibinfo{booktitle}{{\em 2021
  Optical Fiber Communications Conference and Exhibition (OFC)}}. IEEE,
  \bibinfo{pages}{1--3}.
\newblock


\bibitem[\protect\citeauthoryear{Gibson, Hariharan, Lance, McLaren, Montazeri,
  Singh, Wang, Wassel, Wu, Yoo, et~al\mbox{.}}{Gibson et~al\mbox{.}}{2022}]%
        {gibson2022aquila}
\bibfield{author}{\bibinfo{person}{Dan Gibson}, \bibinfo{person}{Hema
  Hariharan}, \bibinfo{person}{Eric Lance}, \bibinfo{person}{Moray McLaren},
  \bibinfo{person}{Behnam Montazeri}, \bibinfo{person}{Arjun Singh},
  \bibinfo{person}{Stephen Wang}, \bibinfo{person}{Hassan~MG Wassel},
  \bibinfo{person}{Zhehua Wu}, \bibinfo{person}{Sunghwan Yoo}, {et~al\mbox{.}}}
  \bibinfo{year}{2022}\natexlab{}.
\newblock \showarticletitle{Aquila: A unified, low-latency fabric for
  datacenter networks}. In \bibinfo{booktitle}{{\em 19th USENIX Symposium on
  Networked Systems Design and Implementation (NSDI 22)}}.
  \bibinfo{pages}{1249--1266}.
\newblock


\bibitem[\protect\citeauthoryear{Greenberg, Hamilton, Jain, Kandula, Kim,
  Lahiri, Maltz, Patel, Sengupta, Rexford, et~al\mbox{.}}{Greenberg
  et~al\mbox{.}}{2009}]%
        {vl2}
\bibfield{author}{\bibinfo{person}{Albert Greenberg}, \bibinfo{person}{James~R
  Hamilton}, \bibinfo{person}{Navendu Jain}, \bibinfo{person}{Srikanth
  Kandula}, \bibinfo{person}{Changhoon Kim}, \bibinfo{person}{Parantap Lahiri},
  \bibinfo{person}{David~A Maltz}, \bibinfo{person}{Parveen Patel},
  \bibinfo{person}{Sushant Sengupta}, \bibinfo{person}{Jennifer Rexford},
  {et~al\mbox{.}}} \bibinfo{year}{2009}\natexlab{}.
\newblock \showarticletitle{VL2: A scalable and flexible data center network}.
  In \bibinfo{booktitle}{{\em ACM SIGCOMM Computer Communication Review}},
  Vol.~\bibinfo{volume}{39}. \bibinfo{publisher}{ACM New York, NY, USA},
  \bibinfo{pages}{51--62}.
\newblock


\bibitem[\protect\citeauthoryear{Guo, Wu, Tan, Shiy, Zhang, and Lu}{Guo
  et~al\mbox{.}}{2009}]%
        {bcube}
\bibfield{author}{\bibinfo{person}{Chuanxiong Guo}, \bibinfo{person}{Hui Wu},
  \bibinfo{person}{Kai Tan}, \bibinfo{person}{Lei Shiy},
  \bibinfo{person}{Yongguang Zhang}, {and} \bibinfo{person}{Songnian Lu}.}
  \bibinfo{year}{2009}\natexlab{}.
\newblock \showarticletitle{Bcube: A high performance, server-centric network
  architecture for modular data centers}. In \bibinfo{booktitle}{{\em ACM
  SIGCOMM Computer Communication Review}}, Vol.~\bibinfo{volume}{39}.
  \bibinfo{publisher}{ACM New York, NY, USA}, \bibinfo{pages}{63--74}.
\newblock


\bibitem[\protect\citeauthoryear{Guo, Wu, Tan, Shiy, Zhang, and Lu}{Guo
  et~al\mbox{.}}{2010}]%
        {dcell}
\bibfield{author}{\bibinfo{person}{Chuanxiong Guo}, \bibinfo{person}{Hui Wu},
  \bibinfo{person}{Kai Tan}, \bibinfo{person}{Lei Shiy},
  \bibinfo{person}{Yongguang Zhang}, {and} \bibinfo{person}{Songnian Lu}.}
  \bibinfo{year}{2010}\natexlab{}.
\newblock \showarticletitle{Dcell: A scalable and fault-tolerant network
  structure for data centers}. In \bibinfo{booktitle}{{\em ACM SIGCOMM Computer
  Communication Review}}, Vol.~\bibinfo{volume}{40}. \bibinfo{publisher}{ACM
  New York, NY, USA}, \bibinfo{pages}{63--74}.
\newblock


\bibitem[\protect\citeauthoryear{Hamedazimi, Qazi, Gupta, Sekar, Das, Longtin,
  Shah, and Tanwer}{Hamedazimi et~al\mbox{.}}{2014}]%
        {firefly}
\bibfield{author}{\bibinfo{person}{Navid Hamedazimi}, \bibinfo{person}{Zafar
  Qazi}, \bibinfo{person}{Himanshu Gupta}, \bibinfo{person}{Vyas Sekar},
  \bibinfo{person}{Samir~R Das}, \bibinfo{person}{Jon~P Longtin},
  \bibinfo{person}{Himanshu Shah}, {and} \bibinfo{person}{Ashish Tanwer}.}
  \bibinfo{year}{2014}\natexlab{}.
\newblock \showarticletitle{Firefly: A reconfigurable wireless data center
  fabric using free-space optics}. In \bibinfo{booktitle}{{\em Proceedings of
  the 2014 ACM conference on SIGCOMM}}. \bibinfo{pages}{319--330}.
\newblock


\bibitem[\protect\citeauthoryear{Jain, Kumar, Mandal, Ong, Poutievski, Singh,
  Venkata, Wanderer, Zhou, Zhu, Zolla, H\"{o}lzle, Stuart, and Vahdat}{Jain
  et~al\mbox{.}}{2013}]%
        {b4}
\bibfield{author}{\bibinfo{person}{Sushant Jain}, \bibinfo{person}{Alok Kumar},
  \bibinfo{person}{Subhasree Mandal}, \bibinfo{person}{Joon Ong},
  \bibinfo{person}{Leon Poutievski}, \bibinfo{person}{Arjun Singh},
  \bibinfo{person}{Subbaiah Venkata}, \bibinfo{person}{Jim Wanderer},
  \bibinfo{person}{Junlan Zhou}, \bibinfo{person}{Min Zhu},
  \bibinfo{person}{Jon Zolla}, \bibinfo{person}{Urs H\"{o}lzle},
  \bibinfo{person}{Stephen Stuart}, {and} \bibinfo{person}{Amin Vahdat}.}
  \bibinfo{year}{2013}\natexlab{}.
\newblock \showarticletitle{B4: experience with a globally-deployed software
  defined wan}. In \bibinfo{booktitle}{{\em Proceedings of the ACM SIGCOMM 2013
  Conference on SIGCOMM}} {\em (\bibinfo{series}{SIGCOMM '13})}.
  \bibinfo{pages}{3–14}.
\newblock
\showURL{%
\url{https://doi.org/10.1145/2486001.2486019}}


\bibitem[\protect\citeauthoryear{Jiang, Zhu, Lan, Yi, Cui, and Guo}{Jiang
  et~al\mbox{.}}{2020}]%
        {byteps}
\bibfield{author}{\bibinfo{person}{Yimin Jiang}, \bibinfo{person}{Yibo Zhu},
  \bibinfo{person}{Chang Lan}, \bibinfo{person}{Bairen Yi},
  \bibinfo{person}{Yong Cui}, {and} \bibinfo{person}{Chuanxiong Guo}.}
  \bibinfo{year}{2020}\natexlab{}.
\newblock \showarticletitle{A unified architecture for accelerating distributed
  $\{$DNN$\}$ training in heterogeneous $\{$GPU/CPU$\}$ clusters}. In
  \bibinfo{booktitle}{{\em 14th USENIX Symposium on Operating Systems Design
  and Implementation (OSDI 20)}}. \bibinfo{pages}{463--479}.
\newblock


\bibitem[\protect\citeauthoryear{Jiang, Lin, Zhong, Huang, Chen, Zhang, Peng,
  Li, Xie, Nong, et~al\mbox{.}}{Jiang et~al\mbox{.}}{2024}]%
        {megascale}
\bibfield{author}{\bibinfo{person}{Ziheng Jiang}, \bibinfo{person}{Haibin Lin},
  \bibinfo{person}{Yinmin Zhong}, \bibinfo{person}{Qi Huang},
  \bibinfo{person}{Yangrui Chen}, \bibinfo{person}{Zhi Zhang},
  \bibinfo{person}{Yanghua Peng}, \bibinfo{person}{Xiang Li},
  \bibinfo{person}{Cong Xie}, \bibinfo{person}{Shibiao Nong}, {et~al\mbox{.}}}
  \bibinfo{year}{2024}\natexlab{}.
\newblock \showarticletitle{$\{$MegaScale$\}$: Scaling large language model
  training to more than 10,000 $\{$GPUs$\}$}. In \bibinfo{booktitle}{{\em 21st
  USENIX Symposium on Networked Systems Design and Implementation (NSDI 24)}}.
  \bibinfo{pages}{745--760}.
\newblock


\bibitem[\protect\citeauthoryear{Jouppi, Kurian, Li, Ma, Nagarajan, Nai, Patil,
  Subramanian, Swing, Towles, et~al\mbox{.}}{Jouppi et~al\mbox{.}}{2023}]%
        {tpuv4}
\bibfield{author}{\bibinfo{person}{Norm Jouppi}, \bibinfo{person}{George
  Kurian}, \bibinfo{person}{Sheng Li}, \bibinfo{person}{Peter Ma},
  \bibinfo{person}{Rahul Nagarajan}, \bibinfo{person}{Lifeng Nai},
  \bibinfo{person}{Nishant Patil}, \bibinfo{person}{Suvinay Subramanian},
  \bibinfo{person}{Andy Swing}, \bibinfo{person}{Brian Towles},
  {et~al\mbox{.}}} \bibinfo{year}{2023}\natexlab{}.
\newblock \showarticletitle{Tpu v4: An optically reconfigurable supercomputer
  for machine learning with hardware support for embeddings}. In
  \bibinfo{booktitle}{{\em Proceedings of the 50th Annual International
  Symposium on Computer Architecture}}. \bibinfo{pages}{1--14}.
\newblock


\bibitem[\protect\citeauthoryear{Khani, Ghobadi, Alizadeh, Zhu, Glick, Bergman,
  Vahdat, Klenk, and Ebrahimi}{Khani et~al\mbox{.}}{2021}]%
        {sip-ml}
\bibfield{author}{\bibinfo{person}{Mehrdad Khani}, \bibinfo{person}{Manya
  Ghobadi}, \bibinfo{person}{Mohammad Alizadeh}, \bibinfo{person}{Ziyi Zhu},
  \bibinfo{person}{Madeleine Glick}, \bibinfo{person}{Keren Bergman},
  \bibinfo{person}{Amin Vahdat}, \bibinfo{person}{Benjamin Klenk}, {and}
  \bibinfo{person}{Eiman Ebrahimi}.} \bibinfo{year}{2021}\natexlab{}.
\newblock \showarticletitle{SiP-ML: high-bandwidth optical network
  interconnects for machine learning training}. In \bibinfo{booktitle}{{\em
  Proceedings of the 2021 ACM SIGCOMM 2021 Conference}}.
  \bibinfo{pages}{657--675}.
\newblock


\bibitem[\protect\citeauthoryear{Kumar, Devraj, Bunandar, and Singh}{Kumar
  et~al\mbox{.}}{2024}]%
        {kumar2024case}
\bibfield{author}{\bibinfo{person}{Abhishek~Vijaya Kumar},
  \bibinfo{person}{Arjun Devraj}, \bibinfo{person}{Darius Bunandar}, {and}
  \bibinfo{person}{Rachee Singh}.} \bibinfo{year}{2024}\natexlab{}.
\newblock \showarticletitle{A case for server-scale photonic connectivity}. In
  \bibinfo{booktitle}{{\em Proceedings of the 23rd ACM Workshop on Hot Topics
  in Networks}}. \bibinfo{pages}{290--299}.
\newblock


\bibitem[\protect\citeauthoryear{Lepikhin, Lee, Xu, Chen, Firat, Huang, Krikun,
  Shazeer, and Chen}{Lepikhin et~al\mbox{.}}{2020}]%
        {gshard}
\bibfield{author}{\bibinfo{person}{Dmitry Lepikhin},
  \bibinfo{person}{HyoukJoong Lee}, \bibinfo{person}{Yuanzhong Xu},
  \bibinfo{person}{Dehao Chen}, \bibinfo{person}{Orhan Firat},
  \bibinfo{person}{Yanping Huang}, \bibinfo{person}{Maxim Krikun},
  \bibinfo{person}{Noam Shazeer}, {and} \bibinfo{person}{Zhifeng Chen}.}
  \bibinfo{year}{2020}\natexlab{}.
\newblock \showarticletitle{Gshard: Scaling giant models with conditional
  computation and automatic sharding}.
\newblock \bibinfo{journal}{{\em arXiv preprint arXiv:2006.16668\/}}
  (\bibinfo{year}{2020}).
\newblock


\bibitem[\protect\citeauthoryear{Li, Zhang, Yadav, Sung, Cheng, Bansal, and
  Chen}{Li et~al\mbox{.}}{2023}]%
        {li2023merge}
\bibfield{author}{\bibinfo{person}{Pingzhi Li}, \bibinfo{person}{Zhenyu Zhang},
  \bibinfo{person}{Prateek Yadav}, \bibinfo{person}{Yi-Lin Sung},
  \bibinfo{person}{Yu Cheng}, \bibinfo{person}{Mohit Bansal}, {and}
  \bibinfo{person}{Tianlong Chen}.} \bibinfo{year}{2023}\natexlab{}.
\newblock \showarticletitle{Merge, then compress: Demystify efficient SMoe with
  hints from its routing policy}.
\newblock \bibinfo{journal}{{\em arXiv preprint arXiv:2310.01334\/}}
  (\bibinfo{year}{2023}).
\newblock


\bibitem[\protect\citeauthoryear{Li, Zhao, Varma, Salpekar, Noordhuis, Li,
  Paszke, Smith, Vaughan, Damania, et~al\mbox{.}}{Li et~al\mbox{.}}{2020}]%
        {pytorch-dist}
\bibfield{author}{\bibinfo{person}{Shen Li}, \bibinfo{person}{Yanli Zhao},
  \bibinfo{person}{Rohan Varma}, \bibinfo{person}{Omkar Salpekar},
  \bibinfo{person}{Pieter Noordhuis}, \bibinfo{person}{Teng Li},
  \bibinfo{person}{Adam Paszke}, \bibinfo{person}{Jeff Smith},
  \bibinfo{person}{Brian Vaughan}, \bibinfo{person}{Pritam Damania},
  {et~al\mbox{.}}} \bibinfo{year}{2020}\natexlab{}.
\newblock \showarticletitle{Pytorch distributed: Experiences on accelerating
  data parallel training}.
\newblock \bibinfo{journal}{{\em arXiv preprint arXiv:2006.15704\/}}
  (\bibinfo{year}{2020}).
\newblock


\bibitem[\protect\citeauthoryear{Li, Liu, Li, Jin, Tian, Zhong, Liu, Zhang, and
  Chen}{Li et~al\mbox{.}}{2024}]%
        {li2024understanding}
\bibfield{author}{\bibinfo{person}{Wenxue Li}, \bibinfo{person}{Xiangzhou Liu},
  \bibinfo{person}{Yuxuan Li}, \bibinfo{person}{Yilun Jin},
  \bibinfo{person}{Han Tian}, \bibinfo{person}{Zhizhen Zhong},
  \bibinfo{person}{Guyue Liu}, \bibinfo{person}{Ying Zhang}, {and}
  \bibinfo{person}{Kai Chen}.} \bibinfo{year}{2024}\natexlab{}.
\newblock \showarticletitle{Understanding communication characteristics of
  distributed training}. In \bibinfo{booktitle}{{\em Proceedings of the 8th
  Asia-Pacific Workshop on Networking}}. \bibinfo{pages}{1--8}.
\newblock


\bibitem[\protect\citeauthoryear{Lin, Tang, Ye, Cui, Zhu, Jin, Zhang, Ning, and
  Yuan}{Lin et~al\mbox{.}}{2024}]%
        {lin2024moe}
\bibfield{author}{\bibinfo{person}{Bin Lin}, \bibinfo{person}{Zhenyu Tang},
  \bibinfo{person}{Yang Ye}, \bibinfo{person}{Jiaxi Cui}, \bibinfo{person}{Bin
  Zhu}, \bibinfo{person}{Peng Jin}, \bibinfo{person}{Junwu Zhang},
  \bibinfo{person}{Munan Ning}, {and} \bibinfo{person}{Li Yuan}.}
  \bibinfo{year}{2024}\natexlab{}.
\newblock \showarticletitle{Moe-llava: Mixture of experts for large
  vision-language models}.
\newblock \bibinfo{journal}{{\em arXiv preprint arXiv:2401.15947\/}}
  (\bibinfo{year}{2024}).
\newblock


\bibitem[\protect\citeauthoryear{Liu, Lu, Forencich, Kapoor, Tewari, Voelker,
  Papen, Snoeren, and Porter}{Liu et~al\mbox{.}}{2014}]%
        {reactor}
\bibfield{author}{\bibinfo{person}{He Liu}, \bibinfo{person}{Feng Lu},
  \bibinfo{person}{Alex Forencich}, \bibinfo{person}{Rishi Kapoor},
  \bibinfo{person}{Malveeka Tewari}, \bibinfo{person}{Geoffrey~M. Voelker},
  \bibinfo{person}{George Papen}, \bibinfo{person}{Alex~C. Snoeren}, {and}
  \bibinfo{person}{George Porter}.} \bibinfo{year}{2014}\natexlab{}.
\newblock \showarticletitle{Circuit Switching Under the Radar with {REACToR}}.
  In \bibinfo{booktitle}{{\em 11th USENIX Symposium on Networked Systems Design
  and Implementation (NSDI 14)}}. \bibinfo{publisher}{USENIX Association},
  \bibinfo{address}{Seattle, WA}, \bibinfo{pages}{1--15}.
\newblock
\showISBNx{978-1-931971-09-6}
\showURL{%
\url{https://www.usenix.org/conference/nsdi14/technical-sessions/presentation/liu_he}}


\bibitem[\protect\citeauthoryear{Liu, Mukerjee, Li, Feltman, Papen, Savage,
  Seshan, Voelker, Andersen, Kaminsky, et~al\mbox{.}}{Liu
  et~al\mbox{.}}{2015}]%
        {liu2015scheduling}
\bibfield{author}{\bibinfo{person}{He Liu}, \bibinfo{person}{Matthew~K
  Mukerjee}, \bibinfo{person}{Conglong Li}, \bibinfo{person}{Nicolas Feltman},
  \bibinfo{person}{George Papen}, \bibinfo{person}{Stefan Savage},
  \bibinfo{person}{Srinivasan Seshan}, \bibinfo{person}{Geoffrey~M Voelker},
  \bibinfo{person}{David~G Andersen}, \bibinfo{person}{Michael Kaminsky},
  {et~al\mbox{.}}} \bibinfo{year}{2015}\natexlab{}.
\newblock \showarticletitle{Scheduling techniques for hybrid circuit/packet
  networks}. In \bibinfo{booktitle}{{\em Proceedings of the 11th ACM Conference
  on Emerging Networking Experiments and Technologies}}.
  \bibinfo{pages}{1--13}.
\newblock


\bibitem[\protect\citeauthoryear{Liu, Urata, Yasumura, Zhou, Bannon, Berger,
  Dashti, Jouppi, Lam, Li, Mao, Nelson, Papen, Tariq, and Vahdat}{Liu
  et~al\mbox{.}}{2023a}]%
        {liu2023lightwave}
\bibfield{author}{\bibinfo{person}{Hong Liu}, \bibinfo{person}{Ryohei Urata},
  \bibinfo{person}{Kevin Yasumura}, \bibinfo{person}{Xiang Zhou},
  \bibinfo{person}{Roy Bannon}, \bibinfo{person}{Jill Berger},
  \bibinfo{person}{Pedram Dashti}, \bibinfo{person}{Norm Jouppi},
  \bibinfo{person}{Cedric Lam}, \bibinfo{person}{Sheng Li},
  \bibinfo{person}{Erji Mao}, \bibinfo{person}{Daniel Nelson},
  \bibinfo{person}{George Papen}, \bibinfo{person}{Mukarram Tariq}, {and}
  \bibinfo{person}{Amin Vahdat}.} \bibinfo{year}{2023}\natexlab{a}.
\newblock \showarticletitle{Lightwave Fabrics: At-Scale Optical Circuit
  Switching for Datacenter and Machine Learning Systems}. In
  \bibinfo{booktitle}{{\em Proceedings of the ACM SIGCOMM 2023 Conference}}
  {\em (\bibinfo{series}{ACM SIGCOMM '23})}.
\newblock
\showURL{%
\url{https://doi.org/10.1145/3603269.3604836}}


\bibitem[\protect\citeauthoryear{Liu, Wang, and Jiang}{Liu
  et~al\mbox{.}}{2023b}]%
        {janus}
\bibfield{author}{\bibinfo{person}{Juncai Liu}, \bibinfo{person}{Jessie~Hui
  Wang}, {and} \bibinfo{person}{Yimin Jiang}.}
  \bibinfo{year}{2023}\natexlab{b}.
\newblock \showarticletitle{Janus: A unified distributed training framework for
  sparse mixture-of-experts models}. In \bibinfo{booktitle}{{\em Proceedings of
  the ACM SIGCOMM 2023 Conference}}. \bibinfo{pages}{486--498}.
\newblock


\bibitem[\protect\citeauthoryear{Lu, Liu, Xu, Zhou, Huang, Zhang, Yan, and
  Li}{Lu et~al\mbox{.}}{2024}]%
        {lu2024not}
\bibfield{author}{\bibinfo{person}{Xudong Lu}, \bibinfo{person}{Qi Liu},
  \bibinfo{person}{Yuhui Xu}, \bibinfo{person}{Aojun Zhou},
  \bibinfo{person}{Siyuan Huang}, \bibinfo{person}{Bo Zhang},
  \bibinfo{person}{Junchi Yan}, {and} \bibinfo{person}{Hongsheng Li}.}
  \bibinfo{year}{2024}\natexlab{}.
\newblock \showarticletitle{Not All Experts are Equal: Efficient Expert Pruning
  and Skipping for Mixture-of-Experts Large Language Models}.
\newblock \bibinfo{journal}{{\em arXiv preprint arXiv:2402.14800\/}}
  (\bibinfo{year}{2024}).
\newblock


\bibitem[\protect\citeauthoryear{Mellette, Das, Guo, McGuinness, Snoeren, and
  Porter}{Mellette et~al\mbox{.}}{2020}]%
        {opera}
\bibfield{author}{\bibinfo{person}{William~M Mellette},
  \bibinfo{person}{Rajdeep Das}, \bibinfo{person}{Yibo Guo},
  \bibinfo{person}{Rob McGuinness}, \bibinfo{person}{Alex~C Snoeren}, {and}
  \bibinfo{person}{George Porter}.} \bibinfo{year}{2020}\natexlab{}.
\newblock \showarticletitle{Expanding across time to deliver bandwidth
  efficiency and low latency}. In \bibinfo{booktitle}{{\em 17th USENIX
  Symposium on Networked Systems Design and Implementation (NSDI 20)}}.
  \bibinfo{pages}{1--18}.
\newblock


\bibitem[\protect\citeauthoryear{Mellette, Forencich, Athapathu, Snoeren,
  Papen, and Porter}{Mellette et~al\mbox{.}}{2024}]%
        {realizing-rotornet}
\bibfield{author}{\bibinfo{person}{William~M Mellette}, \bibinfo{person}{Alex
  Forencich}, \bibinfo{person}{Rukshani Athapathu}, \bibinfo{person}{Alex~C
  Snoeren}, \bibinfo{person}{George Papen}, {and} \bibinfo{person}{George
  Porter}.} \bibinfo{year}{2024}\natexlab{}.
\newblock \showarticletitle{Realizing RotorNet: Toward Practical Microsecond
  Scale Optical Networking}. In \bibinfo{booktitle}{{\em Proceedings of the ACM
  SIGCOMM 2024 Conference}}. \bibinfo{pages}{392--414}.
\newblock


\bibitem[\protect\citeauthoryear{Mellette, McGuinness, Roy, Forencich, Papen,
  Snoeren, and Porter}{Mellette et~al\mbox{.}}{2017}]%
        {rotornet}
\bibfield{author}{\bibinfo{person}{William~M Mellette}, \bibinfo{person}{Rob
  McGuinness}, \bibinfo{person}{Arjun Roy}, \bibinfo{person}{Alex Forencich},
  \bibinfo{person}{George Papen}, \bibinfo{person}{Alex~C Snoeren}, {and}
  \bibinfo{person}{George Porter}.} \bibinfo{year}{2017}\natexlab{}.
\newblock \showarticletitle{Rotornet: A scalable, low-complexity, optical
  datacenter network}. In \bibinfo{booktitle}{{\em Proceedings of the
  Conference of the ACM Special Interest Group on Data Communication}}.
  \bibinfo{pages}{267--280}.
\newblock


\bibitem[\protect\citeauthoryear{Mellette, Schuster, Porter, Papen, and
  Ford}{Mellette et~al\mbox{.}}{2016}]%
        {mellette2016scalable}
\bibfield{author}{\bibinfo{person}{William~Maxwell Mellette},
  \bibinfo{person}{Glenn~M Schuster}, \bibinfo{person}{George Porter},
  \bibinfo{person}{George Papen}, {and} \bibinfo{person}{Joseph~E Ford}.}
  \bibinfo{year}{2016}\natexlab{}.
\newblock \showarticletitle{A scalable, partially configurable optical switch
  for data center networks}.
\newblock \bibinfo{journal}{{\em Journal of Lightwave Technology\/}}
  \bibinfo{volume}{35}, \bibinfo{number}{2} (\bibinfo{year}{2016}),
  \bibinfo{pages}{136--144}.
\newblock


\bibitem[\protect\citeauthoryear{Narayanan, Harlap, Phanishayee, Seshadri,
  Devanur, Ganger, Gibbons, and Zaharia}{Narayanan et~al\mbox{.}}{2019}]%
        {pipedream}
\bibfield{author}{\bibinfo{person}{Deepak Narayanan}, \bibinfo{person}{Aaron
  Harlap}, \bibinfo{person}{Amar Phanishayee}, \bibinfo{person}{Vivek
  Seshadri}, \bibinfo{person}{Nikhil~R Devanur}, \bibinfo{person}{Gregory~R
  Ganger}, \bibinfo{person}{Phillip~B Gibbons}, {and} \bibinfo{person}{Matei
  Zaharia}.} \bibinfo{year}{2019}\natexlab{}.
\newblock \showarticletitle{PipeDream: Generalized pipeline parallelism for DNN
  training}. In \bibinfo{booktitle}{{\em Proceedings of the 27th ACM symposium
  on operating systems principles}}. \bibinfo{pages}{1--15}.
\newblock


\bibitem[\protect\citeauthoryear{Porter, Strong, Farrington, Forencich,
  Chen-Sun, Rosing, Fainman, Papen, and Vahdat}{Porter et~al\mbox{.}}{2013}]%
        {mordia}
\bibfield{author}{\bibinfo{person}{George Porter}, \bibinfo{person}{Richard
  Strong}, \bibinfo{person}{Nathan Farrington}, \bibinfo{person}{Alex
  Forencich}, \bibinfo{person}{Pang Chen-Sun}, \bibinfo{person}{Tajana Rosing},
  \bibinfo{person}{Yeshaiahu Fainman}, \bibinfo{person}{George Papen}, {and}
  \bibinfo{person}{Amin Vahdat}.} \bibinfo{year}{2013}\natexlab{}.
\newblock \showarticletitle{Integrating microsecond circuit switching into the
  data center}.
\newblock \bibinfo{journal}{{\em ACM SIGCOMM Computer Communication Review\/}}
  \bibinfo{volume}{43}, \bibinfo{number}{4} (\bibinfo{year}{2013}),
  \bibinfo{pages}{447--458}.
\newblock


\bibitem[\protect\citeauthoryear{Poutievski, Mashayekhi, Ong, Singh, Tariq,
  Wang, Zhang, Beauregard, Conner, Gribble, Kapoor, Kratzer, Li, Liu, Nagaraj,
  Ornstein, Sawhney, Urata, Vicisano, Yasumura, Zhang, Zhou, and
  Vahdat}{Poutievski et~al\mbox{.}}{2022}]%
        {jupiter-evolving}
\bibfield{author}{\bibinfo{person}{Leon Poutievski}, \bibinfo{person}{Omid
  Mashayekhi}, \bibinfo{person}{Joon Ong}, \bibinfo{person}{Arjun Singh},
  \bibinfo{person}{Mukarram Tariq}, \bibinfo{person}{Rui Wang},
  \bibinfo{person}{Jianan Zhang}, \bibinfo{person}{Virginia Beauregard},
  \bibinfo{person}{Patrick Conner}, \bibinfo{person}{Steve Gribble},
  \bibinfo{person}{Rishi Kapoor}, \bibinfo{person}{Stephen Kratzer},
  \bibinfo{person}{Nanfang Li}, \bibinfo{person}{Hong Liu},
  \bibinfo{person}{Karthik Nagaraj}, \bibinfo{person}{Jason Ornstein},
  \bibinfo{person}{Samir Sawhney}, \bibinfo{person}{Ryohei Urata},
  \bibinfo{person}{Lorenzo Vicisano}, \bibinfo{person}{Kevin Yasumura},
  \bibinfo{person}{Shidong Zhang}, \bibinfo{person}{Junlan Zhou}, {and}
  \bibinfo{person}{Amin Vahdat}.} \bibinfo{year}{2022}\natexlab{}.
\newblock \showarticletitle{Jupiter evolving: transforming google's datacenter
  network via optical circuit switches and software-defined networking}. In
  \bibinfo{booktitle}{{\em Proceedings of the ACM SIGCOMM 2022 Conference}}
  {\em (\bibinfo{series}{SIGCOMM '22})}.
\newblock
\showURL{%
\url{https://doi.org/10.1145/3544216.3544265}}


\bibitem[\protect\citeauthoryear{Qian, Xi, Cao, Gao, Xu, Guan, Fu, Shi, Zhu,
  Miao, Wang, Wang, Zhang, Zeng, Ruan, Yao, Zhai, and Cai}{Qian
  et~al\mbox{.}}{2024}]%
        {alibaba-hpn}
\bibfield{author}{\bibinfo{person}{Kun Qian}, \bibinfo{person}{Yongqing Xi},
  \bibinfo{person}{Jiamin Cao}, \bibinfo{person}{Jiaqi Gao},
  \bibinfo{person}{Yichi Xu}, \bibinfo{person}{Yu Guan},
  \bibinfo{person}{Binzhang Fu}, \bibinfo{person}{Xuemei Shi},
  \bibinfo{person}{Fangbo Zhu}, \bibinfo{person}{Rui Miao},
  \bibinfo{person}{Chao Wang}, \bibinfo{person}{Peng Wang},
  \bibinfo{person}{Pengcheng Zhang}, \bibinfo{person}{Xianlong Zeng},
  \bibinfo{person}{Eddie Ruan}, \bibinfo{person}{Zhiping Yao},
  \bibinfo{person}{Ennan Zhai}, {and} \bibinfo{person}{Dennis Cai}.}
  \bibinfo{year}{2024}\natexlab{}.
\newblock \showarticletitle{Alibaba HPN: A Data Center Network for Large
  Language Model Training}. In \bibinfo{booktitle}{{\em Proceedings of the ACM
  SIGCOMM 2024 Conference}} {\em (\bibinfo{series}{ACM SIGCOMM '24})}.
\newblock


\bibitem[\protect\citeauthoryear{Raja, Lange, Karpov, Shi, Fu, Behrendt,
  Cletheroe, Lukashchuk, Haller, Karinou, Thomsen, Jozwik, Liu, Costa,
  Kippenberg, and Ballani}{Raja et~al\mbox{.}}{2021}]%
        {raja2021ultrafast}
\bibfield{author}{\bibinfo{person}{Arslan~Sajid Raja}, \bibinfo{person}{Sophie
  Lange}, \bibinfo{person}{Maxim Karpov}, \bibinfo{person}{Kai Shi},
  \bibinfo{person}{Xin Fu}, \bibinfo{person}{Raphael Behrendt},
  \bibinfo{person}{Daniel Cletheroe}, \bibinfo{person}{Anton Lukashchuk},
  \bibinfo{person}{Istvan Haller}, \bibinfo{person}{Fotini Karinou},
  \bibinfo{person}{Benn Thomsen}, \bibinfo{person}{Krzysztof Jozwik},
  \bibinfo{person}{Junqiu Liu}, \bibinfo{person}{Paolo Costa},
  \bibinfo{person}{Tobias~Jan Kippenberg}, {and} \bibinfo{person}{Hitesh
  Ballani}.} \bibinfo{year}{2021}\natexlab{}.
\newblock \showarticletitle{Ultrafast optical circuit switching for data
  centers using integrated soliton microcombs}.
\newblock \bibinfo{journal}{{\em Nature Communications\/}}
  \bibinfo{volume}{12}, \bibinfo{number}{1} (\bibinfo{date}{Oct.}
  \bibinfo{year}{2021}).
\newblock
\showISSN{2041-1723}
\showDOI{%
\url{https://doi.org/10.1038/s41467-021-25841-8}}


\bibitem[\protect\citeauthoryear{Ren, Li, Wang, Huang, Li, Xu, Liao, Sun, Liu,
  Tian, Zhang, Wang, Zhong, Liu, Zhang, and Chen}{Ren et~al\mbox{.}}{2025}]%
        {fuselink}
\bibfield{author}{\bibinfo{person}{Zhenghang Ren}, \bibinfo{person}{Yuxuan Li},
  \bibinfo{person}{Zilong Wang}, \bibinfo{person}{Xinyang Huang},
  \bibinfo{person}{Wenxue Li}, \bibinfo{person}{Kaiqiang Xu},
  \bibinfo{person}{Xudong Liao}, \bibinfo{person}{Yijun Sun},
  \bibinfo{person}{Bowen Liu}, \bibinfo{person}{Han Tian},
  \bibinfo{person}{Junxue Zhang}, \bibinfo{person}{Mingfei Wang},
  \bibinfo{person}{Zhizhen Zhong}, \bibinfo{person}{Guyue Liu},
  \bibinfo{person}{Ying Zhang}, {and} \bibinfo{person}{Kai Chen}.}
  \bibinfo{year}{2025}\natexlab{}.
\newblock \showarticletitle{Enabling Efficient GPU Communication over Multiple
  NICs with FuseLink}. In \bibinfo{booktitle}{{\em 19th USENIX Symposium on
  Operating Systems Design and Implementation (OSDI 25)}}.
  \bibinfo{pages}{91--108}.
\newblock


\bibitem[\protect\citeauthoryear{Ryf, Kim, Hickey, Gnauck, Carr, Pardo, Bolle,
  Frahm, Basavanhally, Yoh, et~al\mbox{.}}{Ryf et~al\mbox{.}}{2001}]%
        {1296-mems}
\bibfield{author}{\bibinfo{person}{R Ryf}, \bibinfo{person}{J Kim},
  \bibinfo{person}{JP Hickey}, \bibinfo{person}{A Gnauck}, \bibinfo{person}{D
  Carr}, \bibinfo{person}{F Pardo}, \bibinfo{person}{C Bolle},
  \bibinfo{person}{R Frahm}, \bibinfo{person}{N Basavanhally},
  \bibinfo{person}{C Yoh}, {et~al\mbox{.}}} \bibinfo{year}{2001}\natexlab{}.
\newblock \showarticletitle{1296-port MEMS transparent optical crossconnect
  with 2.07 petabit/s switch capacity}. In \bibinfo{booktitle}{{\em OFC 2001.
  Optical Fiber Communication Conference and Exhibit. Technical Digest
  Postconference Edition (IEEE Cat. 01CH37171)}}, Vol.~\bibinfo{volume}{4}.
  IEEE, \bibinfo{pages}{PD28--PD28}.
\newblock


\bibitem[\protect\citeauthoryear{Saran, Amir, Wilson, Kleinberg, Shrivastav,
  and Weatherspoon}{Saran et~al\mbox{.}}{2024}]%
        {saran2024semi}
\bibfield{author}{\bibinfo{person}{Nitika Saran}, \bibinfo{person}{Daniel
  Amir}, \bibinfo{person}{Tegan Wilson}, \bibinfo{person}{Robert Kleinberg},
  \bibinfo{person}{Vishal Shrivastav}, {and} \bibinfo{person}{Hakim
  Weatherspoon}.} \bibinfo{year}{2024}\natexlab{}.
\newblock \showarticletitle{Semi-Oblivious Reconfigurable Datacenter Networks}.
  In \bibinfo{booktitle}{{\em Proceedings of the 23rd ACM Workshop on Hot
  Topics in Networks}}. \bibinfo{pages}{150--158}.
\newblock


\bibitem[\protect\citeauthoryear{Seok, Quack, Han, Muller, and Wu}{Seok
  et~al\mbox{.}}{2016}]%
        {seok2016large}
\bibfield{author}{\bibinfo{person}{Tae~Joon Seok}, \bibinfo{person}{Niels
  Quack}, \bibinfo{person}{Sangyoon Han}, \bibinfo{person}{Richard~S Muller},
  {and} \bibinfo{person}{Ming~C Wu}.} \bibinfo{year}{2016}\natexlab{}.
\newblock \showarticletitle{Large-scale broadband digital silicon photonic
  switches with vertical adiabatic couplers}.
\newblock \bibinfo{journal}{{\em Optica\/}} \bibinfo{volume}{3},
  \bibinfo{number}{1} (\bibinfo{year}{2016}), \bibinfo{pages}{64--70}.
\newblock


\bibitem[\protect\citeauthoryear{Shazeer, Mirhoseini, Maziarz, Davis, Le,
  Hinton, and Dean}{Shazeer et~al\mbox{.}}{2017}]%
        {shazeer2017outrageously}
\bibfield{author}{\bibinfo{person}{Noam Shazeer}, \bibinfo{person}{Azalia
  Mirhoseini}, \bibinfo{person}{Krzysztof Maziarz}, \bibinfo{person}{Andy
  Davis}, \bibinfo{person}{Quoc Le}, \bibinfo{person}{Geoffrey Hinton}, {and}
  \bibinfo{person}{Jeff Dean}.} \bibinfo{year}{2017}\natexlab{}.
\newblock \showarticletitle{Outrageously large neural networks: The
  sparsely-gated mixture-of-experts layer}.
\newblock \bibinfo{journal}{{\em arXiv preprint arXiv:1701.06538\/}}
  (\bibinfo{year}{2017}).
\newblock


\bibitem[\protect\citeauthoryear{Shoeybi, Patwary, Puri, LeGresley, Casper, and
  Catanzaro}{Shoeybi et~al\mbox{.}}{2020}]%
        {megatron-lm}
\bibfield{author}{\bibinfo{person}{Mohammad Shoeybi}, \bibinfo{person}{Mostofa
  Patwary}, \bibinfo{person}{Raul Puri}, \bibinfo{person}{Patrick LeGresley},
  \bibinfo{person}{Jared Casper}, {and} \bibinfo{person}{Bryan Catanzaro}.}
  \bibinfo{year}{2020}\natexlab{}.
\newblock \bibinfo{title}{Megatron-LM: Training Multi-Billion Parameter
  Language Models Using Model Parallelism}.
\newblock   (\bibinfo{year}{2020}).
\newblock
\showeprint[arxiv]{cs.CL/1909.08053}
\showURL{%
\url{https://arxiv.org/abs/1909.08053}}


\bibitem[\protect\citeauthoryear{Shou, Liu, Nie, Meng, Zhou, Jiang, Lv, Xu, Lu,
  Chen, Yu, Shen, Zhu, and Jiang}{Shou et~al\mbox{.}}{2025}]%
        {infinitehbd}
\bibfield{author}{\bibinfo{person}{Chenchen Shou}, \bibinfo{person}{Guyue Liu},
  \bibinfo{person}{Hao Nie}, \bibinfo{person}{Huaiyu Meng}, \bibinfo{person}{Yu
  Zhou}, \bibinfo{person}{Yimin Jiang}, \bibinfo{person}{Wenqing Lv},
  \bibinfo{person}{Yelong Xu}, \bibinfo{person}{Yuanwei Lu},
  \bibinfo{person}{Zhang Chen}, \bibinfo{person}{Yanbo Yu},
  \bibinfo{person}{Yichen Shen}, \bibinfo{person}{Yibo Zhu}, {and}
  \bibinfo{person}{Daxin Jiang}.} \bibinfo{year}{2025}\natexlab{}.
\newblock \bibinfo{title}{InfiniteHBD: Building Datacenter-Scale High-Bandwidth
  Domain for LLM with Optical Circuit Switching Transceivers}.
\newblock   (\bibinfo{year}{2025}).
\newblock
\showeprint[arxiv]{cs.NI/2502.03885}
\showURL{%
\url{https://arxiv.org/abs/2502.03885}}


\bibitem[\protect\citeauthoryear{Shrivastav, Valadarsky, Ballani, Costa, Lee,
  Wang, Agarwal, and Weatherspoon}{Shrivastav et~al\mbox{.}}{2019}]%
        {shoal}
\bibfield{author}{\bibinfo{person}{Vishal Shrivastav}, \bibinfo{person}{Asaf
  Valadarsky}, \bibinfo{person}{Hitesh Ballani}, \bibinfo{person}{Paolo Costa},
  \bibinfo{person}{Ki~Suh Lee}, \bibinfo{person}{Han Wang},
  \bibinfo{person}{Rachit Agarwal}, {and} \bibinfo{person}{Hakim
  Weatherspoon}.} \bibinfo{year}{2019}\natexlab{}.
\newblock \showarticletitle{Shoal: A network architecture for disaggregated
  racks}. In \bibinfo{booktitle}{{\em 16th USENIX Symposium on Networked
  Systems Design and Implementation (NSDI 19)}}. \bibinfo{pages}{255--270}.
\newblock


\bibitem[\protect\citeauthoryear{Singh, Ong, Agarwal, Anderson, Armistead,
  Bannon, Boving, Desai, Felderman, Germano, Kanagala, Provost, Simmons, Tanda,
  Wanderer, H\"{o}lzle, Stuart, and Vahdat}{Singh et~al\mbox{.}}{2015}]%
        {jupiter-rising}
\bibfield{author}{\bibinfo{person}{Arjun Singh}, \bibinfo{person}{Joon Ong},
  \bibinfo{person}{Amit Agarwal}, \bibinfo{person}{Glen Anderson},
  \bibinfo{person}{Ashby Armistead}, \bibinfo{person}{Roy Bannon},
  \bibinfo{person}{Seb Boving}, \bibinfo{person}{Gaurav Desai},
  \bibinfo{person}{Bob Felderman}, \bibinfo{person}{Paulie Germano},
  \bibinfo{person}{Anand Kanagala}, \bibinfo{person}{Jeff Provost},
  \bibinfo{person}{Jason Simmons}, \bibinfo{person}{Eiichi Tanda},
  \bibinfo{person}{Jim Wanderer}, \bibinfo{person}{Urs H\"{o}lzle},
  \bibinfo{person}{Stephen Stuart}, {and} \bibinfo{person}{Amin Vahdat}.}
  \bibinfo{year}{2015}\natexlab{}.
\newblock \showarticletitle{Jupiter Rising: A Decade of Clos Topologies and
  Centralized Control in Google's Datacenter Network}. In
  \bibinfo{booktitle}{{\em Proceedings of the 2015 ACM Conference on Special
  Interest Group on Data Communication}} {\em (\bibinfo{series}{SIGCOMM '15})}.
\newblock
\showURL{%
\url{https://doi.org/10.1145/2785956.2787508}}


\bibitem[\protect\citeauthoryear{Sukhbaatar, Golovneva, Sharma, Xu, Lin,
  Rozi{\`e}re, Kahn, Li, Yih, Weston, et~al\mbox{.}}{Sukhbaatar
  et~al\mbox{.}}{2024}]%
        {sukhbaatar2024branch}
\bibfield{author}{\bibinfo{person}{Sainbayar Sukhbaatar}, \bibinfo{person}{Olga
  Golovneva}, \bibinfo{person}{Vasu Sharma}, \bibinfo{person}{Hu Xu},
  \bibinfo{person}{Xi~Victoria Lin}, \bibinfo{person}{Baptiste Rozi{\`e}re},
  \bibinfo{person}{Jacob Kahn}, \bibinfo{person}{Daniel Li},
  \bibinfo{person}{Wen-tau Yih}, \bibinfo{person}{Jason Weston},
  {et~al\mbox{.}}} \bibinfo{year}{2024}\natexlab{}.
\newblock \showarticletitle{Branch-Train-MiX: Mixing Expert LLMs into a
  Mixture-of-Experts LLM}.
\newblock \bibinfo{journal}{{\em arXiv preprint arXiv:2403.07816\/}}
  (\bibinfo{year}{2024}).
\newblock


\bibitem[\protect\citeauthoryear{Team}{Team}{2024}]%
        {qwen25}
\bibfield{author}{\bibinfo{person}{Qwen Team}.}
  \bibinfo{year}{2024}\natexlab{}.
\newblock \showarticletitle{Qwen2.5 technical report}.
\newblock \bibinfo{journal}{{\em arXiv preprint arXiv:2412.15115\/}}
  (\bibinfo{year}{2024}).
\newblock


\bibitem[\protect\citeauthoryear{Wan, Zhang, Wang, Hu, Zhang, and Chen}{Wan
  et~al\mbox{.}}{2020}]%
        {rat}
\bibfield{author}{\bibinfo{person}{Xinchen Wan}, \bibinfo{person}{Hong Zhang},
  \bibinfo{person}{Hao Wang}, \bibinfo{person}{Shuihai Hu},
  \bibinfo{person}{Junxue Zhang}, {and} \bibinfo{person}{Kai Chen}.}
  \bibinfo{year}{2020}\natexlab{}.
\newblock \showarticletitle{Rat-resilient allreduce tree for distributed
  machine learning}. In \bibinfo{booktitle}{{\em Proceedings of the 4th
  Asia-Pacific Workshop on Networking}}. \bibinfo{pages}{52--57}.
\newblock


\bibitem[\protect\citeauthoryear{Wang, Andersen, Kaminsky, Papagiannaki, Ng,
  Kozuch, and Ryan}{Wang et~al\mbox{.}}{2010}]%
        {c-through}
\bibfield{author}{\bibinfo{person}{Guohui Wang}, \bibinfo{person}{David~G
  Andersen}, \bibinfo{person}{Michael Kaminsky}, \bibinfo{person}{Konstantina
  Papagiannaki}, \bibinfo{person}{TS~Eugene Ng}, \bibinfo{person}{Michael
  Kozuch}, {and} \bibinfo{person}{Michael Ryan}.}
  \bibinfo{year}{2010}\natexlab{}.
\newblock \showarticletitle{c-Through: Part-time optics in data centers}. In
  \bibinfo{booktitle}{{\em Proceedings of the ACM SIGCOMM 2010 Conference}}.
  \bibinfo{pages}{327--338}.
\newblock


\bibitem[\protect\citeauthoryear{Wang, Ghobadi, Shakeri, Zhang, and
  Hasani}{Wang et~al\mbox{.}}{2024}]%
        {rail-only}
\bibfield{author}{\bibinfo{person}{Weiyang Wang}, \bibinfo{person}{Manya
  Ghobadi}, \bibinfo{person}{Kayvon Shakeri}, \bibinfo{person}{Ying Zhang},
  {and} \bibinfo{person}{Naader Hasani}.} \bibinfo{year}{2024}\natexlab{}.
\newblock \bibinfo{title}{Rail-only: A Low-Cost High-Performance Network for
  Training LLMs with Trillion Parameters}.
\newblock   (\bibinfo{year}{2024}).
\newblock
\showeprint[arxiv]{cs.NI/2307.12169}
\showURL{%
\url{https://arxiv.org/abs/2307.12169}}


\bibitem[\protect\citeauthoryear{Wang, Khazraee, Zhong, Ghobadi, Jia, Mudigere,
  Zhang, and Kewitsch}{Wang et~al\mbox{.}}{2023}]%
        {topoopt}
\bibfield{author}{\bibinfo{person}{Weiyang Wang}, \bibinfo{person}{Moein
  Khazraee}, \bibinfo{person}{Zhizhen Zhong}, \bibinfo{person}{Manya Ghobadi},
  \bibinfo{person}{Zhihao Jia}, \bibinfo{person}{Dheevatsa Mudigere},
  \bibinfo{person}{Ying Zhang}, {and} \bibinfo{person}{Anthony Kewitsch}.}
  \bibinfo{year}{2023}\natexlab{}.
\newblock \showarticletitle{{TopoOpt}: Co-optimizing Network Topology and
  Parallelization Strategy for Distributed Training Jobs}. In
  \bibinfo{booktitle}{{\em 20th USENIX Symposium on Networked Systems Design
  and Implementation (NSDI 23)}}. \bibinfo{publisher}{USENIX Association},
  \bibinfo{address}{Boston, MA}, \bibinfo{pages}{739--767}.
\newblock
\showISBNx{978-1-939133-33-5}
\showURL{%
\url{https://www.usenix.org/conference/nsdi23/presentation/wang-weiyang}}


\bibitem[\protect\citeauthoryear{Wilson, Amir, Saran, Kleinberg, Shrivastav,
  and Weatherspoon}{Wilson et~al\mbox{.}}{2024}]%
        {wilson2024breaking}
\bibfield{author}{\bibinfo{person}{Tegan Wilson}, \bibinfo{person}{Daniel
  Amir}, \bibinfo{person}{Nitika Saran}, \bibinfo{person}{Robert Kleinberg},
  \bibinfo{person}{Vishal Shrivastav}, {and} \bibinfo{person}{Hakim
  Weatherspoon}.} \bibinfo{year}{2024}\natexlab{}.
\newblock \showarticletitle{Breaking the VLB Barrier for Oblivious
  Reconfigurable Networks}. In \bibinfo{booktitle}{{\em Proceedings of the 56th
  Annual ACM Symposium on Theory of Computing}}. \bibinfo{pages}{1865--1876}.
\newblock


\bibitem[\protect\citeauthoryear{Wilson, Amir, Shrivastav, Weatherspoon, and
  Kleinberg}{Wilson et~al\mbox{.}}{2023}]%
        {wilson2023extending}
\bibfield{author}{\bibinfo{person}{Tegan Wilson}, \bibinfo{person}{Daniel
  Amir}, \bibinfo{person}{Vishal Shrivastav}, \bibinfo{person}{Hakim
  Weatherspoon}, {and} \bibinfo{person}{Robert Kleinberg}.}
  \bibinfo{year}{2023}\natexlab{}.
\newblock \showarticletitle{Extending optimal oblivious reconfigurable networks
  to all n}. In \bibinfo{booktitle}{{\em 2023 Symposium on Algorithmic
  Principles of Computer Systems (APOCS)}}. SIAM, \bibinfo{pages}{1--16}.
\newblock


\bibitem[\protect\citeauthoryear{Xia, Schlansker, Ng, and Tourrilhes}{Xia
  et~al\mbox{.}}{2015}]%
        {xia2015enabling}
\bibfield{author}{\bibinfo{person}{Yiting Xia}, \bibinfo{person}{Mike
  Schlansker}, \bibinfo{person}{TS~Eugene Ng}, {and} \bibinfo{person}{Jean
  Tourrilhes}.} \bibinfo{year}{2015}\natexlab{}.
\newblock \showarticletitle{Enabling Topological Flexibility for Data Centers
  Using $\{$OmniSwitch$\}$}. In \bibinfo{booktitle}{{\em 7th USENIX Workshop on
  Hot Topics in Cloud Computing (HotCloud 15)}}.
\newblock


\bibitem[\protect\citeauthoryear{Xu, Sun, Wang, Ren, Wan, Liao, Wang, Zhang,
  and Chen}{Xu et~al\mbox{.}}{2025}]%
        {tacc-asplos}
\bibfield{author}{\bibinfo{person}{Kaiqiang Xu}, \bibinfo{person}{Decang Sun},
  \bibinfo{person}{Hao Wang}, \bibinfo{person}{Zhenghang Ren},
  \bibinfo{person}{Xinchen Wan}, \bibinfo{person}{Xudong Liao},
  \bibinfo{person}{Zilong Wang}, \bibinfo{person}{Junxue Zhang}, {and}
  \bibinfo{person}{Kai Chen}.} \bibinfo{year}{2025}\natexlab{}.
\newblock \showarticletitle{Design and Operation of Shared Machine Learning
  Clusters on Campus}. In \bibinfo{booktitle}{{\em Proceedings of the 30th ACM
  International Conference on Architectural Support for Programming Languages
  and Operating Systems, Volume 1}} {\em (\bibinfo{series}{ASPLOS '25})}.
  \bibinfo{publisher}{Association for Computing Machinery},
  \bibinfo{address}{New York, NY, USA}, \bibinfo{pages}{295--310}.
\newblock
\showISBNx{9798400706981}
\showDOI{%
\url{https://doi.org/10.1145/3669940.3707266}}


\bibitem[\protect\citeauthoryear{Ye, Yoo, and Akella}{Ye et~al\mbox{.}}{2012}]%
        {ye2012awgr}
\bibfield{author}{\bibinfo{person}{Xiaohui Ye}, \bibinfo{person}{SJ~Ben Yoo},
  {and} \bibinfo{person}{Venkatesh Akella}.} \bibinfo{year}{2012}\natexlab{}.
\newblock \showarticletitle{AWGR-based optical topologies for scalable and
  efficient global communications in large-scale multi-processor systems}.
\newblock \bibinfo{journal}{{\em Journal of Optical Communications and
  Networking\/}} \bibinfo{volume}{4}, \bibinfo{number}{9}
  (\bibinfo{year}{2012}), \bibinfo{pages}{651--662}.
\newblock


\bibitem[\protect\citeauthoryear{Yu, Hua, Zhong, Li, Luo, Zheng, and Zheng}{Yu
  et~al\mbox{.}}{2017}]%
        {yu2017fast}
\bibfield{author}{\bibinfo{person}{Yufang Yu}, \bibinfo{person}{Nan Hua},
  \bibinfo{person}{Zhizhen Zhong}, \bibinfo{person}{Jialong Li},
  \bibinfo{person}{Ruijie Luo}, \bibinfo{person}{Zelin Zheng}, {and}
  \bibinfo{person}{Xiaoping Zheng}.} \bibinfo{year}{2017}\natexlab{}.
\newblock \showarticletitle{Fast-Reconfigurable Optical Interconnect
  Architecture Based on Time-Synchronized Node Coordination for High
  Performance Computing}, In \bibinfo{booktitle}{Asia Communications and
  Photonics Conference}.
\newblock \bibinfo{journal}{{\em Asia Communications and Photonics
  Conference\/}}, \bibinfo{pages}{S4C.6}.
\newblock
\showDOI{%
\url{https://doi.org/10.1364/ACPC.2017.S4C.6}}


\bibitem[\protect\citeauthoryear{Zhang, Zhang, Wang, Govindan, Mogul, and
  Vahdat}{Zhang et~al\mbox{.}}{2021}]%
        {zhang2021gemini}
\bibfield{author}{\bibinfo{person}{Mingyang Zhang}, \bibinfo{person}{Jianan
  Zhang}, \bibinfo{person}{Rui Wang}, \bibinfo{person}{Ramesh Govindan},
  \bibinfo{person}{Jeffrey~C Mogul}, {and} \bibinfo{person}{Amin Vahdat}.}
  \bibinfo{year}{2021}\natexlab{}.
\newblock \showarticletitle{Gemini: Practical reconfigurable datacenter
  networks with topology and traffic engineering}.
\newblock \bibinfo{journal}{{\em arXiv preprint arXiv:2110.08374\/}}
  (\bibinfo{year}{2021}).
\newblock


\bibitem[\protect\citeauthoryear{Zhang, Xue, Guo, Li, Li, Shen, Qian, Yin, Wei,
  Yuan, et~al\mbox{.}}{Zhang et~al\mbox{.}}{2024}]%
        {zhang2024fast}
\bibfield{author}{\bibinfo{person}{Shicheng Zhang}, \bibinfo{person}{Xuwei
  Xue}, \bibinfo{person}{Bingli Guo}, \bibinfo{person}{Yixuan Li},
  \bibinfo{person}{Wenzhe Li}, \bibinfo{person}{Shikui Shen},
  \bibinfo{person}{Haoze Qian}, \bibinfo{person}{Xiaojie Yin},
  \bibinfo{person}{Buzheng Wei}, \bibinfo{person}{Guojun Yuan},
  {et~al\mbox{.}}} \bibinfo{year}{2024}\natexlab{}.
\newblock \showarticletitle{Fast-tunable Graphene-based AWGR for Deep Learning
  Training Networks}. In \bibinfo{booktitle}{{\em Proceedings of the 1st
  SIGCOMM Workshop on Hot Topics in Optical Technologies and Applications in
  Networking}}. \bibinfo{pages}{14--20}.
\newblock


\bibitem[\protect\citeauthoryear{Zhang, Wu, Michaels, and Henriksson}{Zhang
  et~al\mbox{.}}{2022}]%
        {zhang2022beam}
\bibfield{author}{\bibinfo{person}{Xiaosheng Zhang}, \bibinfo{person}{Ming
  Chiang~A Wu}, \bibinfo{person}{Andrew~S Michaels}, {and}
  \bibinfo{person}{Johannes Henriksson}.} \bibinfo{year}{2022}\natexlab{}.
\newblock \bibinfo{title}{Beam-steering system based on a MEMS-actuated
  vertical-coupler array}.
\newblock   (\bibinfo{date}{Sept.~13} \bibinfo{year}{2022}).
\newblock
\newblock
\shownote{US Patent 11,441,353.}


\bibitem[\protect\citeauthoryear{Zhu, Qu, Dong, Ruan, Tong, He, and Cheng}{Zhu
  et~al\mbox{.}}{2024}]%
        {llama-moe}
\bibfield{author}{\bibinfo{person}{Tong Zhu}, \bibinfo{person}{Xiaoye Qu},
  \bibinfo{person}{Daize Dong}, \bibinfo{person}{Jiacheng Ruan},
  \bibinfo{person}{Jingqi Tong}, \bibinfo{person}{Conghui He}, {and}
  \bibinfo{person}{Yu Cheng}.} \bibinfo{year}{2024}\natexlab{}.
\newblock \showarticletitle{LLaMA-MoE: Building Mixture-of-Experts from LLaMA
  with Continual Pre-training}.
\newblock \bibinfo{journal}{{\em arXiv preprint arXiv:2406.16554\/}}
  (\bibinfo{year}{2024}).
\newblock
\showURL{%
\url{https://arxiv.org/abs/2406.16554}}


\bibitem[\protect\citeauthoryear{Zu, Ghaffarkhah, Dang, Towles, Hand, Huda,
  Bello, Kolbasov, Rezaei, Du, et~al\mbox{.}}{Zu et~al\mbox{.}}{2024}]%
        {zu2024resiliency}
\bibfield{author}{\bibinfo{person}{Yazhou Zu}, \bibinfo{person}{Alireza
  Ghaffarkhah}, \bibinfo{person}{Hoang-Vu Dang}, \bibinfo{person}{Brian
  Towles}, \bibinfo{person}{Steven Hand}, \bibinfo{person}{Safeen Huda},
  \bibinfo{person}{Adekunle Bello}, \bibinfo{person}{Alexander Kolbasov},
  \bibinfo{person}{Arash Rezaei}, \bibinfo{person}{Dayou Du}, {et~al\mbox{.}}}
  \bibinfo{year}{2024}\natexlab{}.
\newblock \showarticletitle{Resiliency at Scale: Managing
  $\{$Google's$\}$$\{$TPUv4$\}$ Machine Learning Supercomputer}. In
  \bibinfo{booktitle}{{\em 21st USENIX Symposium on Networked Systems Design
  and Implementation (NSDI 24)}}. \bibinfo{pages}{761--774}.
\newblock


\end{thebibliography}

\appendix 
\appendixpage
Appendices are supporting material that has not been peer-reviewed.

\section{Production Measurement Details}\label{sec:appendix:measure-setting}

\subsection{Profiling of MoE Models}\label{sec:appendix:timeline}
\begin{figure}[h!]
    \begin{subfigure}[t]{0.9\linewidth}
        \centering
        \includegraphics[width=\linewidth]{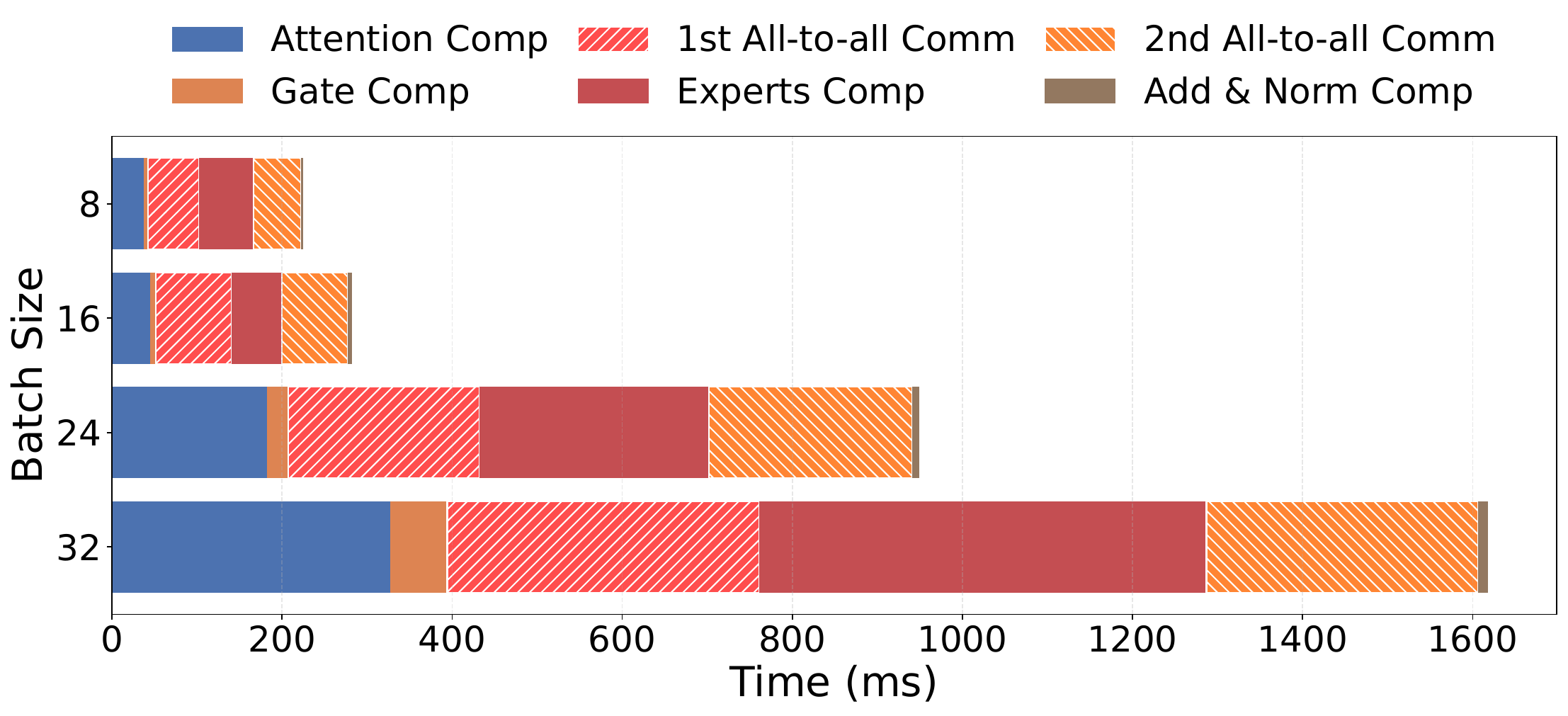}
        \caption{\llamamoe{}.}
        \label{fig:timeline:llama}
    \end{subfigure}
    \begin{subfigure}[t]{0.9\linewidth}
        \centering
        \includegraphics[width=\linewidth]{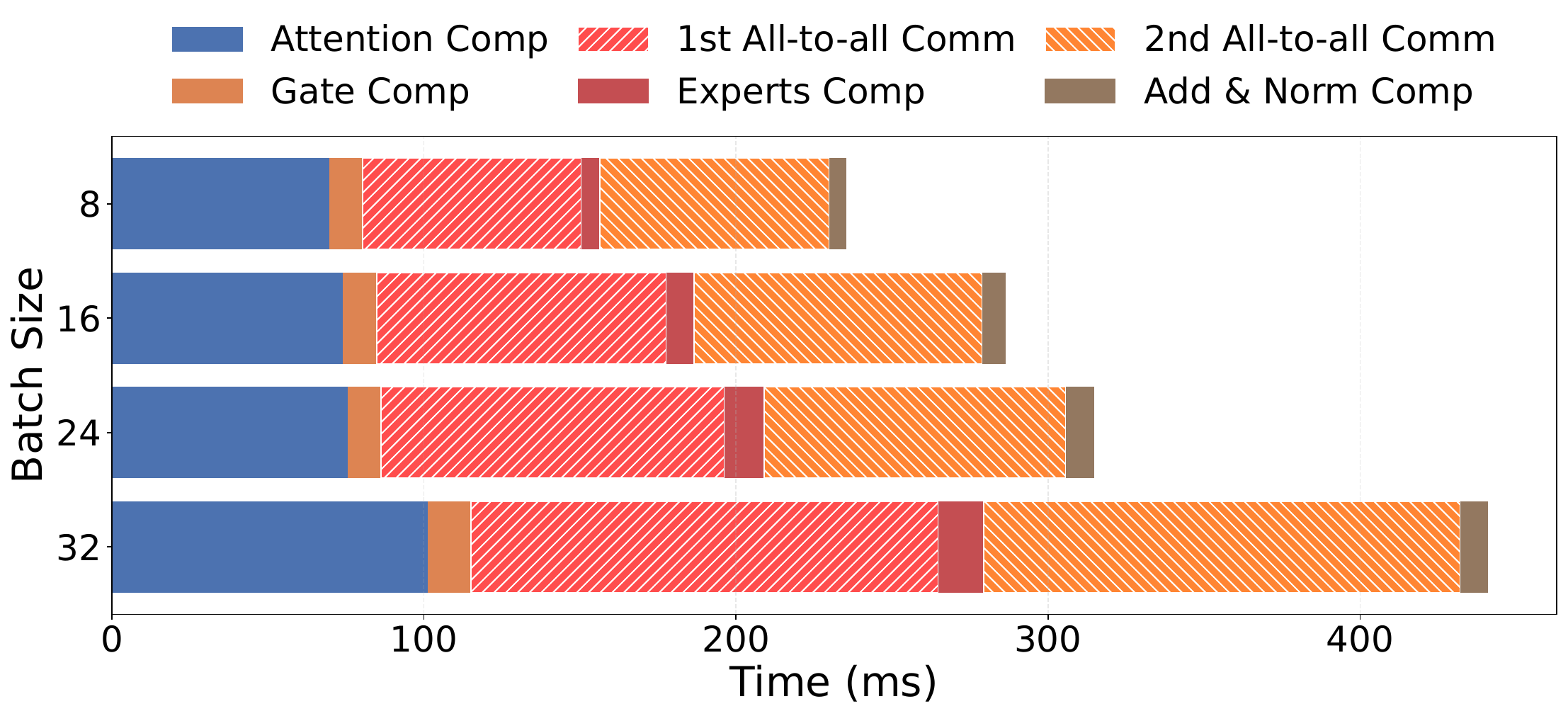}
        \caption{\qwenmoe{}.}
        \label{fig:timeline:qwen}
    \end{subfigure}
    \caption{Timeline of MoE models.}
    \label{fig:timeline:other}
\end{figure}

\figref{fig:timeline:other} presents the profiling results of an MoE layer for \llamamoe{} and \qwenmoe{}. EP communication constitutes a more significant portion in these models compared to Mixtral models. For instance, in \llamamoe{}, the two all-to-all communication phases account for 42\%–58\% of the iteration time. In comparison, EP communication dominates even more in \qwenmoe{}, reaching up to 68\%.

\subsection{Non-uniform token distribution in trained MoE model}\label{sec:appendix:infer-dist}
\begin{figure}[t!]
    \centering
    \includegraphics[width=0.8\linewidth]{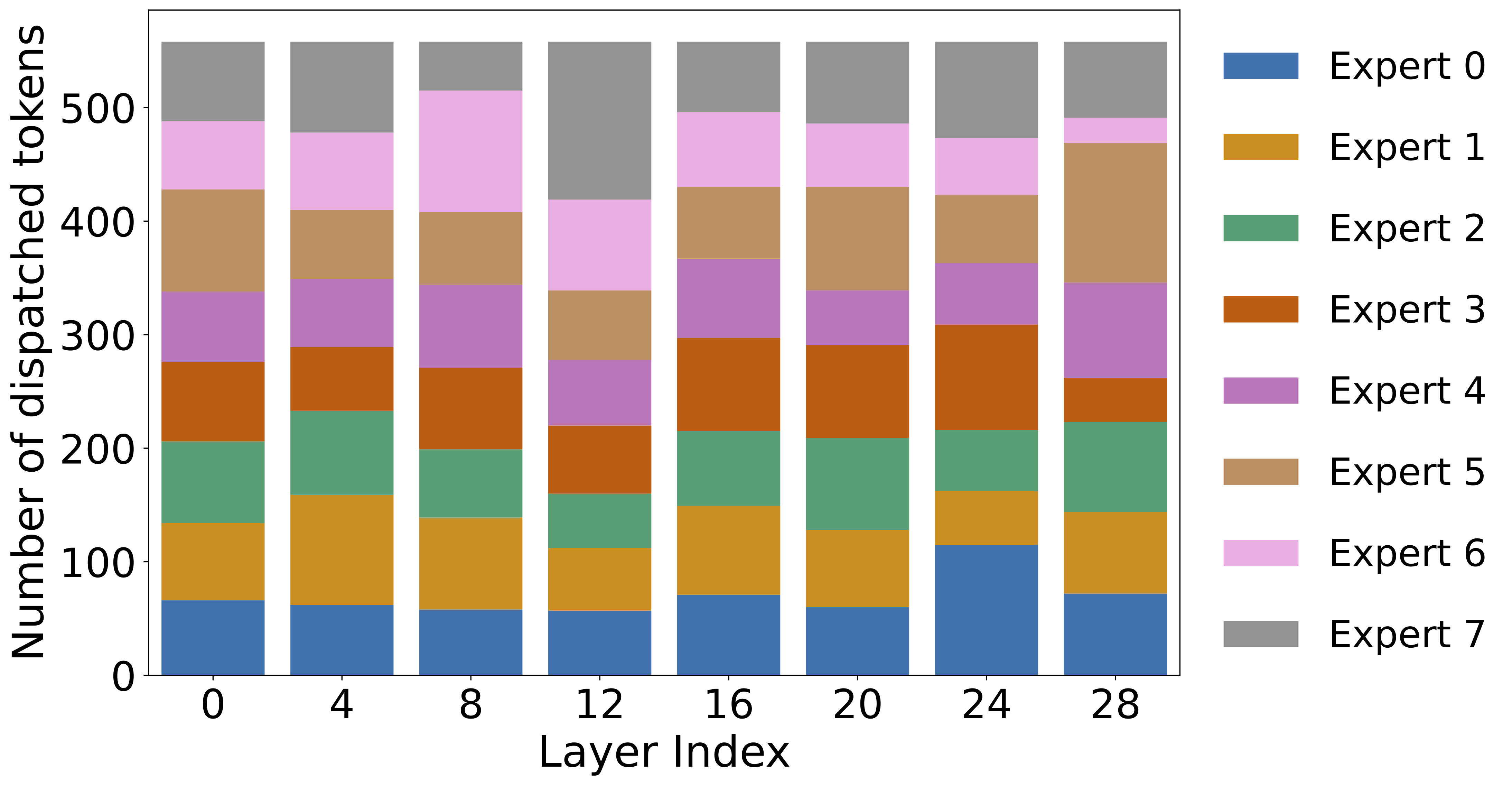}
    \caption{[Mixtral 8$\times$7B in production] Non-uniform token distribution across MoE blocks.}
    \label{fig:motiv:mixtral-infer-dist}
\end{figure}

We measured the all-to-all token distribution in the forward pass of pre-trained \mixtral{}~\cite{mixtral-huggingface}, as shown in \figref{fig:motiv:mixtral-infer-dist}. We observe that the number of tokens dispatched to each expert is non-uniform and varies across different MoE blocks, 
which argues for a necessary mechanism to adapt to the dynamic traffic in EP even when the model has largely converged.

\begin{figure}[t!]
    \centering
    \includegraphics[width=0.8\linewidth]{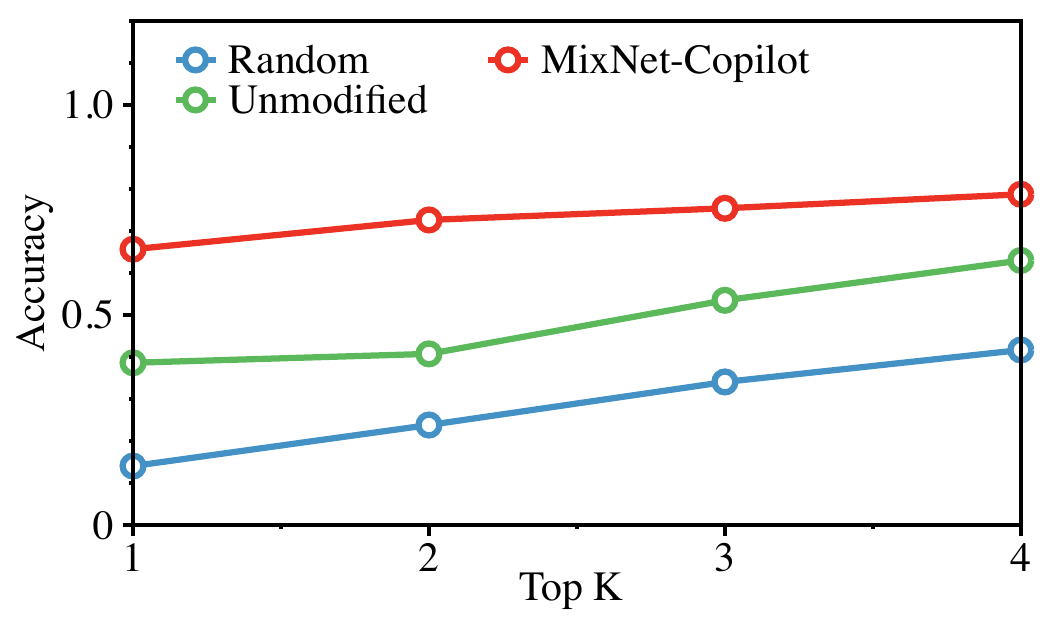}
    \caption{Average prediction accuracy of \textsc{MixNet-Copilot}.}
    \label{fig:design:prediction}
\end{figure}

\begin{figure*}[t!]
    \centering
    \includegraphics[width=0.85\linewidth]{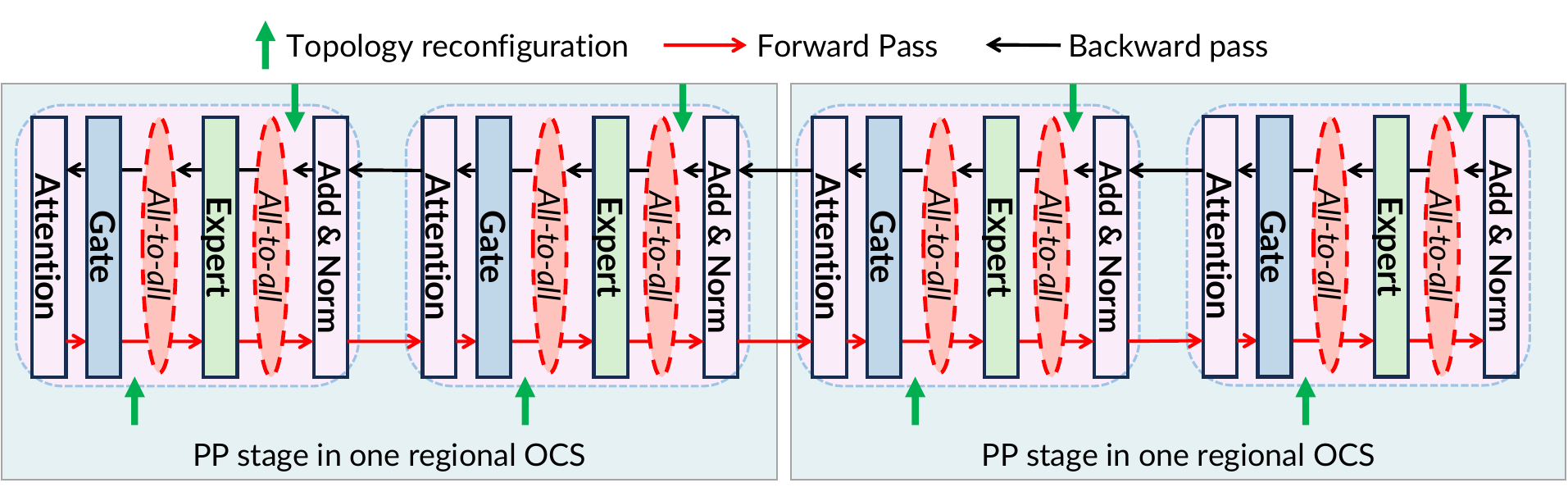}
    \caption{Reconfiguration timeline during runtime.}
    \label{fig:design:reconfig-timeline}
\end{figure*}

\section{Implementation Details}
\subsection{Traffic Demand Prediction}\label{sec:appendix:predict}
\sys{} aims to handle the first all-to-all communication in the forward pass with a predictive approach. By default, the OCS topology for this initial communication is either randomly generated (e.g., for the first all-to-all in the first layer) or remains unchanged from the previously used topology (e.g., the first all-to-all in the second layer). The traffic demand prediction algorithm predicts the \emph{conditional probability} of the traffic matrix, denoting the conditional probability of a token gated to expert $j$ given that it is gated to the expert $i$ in the last layer. With the conditional probability matrix and the empirical token distribution in the previous layer, we can predict the traffic distribution in the current layer.                 

\parab{Matrix Estimation:}  For each layer, \sys{} estimates the conditional probability matrix with the traffic demand records in recent iterations. Focusing on the recent expert load distributions, we employ a weighted average within a fixed window of traffic records in time series. For each layer, the optimization objective is to minimize the square error between the predicted load distribution and the ground truth (for simplicity, we omit the layer index here):
\begin{equation}
  \min_{P} \quad \sum_{i=1}^{k} w_i \cdot \Sigma_i\left((Y_i - P X_i)^2\right),
\end{equation}
where $k$ is the window size. The transition matrix $P$ is of size $N \times N$, representing the conditional probability of the current layer's expert load distribution, given the expert load distribution of the previous layer. \( X_i \), \( Y_i \) are normalized expert load distribution vectors of two neighboring layers. Each element in \(P\) is constrained to be in the range \([0,1]\), and the sum of each column is constrained to be 1 to ensure that \(P\) is a valid probability matrix.

We employ the Sequential Least Squares Programming (SLAP) method for optimization, as it is suitable for nonlinear problems with linear constraints. The algorithm is implemented via the \ct{scipy\allowbreak.optimize} library in Python. During the inference, with the load distribution in layer $i$ given, we can predict the expert load distribution for the next layer in advance for its first all-to-all communication.

We name this method \copilot{}. \figref{fig:design:prediction} compares the prediction accuracy of \copilot{} against the aforementioned methods, \ie{} the randomly assigned token distribution (uniform bandwidth allocation),  and the unmodified token distribution from the previous layers (unchanged topology) on collected traces from measurements. Top $K$ accuracy measures whether \sys{} is able to find the top-k activation-intensive experts. We find that \copilot{} exhibits significantly higher accuracy than other counterparts, which implies that \copilot{} can find the most intensive pairs in all-to-all communication with high probability. Therefore, \copilot{} offers an opportunity to proactively reconfigure the topology for the FP's first all-to-all in advance.

\subsection{Details of Topology Reconfigurations}\label{sec:appendix:reconfig-timeline}

\figref{fig:design:reconfig-timeline} details the runtime reconfiguration methodology of \sys{}. Each MoE layer involves four all-to-all communication phases. As discussed in \secref{sec:design:traffic-demand-prediction}, \sys{} can deterministically characterize the traffic patterns for the second all-to-all communication in the forward pass (FP) and both all-to-all communications in the backward pass (BP). 

Consequently, \sys{} conceptually reconfigures the topology twice per MoE layer—once during FP and once during BP. However, for the first all-to-all communication in FP, \sys{} cannot fully characterize the traffic matrix in advance due to the absence of runtime information at that stage of the iteration. To tackle this issue, \sys{} performs an inaccurate reconfiguration based on partial estimates or reuses the topology from the previous layer. It then recalibrates the topology for the second all-to-all in FP with minor OCS adjustments, ensuring a more accurate configuration for subsequent communication phases.

\section{ProtoType Details}\label{sec:appendix:testbed}
\begin{figure}[t!]
    \centering
    \includegraphics[width=0.85\linewidth]{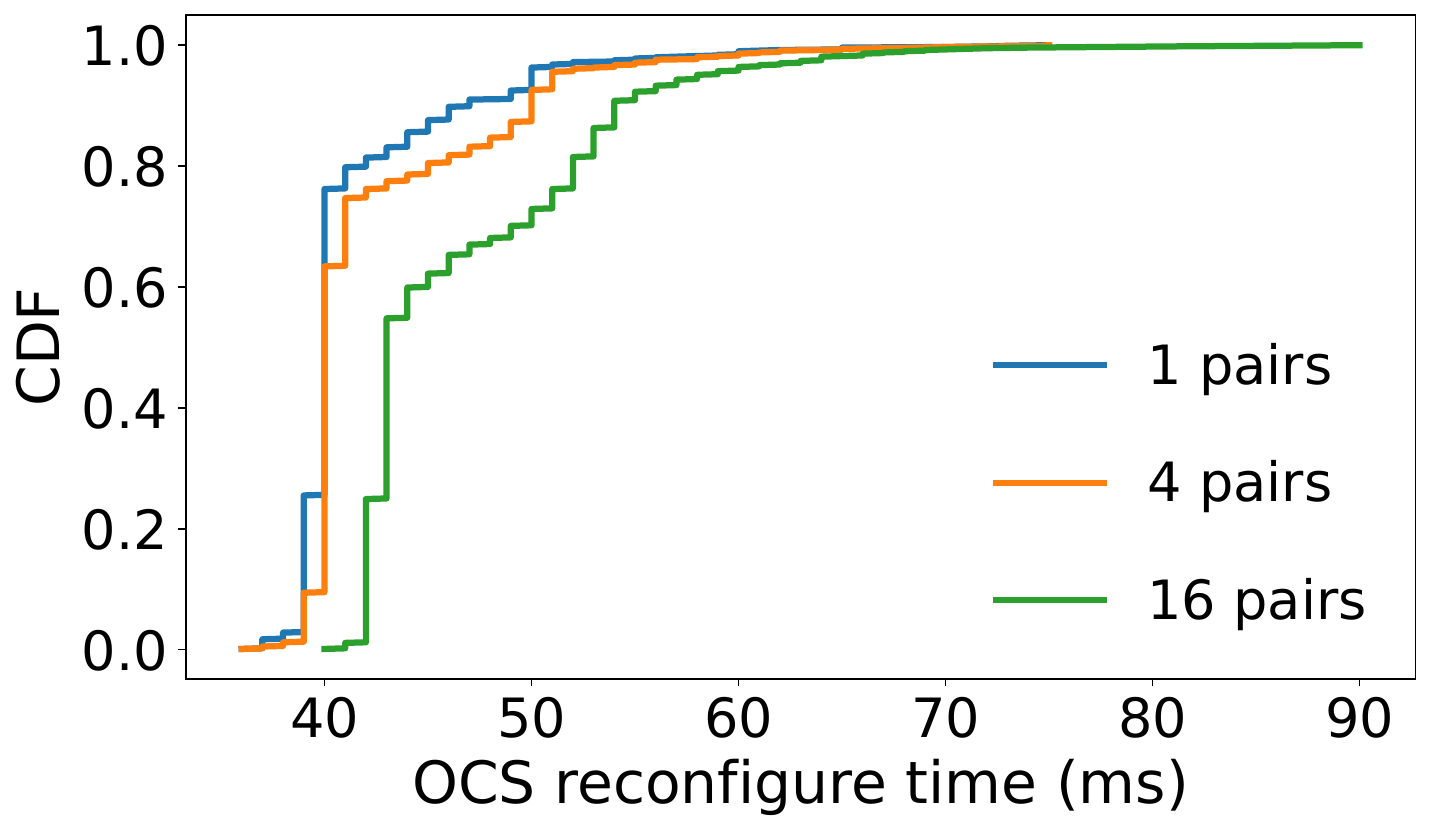}
    \caption{[Testbed] Reconfiguration delay across different number of pairs.}
    \label{fig:testbed:ocs-cdf}
\end{figure}

\begin{figure}[t!]
    \centering
    \includegraphics[width=0.85\linewidth]{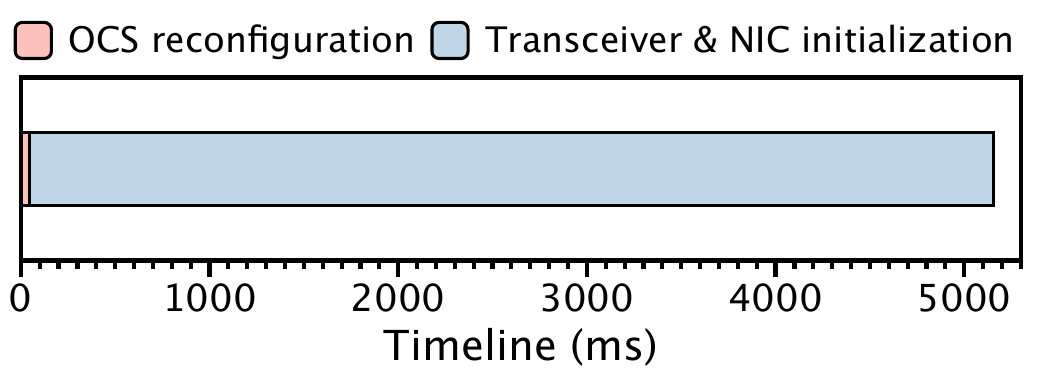}
    \caption{[Testbed] Overall timeline of one OCS control.}
    \label{fig:testbed:ocs-timeline}
\end{figure}

\begin{figure}[t!]
    \centering
    \includegraphics[width=0.85\linewidth]{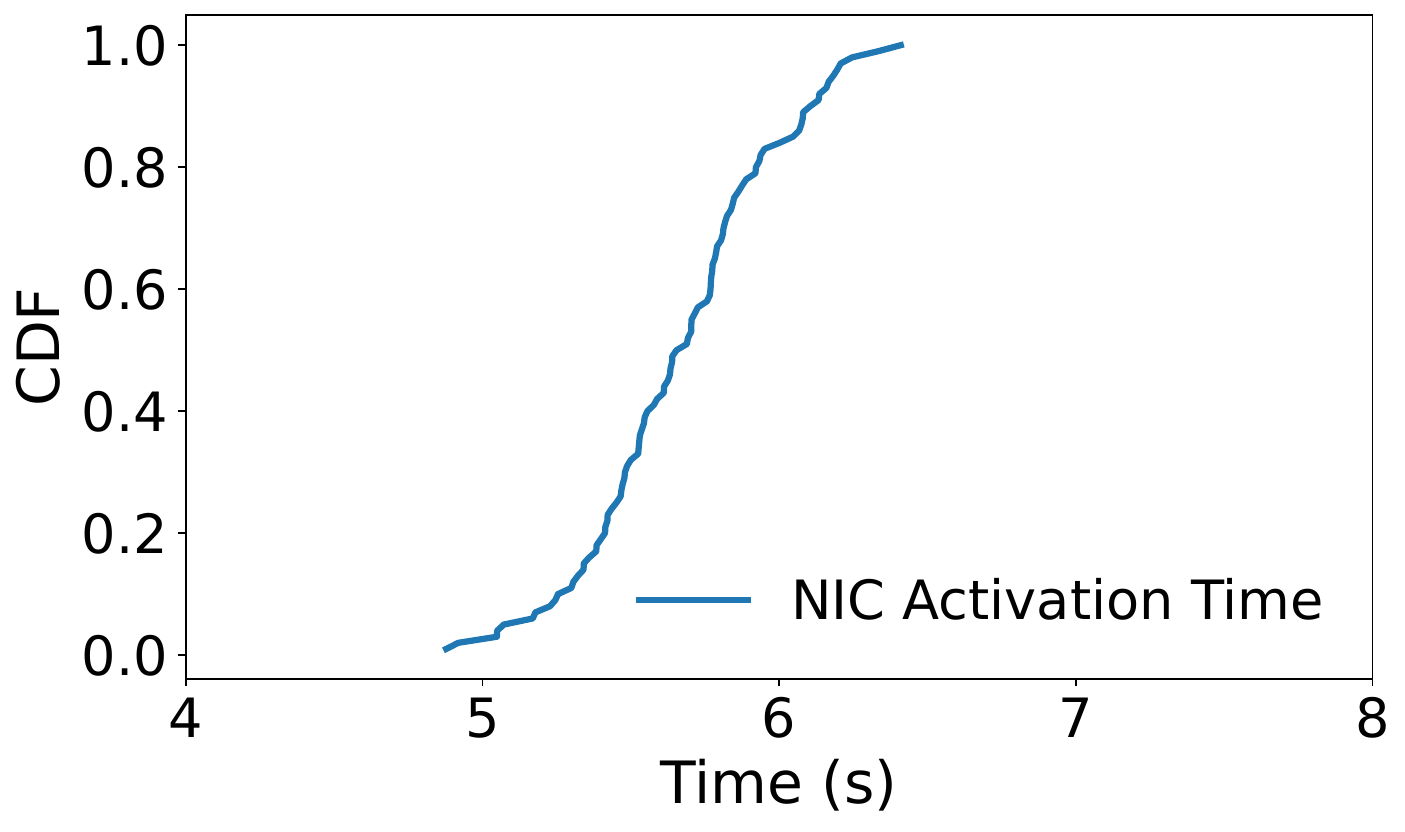}
    \caption{[Testbed] CDF of time elapsed from OCS reconfiguration completion to NIC becoming active.}
    \label{fig:testbed:nic-up-cdf}
\end{figure}

\parab{Prototype profiling.} We profiled the overall reconfiguration turnaround time of our OCS, and the results are shown in \figref{fig:testbed:ocs-cdf}. The OCS is controlled by issuing TL1 commands over Ethernet. We observed that as the number of pairs increases, the reconfiguration time slightly rises. The average reconfiguration time is approximately 41.44 ms for 1 pair, 42.44 ms for 4 pairs, and 46.75 ms for 16 pairs. The 99th percentile reconfiguration times are around 60 ms for 1 pair, 62 ms for 4 pairs, and 68 ms for 16 pairs. Notably, 99\% of the reconfiguration times are under 70 ms, which is acceptable for MoE training, given the relatively long expert computation times typically used in practice (e.g., 122 ms for a batch size of 16).

\figref{fig:testbed:ocs-timeline} illustrates the overall timeline from issuing an OCS reconfiguration control command to the successful completion of an RDMA send, providing a detailed view of the control process. The process consists of two main stages: (1) the control server sends a reconfiguration command to the OCS; (2) the transceiver and NIC initialize the physical link and set up the network device. Our observations indicate that the overall turnaround time of one reconfiguration is predominantly influenced by the physical link initialization and NIC device initialization.

We further plot the CDF of the time elapsed from the OCS reconfiguration completion to the NIC becoming active. The results are shown in \figref{fig:testbed:nic-up-cdf}. The average NIC activation time is approximately 5.67 s and the 99 percentile is around 6.33 s. These findings align with previous observations from \cite{realizing-rotornet}, highlighting that some commodity transceivers and NICs are not well optimized for fast reconfiguration. 
We have discussed this issue with a transceiver vendor, who confirmed that the observed multi-second NIC reactivation latency is not a fundamental limitation, but rather a consequence of current commercial transceiver modules not being optimized for fast reconfigurable optical switching in datacenter environments. Note that the burst-mode transceiver (\eg \cite{burst-mode-transceiver1,burst-mode-transceiver2}) has already been deployed with passive optical networks (PONs) in access networks, which are designed to handle intermittent upstream transmissions with fast CDR locking and signal recovery. Therefore, extending datacenter transceivers with burst-mode features supported in PON transceivers is an engineering problem rather than an architectural barrier. In particular, these engineering efforts include:
1) Classical receiver-side CDR circuits require continuous data streams to maintain CDR state. To prevent loss-of-signal (LOS) during OCS reconfiguration, the optical transceiver can be configured into a local Tx/Rx loopback mode~\cite{infinitehbd} prior to the switching event, ensuring that the receiver continues to observe a valid signal throughout the transition. 2) After the OCS reconfiguration, the CDR logic can be optimized using fast-locking CDR designs (\eg \cite{clark2018sub}), which enable rapid recovery and re-synchronization upon detection of the new signal.

As a result, we currently exclude this NIC activation time to calculate the actual training time in \sys{} testbed experiments. Due to the limited number of GPUs, we cannot train the full MoE models as shown in \tabref{tab:measure-setting}. We only run 7 layers of \mixtral{}, 16 layers of \llamamoe{}, and 12 layers of \qwenmoe{}.

\begin{table}[t!]
    \centering
    \begin{threeparttable}
    \fontsize{8}{10}\selectfont
    \renewcommand{\arraystretch}{1.0}
    \begin{tabular}{p{1.2cm}p{0.8cm}p{1.0cm}p{1.0cm}p{0.8cm}p{1.0cm}}
         \toprule
            \textbf{Link Bandwidth} & \textbf{Trans-ceiver (\$)}  & \textbf{NIC (\$)}  & \textbf{Elec. switch port (\$)} & \textbf{OCS port (\$)} & \textbf{Patch panel port (\$)} \\
         \midrule
         100 Gbps & 99~\cite{100g-transceiver} & 659~\cite{100g-nic} & 187~\cite{topoopt} & 520~\cite{ocs-price} & 100~\cite{patch-panel-price}\\
         200 Gbps & 239~\cite{200g-transceiver}  & 1079~\cite{200g-nic} & 374~\cite{topoopt} & 520~\cite{ocs-price} & 100~\cite{patch-panel-price}\\
         400 Gbps & 659~\cite{400g-transceiver}  & 1499~\cite{400g-nic} & 1090~\cite{400g-switch} & 520~\cite{ocs-price} & 100~\cite{patch-panel-price}\\
         800 Gbps & 1399~\cite{800g-transceiver} & 2248\tnote{1}~~\cite{400g-nic} & 1400~\cite{800g-switch} & 520~\cite{ocs-price}  & 100~\cite{patch-panel-price} \\
         \bottomrule
    \end{tabular}
    \begin{tablenotes}
        \scriptsize
        \item[1] Conservatively estimated as 1.5 times the price of 400G NIC, as 800G products are not yet commercially available.
    \end{tablenotes}
    \caption{Cost of network components.}
    \label{tab:cost-component}
\end{threeparttable}
\end{table}

\begin{figure}[t!]
    \centering
    \includegraphics[width=0.8\linewidth]{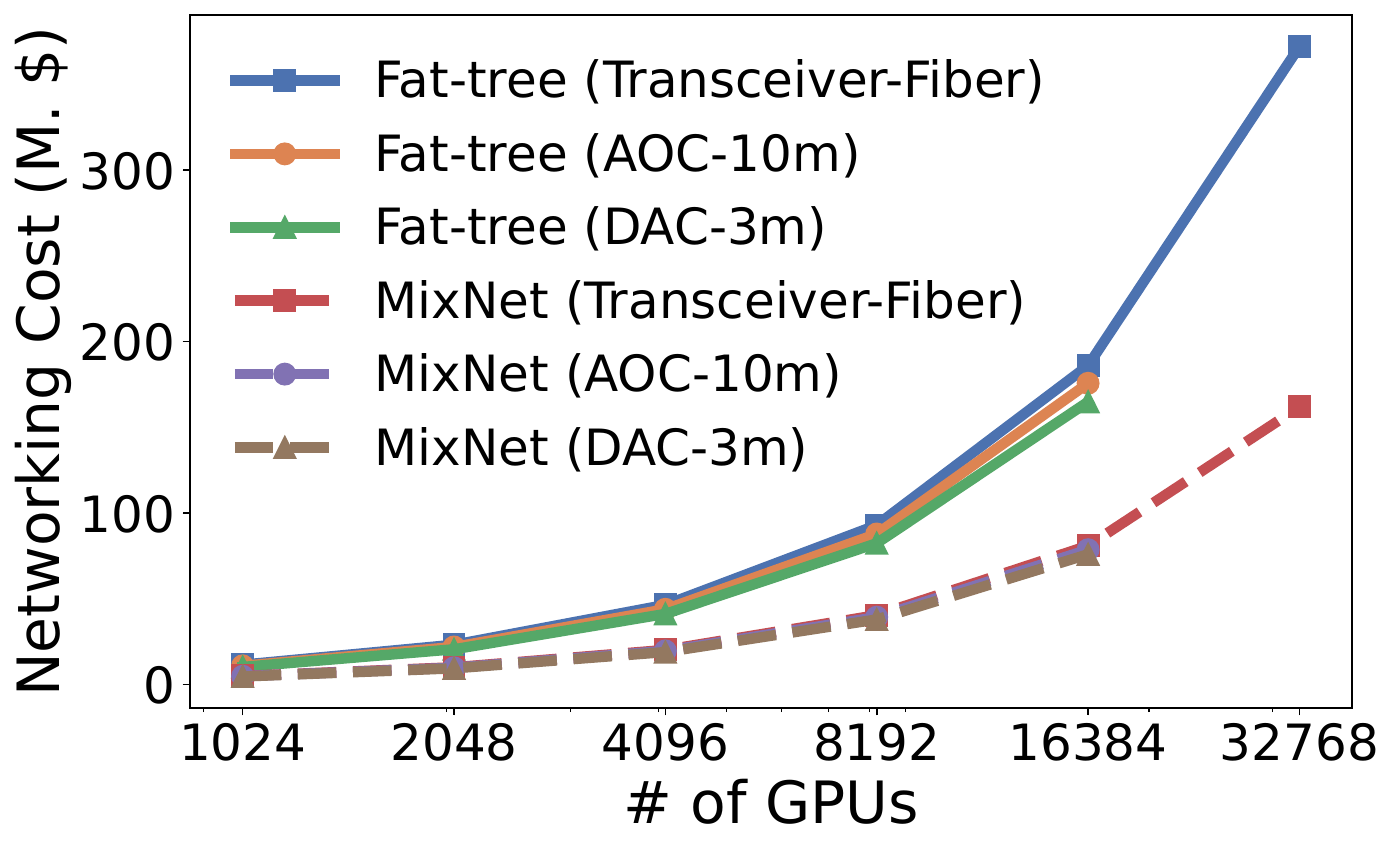}
    \caption{[Simulation] Cost comparison of different EPS links at 400 Gbps bandwidth.}
    \label{fig:sim:cable-type}
\end{figure}

\begin{figure*}[t!]
    \begin{subfigure}[t]{0.24\linewidth}
        \centering
        \includegraphics[width=\linewidth]{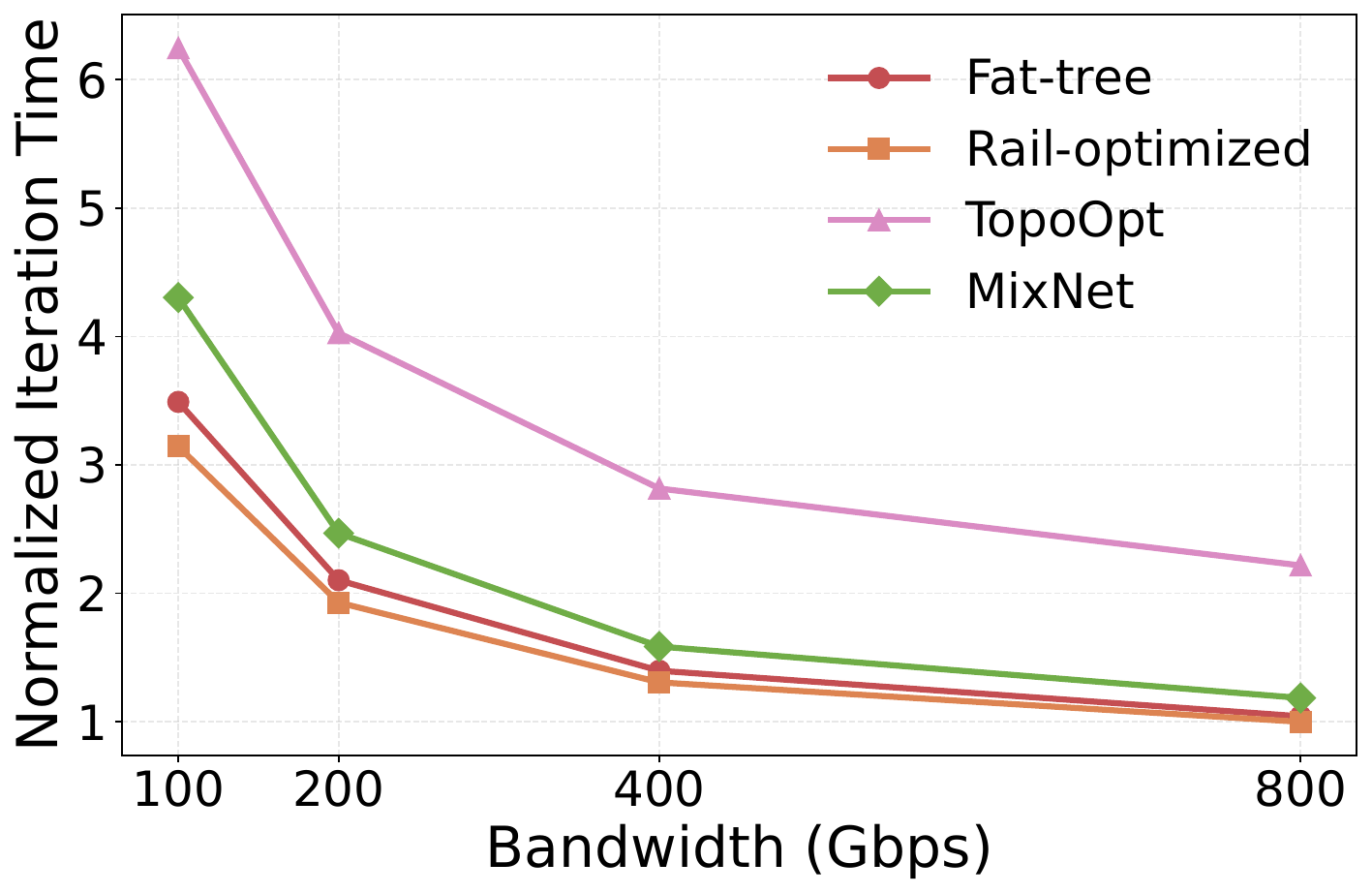}
        \caption{\mixtrallarge{} (batch size 32)}
    \end{subfigure}
    \hspace{0.2em}
    \begin{subfigure}[t]{0.24\linewidth}
        \centering
        \includegraphics[width=\linewidth]{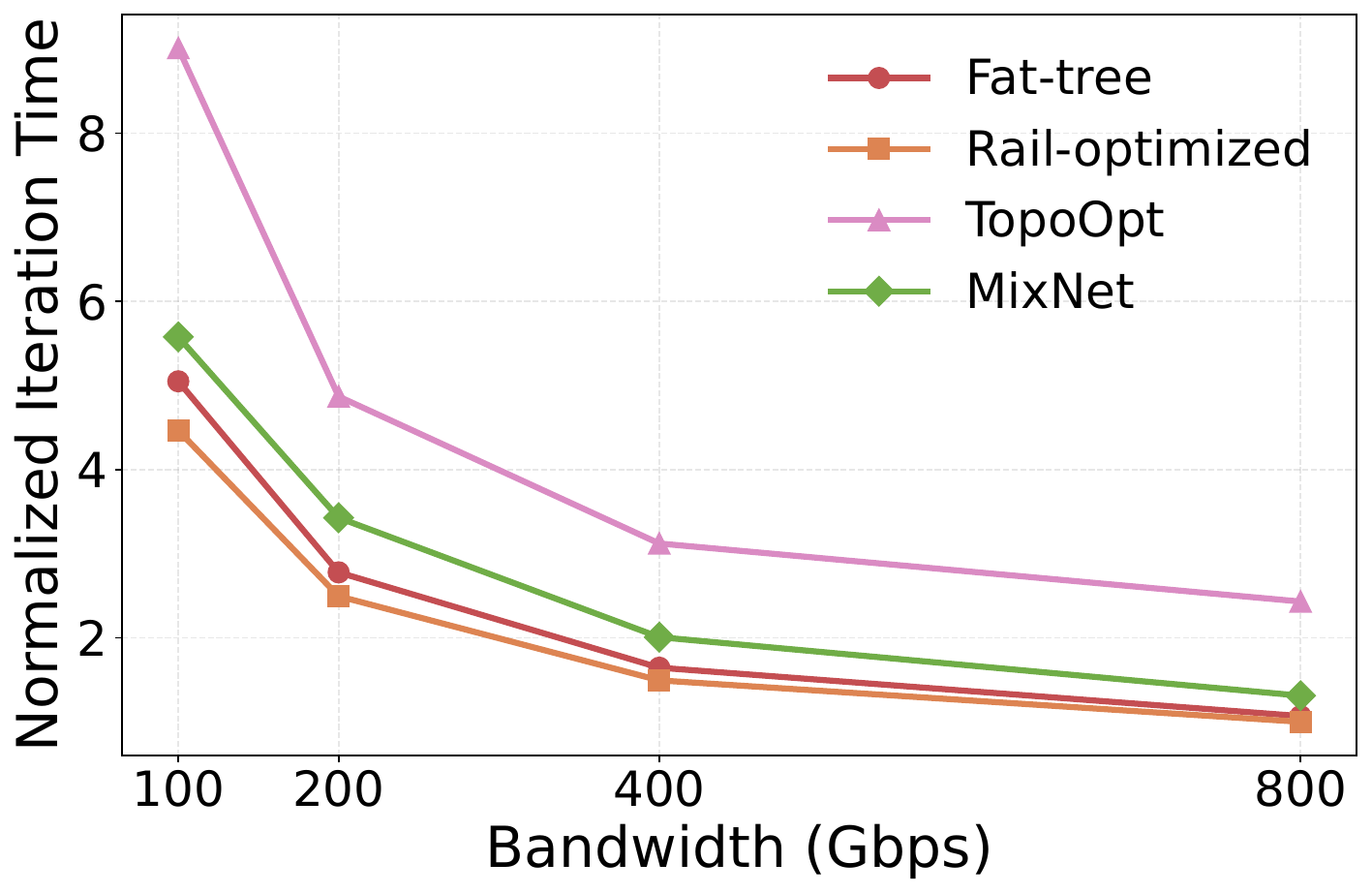}
        \caption{\mixtrallarge{} (batch size 64)}
    \end{subfigure}
    \hspace{0.2em}
    \begin{subfigure}[t]{0.24\linewidth}
        \centering
        \includegraphics[width=\linewidth]{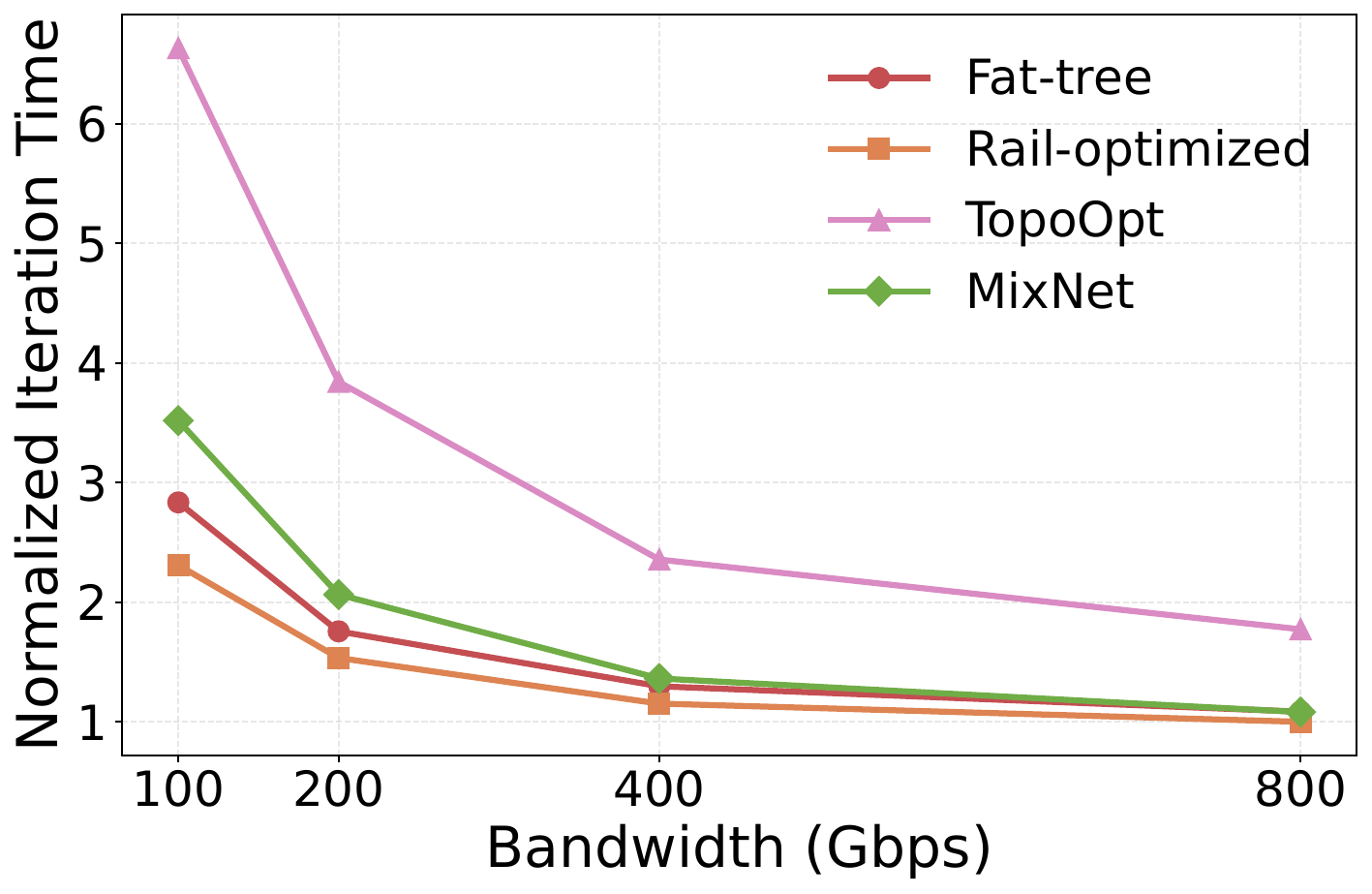}
        \caption{\mixtral{} (batch size 32)}
    \end{subfigure}
    \hspace{0.2em}
    \begin{subfigure}[t]{0.24\linewidth}
        \centering
        \includegraphics[width=\linewidth]{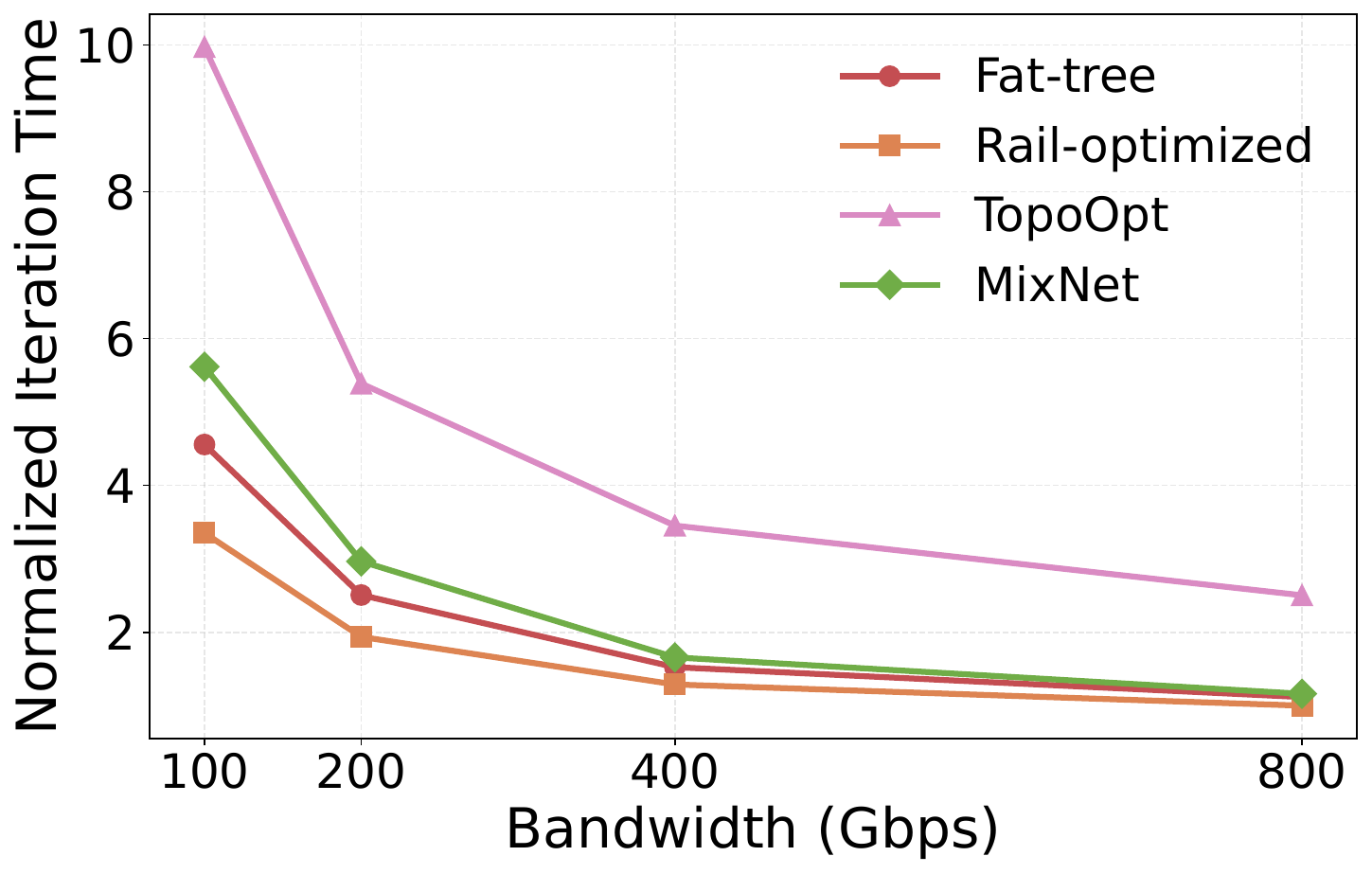}
        \caption{\mixtral{} (batch size 64)}
    \end{subfigure}
    \caption{[Simulation] Training speed ups of Mixtral models with large batch sizes.}
    \label{fig:sim:mixtral-large}
\end{figure*}

\section{Large-Scale Simulation Details}\label{sec:appendix:sim-detail}

\subsection{Simulated MoE Models and Parallelization Strategies}\label{sec:appendix:sim:setup}
We simulate the training process of four MoE models: \mixtrallarge{}~\cite{mixtral22b}, \mixtral{}~\cite{mixtral}, \qwenmoe{}~\cite{qwen-moe} and \deepseek{}~\cite{deepseek-r1}. For \mixtrallarge{}, we use a hybrid parallelism that combines an EP degree of 8, TP degree of 8, PP degree of 8 at a sequence length of 4096, and micro-batch size of 8. For \deepseek{}, we follow the default training parallelisms in \cite{deepseek-r1} with 64-way EP and 16-way PP. For other models, we reuse the same configurations in \tabref{tab:measure-setting}. 

\subsection{Cost Analysis Details}\label{sec:appendix:cost}

\tabref{tab:cost-component} lists the costs of network components used in \secref{sec:sim:cost}. We reuse the prices for electronic switches at 100G, 200G as well as for NICs, OCS ports, and patch panel ports from TopoOpt~\cite{topoopt}, and we add the prices of transceivers, NICs, and electronic switch ports for 400 Gbps and 800 Gbps link accordingly. We also follow the same methodology as in TopoOpt when calculating the fiber costs.

\subsection{Different EPS Link Options}
The OCS portion of \sys{} requires optical transceivers with pluggable fibers to allow optical switching.
For the EPS part of \sys{}, especially short-reach rack-scale links between the servers and ToR switches, Direct Attach Copper (DAC) cables or Active Optical Cables (AOC)  are more cost-effective alternatives to optical transceivers plus fibers (typically used for long-reach links). We analyze the cost implications of these link options in \figref{fig:sim:cable-type}. The results show that replacing the EPS links with DAC or AOC slightly reduce the costs for both fat-tree interconnect and \sys{}. Most importantly, the cost effectiveness of \sys{} is orthogonal to the choices of EPS links, and maintains significant cost advantages over fat-tree topology. For example, with 400 Gbps DAC cables option in a 4096-GPU cluster, \sys{} achieves 2.2$\times$ lower total cost compared to fat-tree topology.

\subsection{Training Speed Ups of Mixtral Models with Larger Batch Sizes}\label{sec:appendix:mixtral-large}

We further evaluate \sys{}'s performance with larger batch sizes using two Mixtral MoE models (\mixtral{} and \mixtrallarge{}). For each model, we test batch sizes of 32 and 64 across varying network bandwidths (100-800 Gbps). As shown in \figref{fig:sim:mixtral-large}, \sys{} consistently outperforms TopoOpt under all configurations. Specifically, \sys{} achieves an average speedup of 1.8$\times$ for Mixtral-8x7B with a batch size of 32 and 2.0$\times$ with a batch size of 64, as training becomes more communication-intensive compared to the settings in \figref{fig:sim:speed}. Furthermore, we observe that as link bandwidth increases, \sys{}'s performance gradually approaches that of Fat-tree and Rail-optimized architectures.

\begin{figure}[t!]
    \begin{subfigure}[t]{0.8\linewidth}
        \centering
        \includegraphics[width=\linewidth]{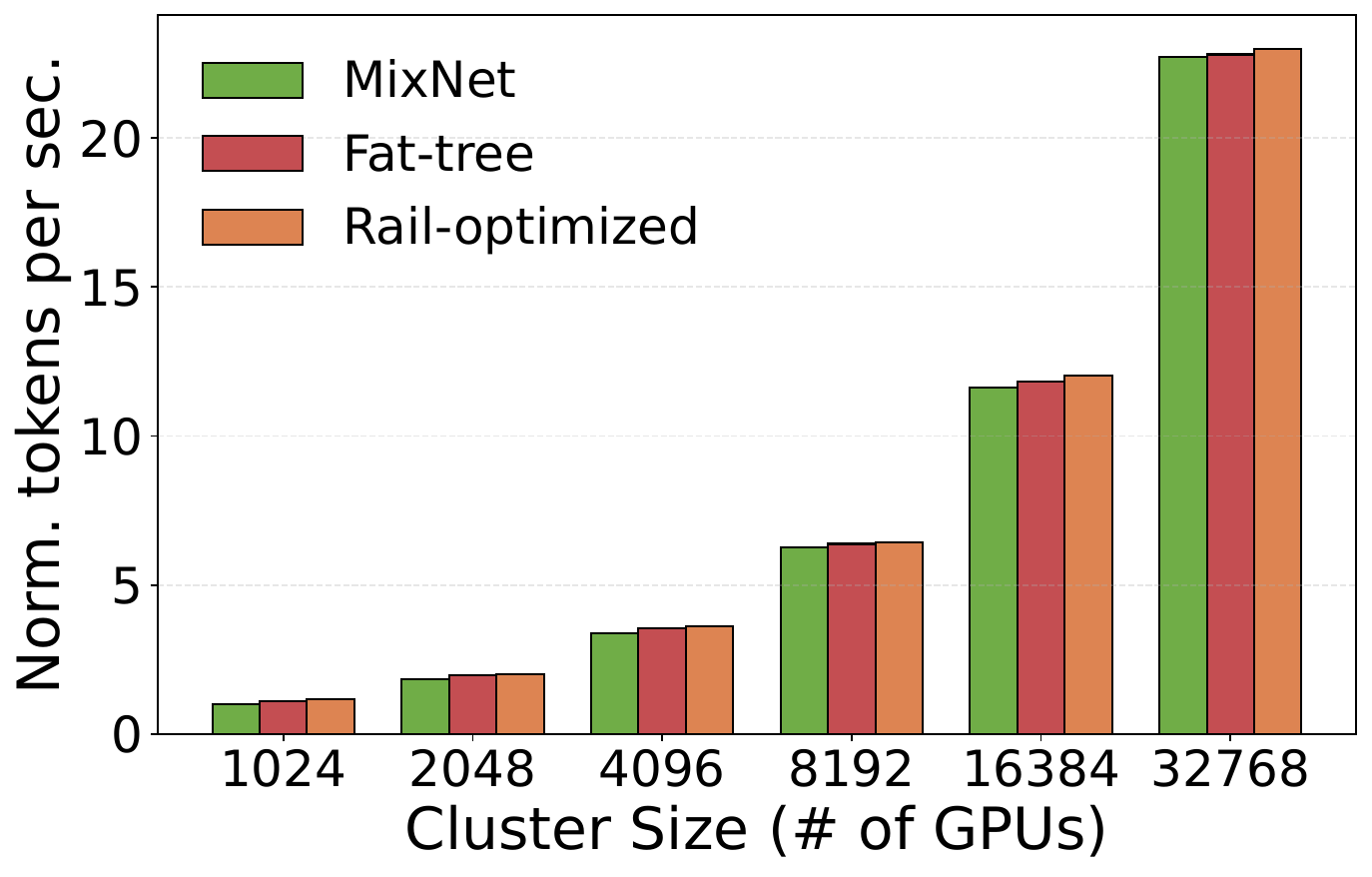}
        \caption{Performance comparison.}
        \label{fig:sim:scalability}
    \end{subfigure}
    \begin{subfigure}[t]{0.8\linewidth}
        \centering
        \includegraphics[width=\linewidth]{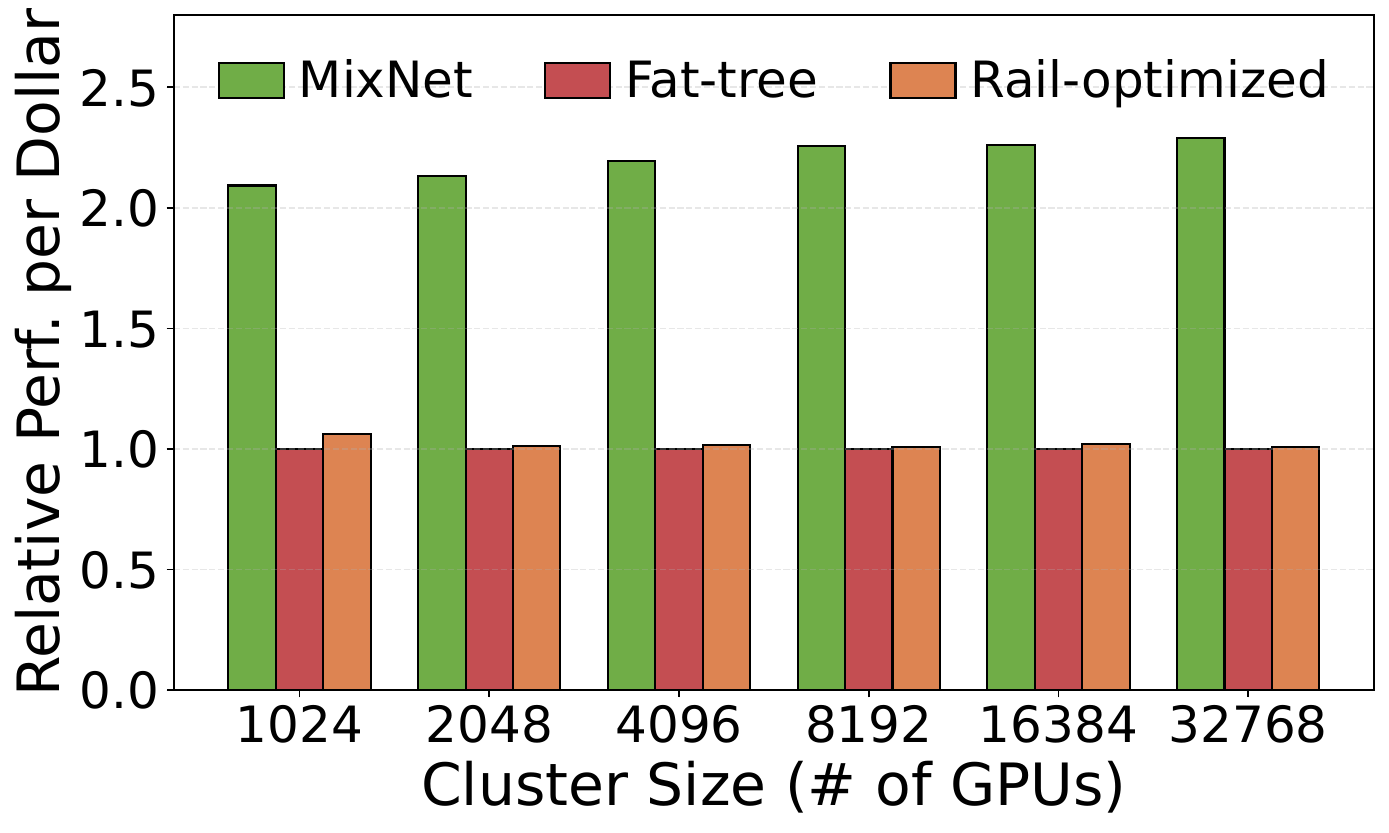}
        \caption{Performance-cost comparison.}
        \label{fig:sim:scalability-cost}
    \end{subfigure}
    \caption{[Simulation] Scalability analysis of \sysbf{} with different cluster sizes.}
\end{figure}

\begin{figure}[t!]
    \centering
    \includegraphics[width=0.8\linewidth]{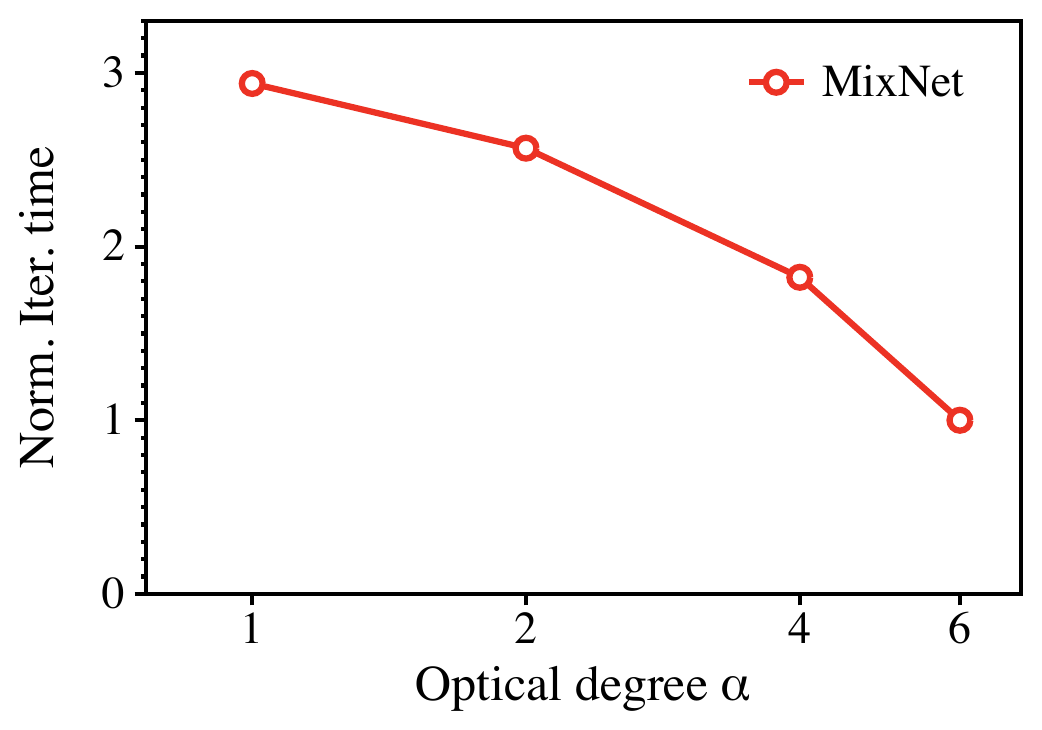}
    \caption{[Simulation] Impact of optical degree $\alpha$ in \sysbf{}.}
    \label{fig:sim:optical-degree}
\end{figure}

\subsection{Scalability}
We demonstrate the scalability of \sys{} in \figref{fig:sim:scalability}. The \mixtral{} model is evaluated at 400 Gbps bandwidth, with the cluster size varying from 128 servers to 4,096 servers, covering up to 32768 GPUs. 
\sys{} demonstrates scalability by fundamentally relaxing the port limits of OCS through the design of several decentralized regionally reconfigurable domains, allowing it to scale similarly to a fat-tree topology. 
Our results show that \sys{} scales effectively with increasing number of GPUs, achieving training throughput comparable to both non-blocking Fat-tree and Rail-optimized topologies in terms of tokens processed per second.
We further present the performance-cost comparison in \figref{fig:sim:scalability-cost}, which shows that \sys{} consistently achieves a superior performance-cost trade-off—approximately 2$\times$ higher performance-per-dollar—compared to Fat-tree and Rail-optimized topologies as the number of GPUs increases. This suggests that \sys{} maintains the training cost-effectiveness even as the cluster size grows.

\subsection{Impact of Optical Degree}
We show the impact of the optical degree on \sys{}'s performance in \figref{fig:sim:optical-degree}. We evaluate the \mixtrallarge{} model on a cluster of 128 servers with 100 Gbps link bandwidth. The optical degree $\alpha$ in \sys{} is varied to adjust its connectivity in the OCS. 
We reduce the bandwidth of each electronic port when increasing their number, to ensure a cost-equivalent comparison.
Our findings show that, as the optical degree increases, \sys{} further reduces iteration time, as more communication-intensive GPU pairs can be provisioned with dedicated high-bandwidth optical circuits.

\subsection{Impact of Reconfiguration Latency}
\begin{figure}[t!]
    \centering
    \includegraphics[width=0.8\linewidth]{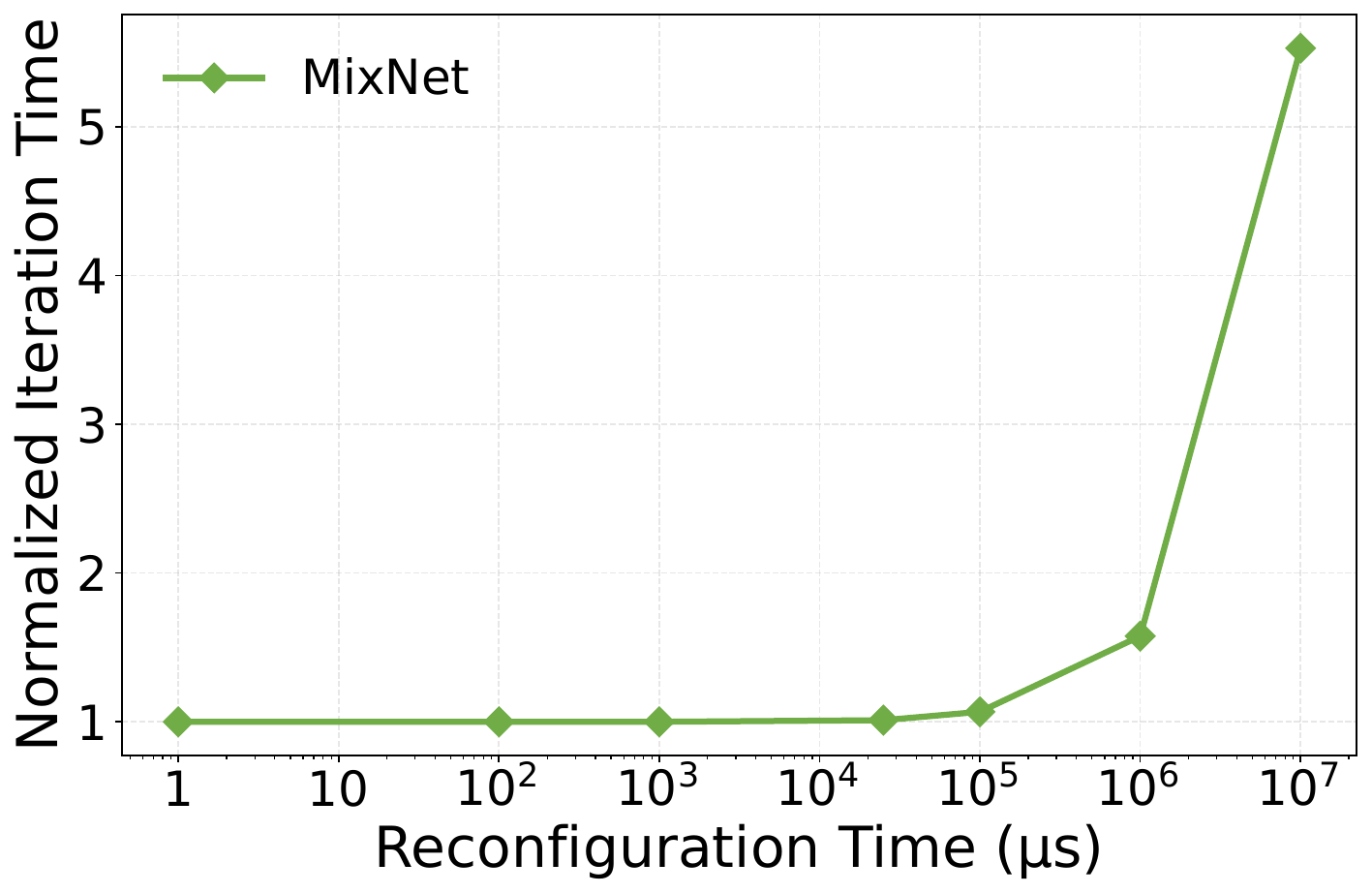}
    \caption{[Simulation] Impact of reconfiguration latency.}
    \label{fig:sim:ocs-reconfig}
\end{figure}

To investigate \sys{}'s sensitivity to OCS reconfiguration latency, we evaluate the \mixtrallarge{} model on a cluster of 128 servers with 400 Gbps link bandwidth, varying the reconfiguration latency from 1 µs to 10 s. 
\figref{fig:sim:ocs-reconfig} shows the normalized iteration time. \sys{} assumes the use of a millisecond-scale reconfigurable OCS (25 ms) in its current implementation. 
We observe that further reducing reconfiguration latency does not yield significant performance gains, as the OCS reconfiguration process can already be fully hidden during the computation phase. However, provisioned with microsecond-scale reconfigurable OCS, \sys{} can enable fully accurate topology reconfigurations for the first all-to-all communication in the forward pass (FP), resulting in marginal performance improvements in this specific phase.
On the other hand, when the reconfiguration latency exceeds 1000 ms, performance degrades obviously, as the OCS reconfiguration may not be hidden and starts to block the training process. As a result, \sys{} does not perform well with second-scale reconfigurable OCS systems.

\label{lastpage}

\end{document}